\documentclass[authoryear,preprint,review,11pt]{elsarticle}

\usepackage{amsfonts, amssymb, amsmath, amscd}
\usepackage{enumerate}
\usepackage{graphicx}
\usepackage[dvips]{epsfig}
\usepackage{color}
\usepackage{booktabs}
\usepackage{subfigure}
\usepackage{soul}

\journal{Experimental Thermal and Fluid Science}

\begin{document}

\begin{frontmatter}



\title{Bayesian experimental design for the active nitridation of graphite by atomic nitrogen}


\author{Gabriel Terejanu, Rochan R. Upadhyay, Kenji Miki}
\cortext[eep1]{Corresponding author: Gabriel Terejanu. E-mail address: terejanu@ices.utexas.edu. Address: ICES, The University of Texas at Austin, 1 University Station, C0200 Austin, TX 78712, USA.}

\address{Center for Predictive Engineering and Computational Sciences (PECOS),\\
Institute for Computational Engineering and Sciences (ICES),\\
The University of Texas at Austin}

\begin{abstract}
The problem of optimal data collection to efficiently learn the model parameters of a 
graphite nitridation experiment is studied in the context of Bayesian analysis using both 
synthetic and real experimental data. The paper emphasizes that the optimal design can be 
obtained as a result of an information theoretic sensitivity analysis. Thus, the preferred 
design is where the statistical dependence between the model parameters and observables is 
the highest possible. In this paper, the statistical dependence between random variables 
is quantified by mutual information and estimated using a $k$ nearest neighbor based 
approximation. It is shown, that by monitoring the inference process via measures such as 
entropy or Kullback-Leibler divergence, one can determine when to stop the data collection 
process. The methodology is applied to select the most informative designs on both a 
simulated data set and on an experimental data set, previously published in the literature. 
It is also shown that the 
sequential Bayesian analysis used in the experimental design can also be useful in 
detecting conflicting information between measurements and model predictions.
\end{abstract}


\end{frontmatter}


\section{Introduction}

The paper examines the problem of optimal data collection in order to calibrate
a graphite nitridation model. The calibration of the model is done 
in the context of Bayesian framework, where the probability distribution of the 
uncertain parameters is inferred from observations using Bayes' theorem. Here,
the goal of the experimental design is to select the optimal values for the control
variables which can be tuned by the experimentalist such that the post-experimental
uncertainty in the model parameters is reduced. 

Since pre-experimental decisions have to be made before any measurements are taken, it 
is convenient to frame the design problem also in the Bayesian framework, in order to 
average over the unknown future observations. A comprehensive and unified view of 
Bayesian experimental design is given in \cite{Chaloner1995}. In this paper, the 
optimal design is found to maximize the expected utility when the purpose of the experiment
is to estimate the parameters of the mathematical model. Thus the utility function employed
in this work is based on Shannon's measure of information, \cite{Shannon1948}, which reflects 
our goal for parameter estimation as described by \cite{Lindley1956}. Other utility functions 
can be tailored when the purpose of the experiment is prediction, hypothesis testing or 
mixed objectives including minimizing the financial cost of the experiment.

The proposed approach is applied to a sequential design when the acquisition of the data
can be done in the context of experiments on-demand. Such a sequential approach
is described by \cite{Fedorov1972} and \cite{Loredo2003} which clearly identify the three 
stages of the experimental process: experiment - inference - design. The data already collected is used 
to update the probability density function (pdf) of the parameters, and the result of this 
inference can be further used in identifying the next design which best resolves our 
questions of interest in the model. After the experiment is performed, the cycle continues
with a new inference step, followed by a design step and so on until the experimental
objective is reached. 

Due to the high computational complexity of the Bayesian experimental design, its 
adoption is rather low especially when dealing with nonlinear models. A large body of work
can be found in the literature on optimal design for linear or linearized models under
the name of Bayesian alphabetic criteria, see \cite{DasGupta1995}. However in the recent
years a revived interest in Bayes optimal designs for nonlinear models can be attributed
to Bayesian recursive update, efficient estimators of information-theoretic measures
and to 
Markov chain Monte Carlo (MCMC) algorithms which can efficiently sample complicated 
posterior distributions. In \cite{Muller1999}, the author makes use of MCMC to 
create a new simulation-based design, approach which jointly samples from an artificial 
probability model on the design, data and parameters.

By exploiting the independence property of the noise from the design, \cite{Sebastiani1997} 
extended the maximum entropy sampling proposed by \cite{Shewry1987} to estimation problems.
They show that the experiment which provides the maximum amount of information for
model parameters is the one for which the predictive distribution has the largest entropy. 
In other words, the most informative experiment is the one where we know the least. A similar
information-theoretic approach is presented by \cite{Mehr2002}, which maximizes the
entropy of Gaussian priors. These methods are aligned with previous ideas found in \cite{Lindley1956}
of incorporating information-theoretic approaches in the experimental design process.

The main contribution of this paper is to emphasize the intuitive interpretation of the Bayesian
experimental design as an information-theoretic sensitivity analysis, methodology which is 
applied to an engineering problem for which real experimental measurements exist. It is shown that the 
optimal design for estimation problems is the one which maximizes the mutual information of 
the parameters and the future observations. Given the connection between mutual information 
and copula functions used to model statistical dependencies, see \cite{Calsaverini2009}, this 
new interpretation of Bayesian experimental design reveals that optimal sampling for parameter 
estimation can be yielded by an information-theoretic sensitivity analysis, see also Appendix B. 
Thus the optimal 
design occurs where the statistical dependence between observables and parameters is 
maximized. In contrast to the maximum entropy sampling, no assumptions are made about the 
functional dependence of the entropy of conditional distribution on the design. 

In the inference stage, we use an adaptive MCMC algorithms proposed by \cite{Cheung_2009C} 
to obtain samples from the posterior distribution 
of model parameters. Estimators based on $k$-nearest neighbor are used to compute the information 
theoretic measures required in the design stage. The use of these estimators is advantageous when 
only samples are available to describe the underlying distributions. The mutual information is estimated 
using Kraskov's approach, see \cite{Kraskov2004}, which extends the $k$-nearest neighbor based 
estimator for differential Shannon entropy developed by \cite{Kozachenko1987}. 

While papers on experimental design can be found in geoscience - \cite{Guest2009}, neuroscience 
- \cite{Paninski2005}, biomedical applications - \cite{Clyde1996}, \cite{Chung2011}, \cite{Horesh2011}, 
engineering - \cite{Tucker2008}, just to name a few, 
the application of Bayesian design principles to actual experiments still lags far behind 
the theoretical advancements, see \cite{Chaloner1995}. In \cite{Curtis2000}, the authors compare 
the small number of papers published on average per year on experimental design to the large 
number of papers published on inverse methods, emphasizing the disconnect between the amount of 
data and the amount of information contained in the data. A more efficient learning of model
parameters can be accomplished by using Bayesian experimental design which tightly couples the
computational modeling, experimental endeavors and data analysis.

In this work, we consider an experiment for the nitridation of graphite conducted by  \cite{Marschall_nitridation_AIAA}.
The main objective of the experiment is to measure the reaction rate of the graphite with active nitrogen. 
This quantity is of great importance for assessment of the effectiveness of the thermal protective system
of space vehicles. The main parameter of interest is the reaction probability of the graphite nitridation reaction. 
The values of this quantity presented in the literature (\cite{Park-Bogdanoff_Nitridation_paper}, 
\cite{Goldstein_paper}, \cite{Japanese_Nitridation_paper}, \cite{Marschall_nitridation_AIAA}) vary by several orders of magnitude. 
The experiments reported in \cite{Marschall_nitridation_AIAA} have a detailed description of the scenarios with readily available data. 
The authors conducted a series of runs at different conditions that are not based on experimental design considerations. 
Hence the effectiveness of a rigorous experimental design strategy can be demonstrated. 

In Section \ref{sec:BayesExpDes} the problem of optimal experimental design is described in 
the Bayesian framework. The experimental setup for the nitridation of graphite is detailed 
in Section \ref{Expdata}, followed by the model description in Section \ref{sec:model}.
The numerical results for both simulated data and real experimental data are presented
in Section \ref{sec:result}, and the concluding remarks are given in Section~\ref{Conclusion}.

\section{Description of the experimental design}
\label{sec:BayesExpDes}

Given a set of observations $D_{n-1}=\{\tilde{\mathbf{d}}_1,\tilde{\mathbf{d}}_2, \ldots, \tilde{\mathbf{d}}_{n-1}\}$ 
we are concerned with finding the next experimental design $\boldsymbol{\xi}_n \in \boldsymbol{\Xi}$ 
such that the model parametric uncertainty is reduced after the experiment is performed and 
the associated measurement data $\tilde{\mathbf{d}}_n$ is collected. The mathematical models used in 
this paper are generally represented by the following abstract model:
\begin{eqnarray}
\mathbf{r}(\mathbf{u},\boldsymbol{\theta},\boldsymbol{\xi},\boldsymbol{\epsilon}_s) & = & \mathbf{0} \label{eq:model_system} \\
\mathbf{d} & = & \mathbf{d}(\mathbf{u},\boldsymbol{\theta},\boldsymbol{\xi},\boldsymbol{\epsilon}_m) \label{eq:discrepancy_model}
\end{eqnarray}
Here $\mathbf{u}$ is the state of the system which obeys the governing equations defined by
$\mathbf{r}(\cdot)$, $\boldsymbol{\xi} \in \boldsymbol{\Xi}$ is the control variables associated
with the experimental scenario, $\boldsymbol{\Xi}$ is the design space and $\boldsymbol{\theta} \in \Theta$
are the model parameters, where $\Theta$ is the parameter space of the model. Here,
the model prediction $\mathbf{d}_n$ calculated using the measurement model $\mathbf{d}(\cdot)$ 
is comparable with the experimental data $\tilde{\mathbf{d}}_n$ for a particular scenario input 
$\boldsymbol{\xi}_n$. The random vector or random field $\boldsymbol{\epsilon}_s$ captures 
any stochastic forcing present in the governing equations such as unmodeled dynamics
and $\boldsymbol{\epsilon}_m$ models the discrepancy between model predictions and 
experimental data.


\subsection{Bayesian experimental design}

In the followings it is assumed that we can afford to perform up to $N$ experiments
to obtain the desired measurements. Preferably, one will want to perform as few informative
experiments as possible in order to calibrate the model.
The experimental design process employed in this paper
generates an optimal sequence of designs such that the information gain is maximized
after each experiment. More about this sequential process of designing experiments 
can be found in \cite{Fedorov1972}. Note that since this decision process is sequential, 
and the next design depends on the previous experiments, it does not guarantee to find the 
optimal $N$ designs among all possible combinations of $N$ designs. This would correspond 
to a batch update of the prior parametric uncertainty and an expensive combinatorial 
decision process, which would require a large number of Monte Carlo samples to estimate 
the necessary expected utilities for each possible combination of designs.

\begin{figure}
  \begin{center}
    \includegraphics[width=4.0in]{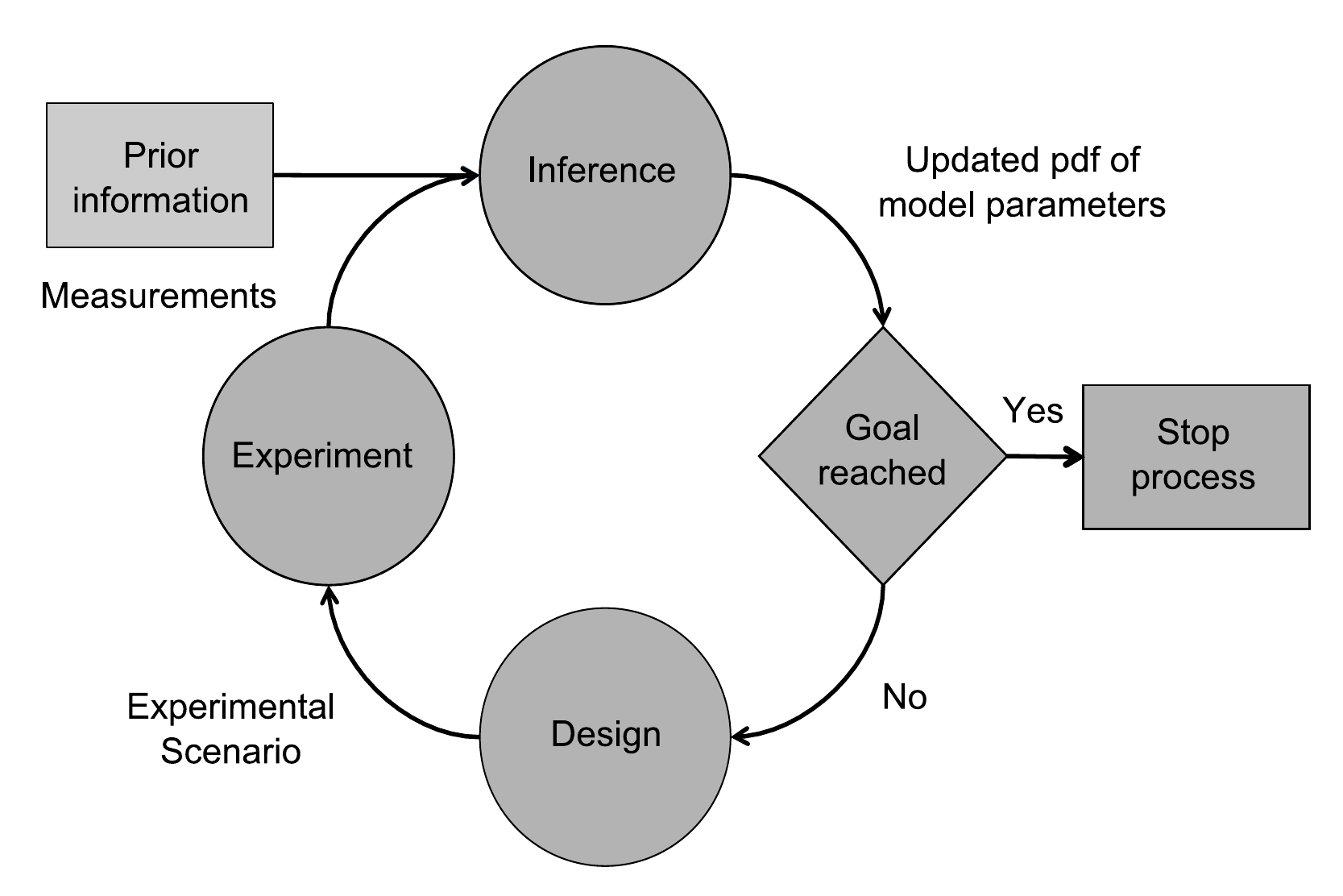}
  \end{center}
  \caption{Schematic of sequential experimental design process}
  \label{fig:exp_design}
\end{figure}

The entire experimental process is divided in three stages: experiment stage, inference
stage and, design stage. In the experiment stage, new data is collected according to the 
strategy obtained in the previous design stage. Initially, this data can also be obtained
from any experiments documented in the literature that are related with the underlying modeling 
problem. In the inference stage, the newly obtained experimental data is used to update the 
prior pdf of the model parameters using Bayes rule. The resulting pdf is further used to 
predict the distribution of the future observations for a variety of scenarios which cover 
the design space. The best design is chosen by maximizing the expected Shannon information 
gain, expectation taking with respect to the distribution of future data. The experimental 
process is depicted in Fig.\ref{fig:exp_design}.


\subsection{Bayesian inference step}

The inference stage consists in applying Bayes's theorem to calculate the distribution of the model
parameters conditioned on all the available data, $p(\boldsymbol{\theta}|D_n)$. 
Using the conditioning rule, one can derive the following recursive Bayesian update:
\begin{eqnarray}
p(\boldsymbol{\theta}|D_n) &=& \frac{p(D_n | \boldsymbol{\theta})p(\boldsymbol{\theta})}{p(D_n)} \nonumber \\
&=& \frac{p(\tilde{\mathbf{d}}_n, D_{n-1} | \boldsymbol{\theta})p(\boldsymbol{\theta})}
    {p(\tilde{\mathbf{d}}_n, D_{n-1})} \nonumber \\
&=& \frac{p(\tilde{\mathbf{d}}_n | D_{n-1} , \boldsymbol{\theta}) p(D_{n-1} | \boldsymbol{\theta}) p(\boldsymbol{\theta})}
    {p(\tilde{\mathbf{d}}_n | D_{n-1}) p(D_{n-1})} \nonumber \\
&=& \frac{p(\tilde{\mathbf{d}}_n| D_{n-1}, \boldsymbol{\theta},\boldsymbol{\xi}_n) p(\boldsymbol{\theta}|D_{n-1})}
    {p(\tilde{\mathbf{d}}_n | D_{n-1}, \boldsymbol{\xi}_n)}
\end{eqnarray}

Under the assumption of conditionally independent measurements given model parameters, then the above expression
can be simplified using the following relation: $p(\tilde{\mathbf{d}}_n| D_{n-1}, \boldsymbol{\theta},\boldsymbol{\xi}_n) = 
p(\tilde{\mathbf{d}}_n|\boldsymbol{\theta},\boldsymbol{\xi}_n)$. The denominator is given by,
\begin{equation}
p(\tilde{\mathbf{d}}_n | D_{n-1}, \boldsymbol{\xi}_n) = \int_{\Theta} p(\tilde{\mathbf{d}}_n|\boldsymbol{\theta},\boldsymbol{\xi}_n)
    p(\boldsymbol{\theta}|D_{n-1}) \mathrm{d} \boldsymbol{\theta}
\end{equation}
and quantifies the evidence provided by the new experimental data, $\tilde{\mathbf{d}}_n$, 
in support of our model conditioned on all the previous measured data. The use of Bayesian 
inference in the sequential experimental design is advantageous as it permits the iterative 
accumulation of information. The posterior distribution obtained in this stage becomes the 
prior distribution in the next stage of inference. After all the $N$ experiments have been 
performed, the last posterior distribution summarizes the information contained in all the 
available measurements.

Instead of performing $N$ experiments, the data collection process can be stopped earlier, 
when precision measures computed in the interim stages, such as the determinant of the sample 
covariance matrix, satisfy the thresholds predefined by the user. The process can also be 
ceased when the rate of reducing the uncertainty in the parameters has slowed enough or when
indication exists that model predictions and the corresponding experimental observations
are in disagreement.

The inverse problem of calibrating the model parameters from the measurement data is solved
using Markov chain Monte Carlo simulations. In our simulations, samples from the posterior 
distribution $\boldsymbol{\theta} \sim p(\boldsymbol{\theta}|D_n)$ are obtained using the
statistical library QUESO \cite{PrSc11} equipped with the Hybrid Gibbs
Transitional Markov Chain Monte Carlo method proposed by 
\cite{Cheung_2008B, Cheung_2009C}.


\subsection{Optimal experimental design step}

According to \cite{Lindley1956} when the objective of the experiment is to learn about
the model parameters $\boldsymbol{\theta}$, then the utility function is given by the
amount of information provided by the measurement $\tilde{\mathbf{d}}_n$ as result of 
performing the experiment with the design $\boldsymbol{\xi}_n$. Since $\tilde{\mathbf{d}}_n$
has not yet been observed, and we have to make a decision regarding the design $\boldsymbol{\xi}_n$
prior to the experiment, in the following, we are using the unknown measurement $\mathbf{d}_n$ 
and its corresponding predictive distribution.
\begin{eqnarray} \label{utilFunc}
U(\mathbf{d}_n,\boldsymbol{\xi}_n) &=& \int_{\boldsymbol{\Theta}} p(\boldsymbol{\theta}|\mathbf{d}_n, D_{n-1})
    \log p(\boldsymbol{\theta}|\mathbf{d}_n,D_{n-1}) \mathrm{d}\boldsymbol{\theta} \nonumber \\
&& - \int_{\boldsymbol{\Theta}} p(\boldsymbol{\theta}|D_{n-1}) \log p(\boldsymbol{\theta}|D_{n-1}) 
    \mathrm{d}\boldsymbol{\theta}
\end{eqnarray}
Here, the prior pdf for model parameters is given by $p(\boldsymbol{\theta}|D_{n-1})$, and it 
is used to compute the following predictive distribution for the observables when the input scenario is
$\boldsymbol{\xi}_n$:
\begin{equation}
p(\mathbf{d}_n|D_{n-1},\boldsymbol{\xi}_n) = \int_{\Theta} p(\mathbf{d}_n|\boldsymbol{\theta},\boldsymbol{\xi}_n)
    p(\boldsymbol{\theta}|D_{n-1}) \mathrm{d} \boldsymbol{\theta}
\end{equation}

Since we are in the pre-experimental stage, we can compute the average amount of information
provided by an experiment $\boldsymbol{\xi}_n$ by marginalizing over all the unknown future
observations:
\begin{equation} \label{expUtil}
\mathrm{E}_{\mathbf{d}_n}[U(\mathbf{d}_n,\boldsymbol{\xi}_n)] = \int_{\mathcal{D}} U(\mathbf{d}_n,\boldsymbol{\xi}_n)
    p(\mathbf{d}_n|D_{n-1},\boldsymbol{\xi}_n) \mathrm{d} \mathbf{d}_n
\end{equation}

The optimal experiment is obtained by solving the following optimization problem:
\begin{equation}
\boldsymbol{\xi}_n^* = \arg\max_{\boldsymbol{\xi}_n \in \boldsymbol{\Xi}} 
    \mathrm{E}_{\mathbf{d}_n}[U(\mathbf{d}_n,\boldsymbol{\xi}_n)]
\end{equation}

By substituting Eq.\eqref{utilFunc} into Eq.\eqref{expUtil}, the expected utility can be 
written in the following form:
\begin{eqnarray}
\mathrm{E}_{\mathbf{d}_n}[U(\mathbf{d}_n,\boldsymbol{\xi}_n)] &=& \int_{\mathcal{D}}
    \int_{\boldsymbol{\Theta}} p(\boldsymbol{\theta},\mathbf{d}_n| D_{n-1},\boldsymbol{\xi}_n)
    \log \frac{p(\boldsymbol{\theta},\mathbf{d}_n| D_{n-1}, \boldsymbol{\xi}_n)}
    {p(\mathbf{d}_n|D_{n-1},\boldsymbol{\xi}_n)} 
    \mathrm{d}\boldsymbol{\theta} \mathrm{d}\mathbf{d}_n \nonumber \\
&& - \int_{\mathcal{D}} \int_{\boldsymbol{\Theta}} p(\boldsymbol{\theta},\mathbf{d}_n| D_{n-1}, \boldsymbol{\xi}_n) 
    \log p(\boldsymbol{\theta}|D_{n-1}) \mathrm{d}\boldsymbol{\theta} \mathrm{d}\mathbf{d}_n \\
&=& \int_{\mathcal{D}} \int_{\boldsymbol{\Theta}} p(\boldsymbol{\theta},\mathbf{d}_n| D_{n-1}, \boldsymbol{\xi}_n)
    \log \frac{p(\boldsymbol{\theta},\mathbf{d}_n| D_{n-1}, \boldsymbol{\xi}_n)}
    {p(\boldsymbol{\theta}|D_{n-1})p(\mathbf{d}_n|D_{n-1},\boldsymbol{\xi}_n)} 
    \mathrm{d}\boldsymbol{\theta} \mathrm{d}\mathbf{d}_n \nonumber \\
&=& \mathrm{I}(\boldsymbol{\theta};\mathbf{d}_n| D_{n-1}, \boldsymbol{\xi}_n)
\end{eqnarray}

Therefore the optimal experiment is the one which maximizes the mutual information between
the model parameters and the model predictions. In other words we would like to sample where 
the expected observations have the highest impact on model parameters.
\begin{align}
\boldsymbol{\xi}_n^* = \arg\max_{\boldsymbol{\xi}_n \in \boldsymbol{\Xi}} 
    \mathrm{I}(\boldsymbol{\theta};\mathbf{d}_n| D_{n-1}, \boldsymbol{\xi}_n)
\end{align}

From information theory, see, for example \cite{Cover1991}, mutual information quantifies
the reduction in uncertainty that knowing the model parameters provides about the model 
predictions and vice-versa. Being written as the Kullback-Leibler divergence between the 
joint distribution of parameters and predictions, and the product of their marginals, 
we can also say that mutual information provides a measure of statistical dependence between
the two random variables. This has been shown more formally in \cite{Calsaverini2009} by 
making the connection between mutual information and copula functions which are used to 
model the dependence between random variables. For completeness this connection is also
presented in Appendix B. It turns out that the mutual information 
is just the negative copula entropy. Hence, mutual information is independent on the marginal
distributions and it quantifies only the dependence information contained in the copula function.

In neuroscience, \cite{Paninski2005} uses the same Bayesian optimal design based on 
maximizing the mutual information between model parameters and model predictions. It is shown
that the information-maximization sampling is asymptotically more efficient than an
independent and identically distributed sampling strategy. Compared with maximum entropy
sampling for parameter estimation proposed by \cite{Sebastini2000}, the sampling based
on maximizing the mutual information is more general, no assumption about the independence
assumption is made about the discrepancy model with respect to the design.

Finally, in the design step, the experimental space $\boldsymbol{\Xi}$ is discretized and 
the corresponding marginal distributions of model predictions and parameters, as well as the 
joint pdf of the two are computed. The mutual information corresponding to
each experimental design is calculated as in Subsection \ref{subsec:estMI}, and the design
corresponding to the highest MI is selected for the next experiment. Here, samples from the 
joint distribution of parameters and model predictions are obtained using Monte Carlo 
simulations according to the following relation:
\begin{equation}
p(\mathbf{d}_n, \boldsymbol{\theta} | D_{n-1},\boldsymbol{\xi}_n) =
    p(\mathbf{d}_n|\boldsymbol{\theta},\boldsymbol{\xi}_n) p(\boldsymbol{\theta}|D_{n-1})
\end{equation}


\subsection{Estimating the mutual information}
\label{subsec:estMI}

The estimation of the mutual information in this paper is done using the $k$NN (Nearest Neighbor) method 
proposed by \cite{Kraskov2004}. This is based on the $k$NN estimator for the entropy 
proposed by \cite{Kozachenko1987}. The entropy of a random variable can be approximated 
from $N$ samples using Kozachenko-Leonenko's estimator:
\begin{align} \label{entropy}
\mathrm{H}(X) \approx -\psi(k) + \psi(N) + \frac{d_X}{N}\sum_{i=1}^N \log \epsilon(i)
\end{align}
where $\psi(\cdot)$ is the digamma function, $d_X$ is the dimensionality of the random
variable $X$, and $\epsilon(i)$ is the distance 
from sample $X_i$ to its $k$NN, $X_{k(i)}$, and is given by the maximum norm:
\begin{equation}
\epsilon(i) = \| X_i - X_{k(i)} \|_\infty
\end{equation}

This is a biased estimator due to the assumption of local uniformity of the density. By 
definition, the MI can be decomposed as the sum of marginal entropies minus the joint
entropy. Therefore, the above estimator can be used to approximate all three entropies and thus
the mutual information. However the errors made in estimating individual entropies will 
not cancel out in general. To partially alleviate this problem, Kraskov's approach is not to 
fix $k$ when estimating the marginal entropies. It can be shown, see \cite{Kraskov2004}, 
that the MI of the parameters and model prediction can be computed as follows:
\small
\begin{align}
\mathrm{I}(\boldsymbol{\theta};\mathbf{d}_n| D_{n-1}, \boldsymbol{\xi}_n) \approx 
    \psi(k) - \frac{1}{N} \sum_{i=1}^N \bigg( \psi(n_{\theta}(i)+1) + \psi(n_d(i)+1) \bigg) + \psi(N)
\end{align}
\normalsize
where $n_d(i)$ and $n_\theta(i)$ are the number of points in the marginal space within $\epsilon(i)$ distance from the $i$th sample.
Here, $\epsilon(i)$ is the distance from the $i$th sample to its $k$NN in the joint space.

Based on numerical studies, \cite{Khan2007} shows that the estimators based on $k$NN and 
kernel density estimation (KDE) outperform commonly used dependence 
measures such as linear 
correlation, cross-correlogram or Kendall's $\tau$. Furthermore, it is also shown that 
in general the $k$NN estimator captures better the nonlinear dependence than the KDE.
However the Kraskov's MI estimator has to be used with care. 
A small value for $k$ will result
in a small bias but a larger variance and vice-versa. Also, the efficiency of the estimator
decreases as the dimensionality of the joint space increases. 

For completeness, an example test
has been carried out to show the performance of the $k$NN estimator for the MI as function
of the dimensionality for the joint space, the number of the samples and the number
of $k$NN's. The joint pdf of two random variables of known dimensions
is modeled using a Gaussian density function for which we can compute the exact MI.
For each combination of joint dimensionality, $k$NN, and sample size, a number of
$1000$ trial runs have been used to generate samples according to the
Gaussian density function. The entries in the covariance matrix have been considered
to be $1$ on the diagonal and $0.9$ for the off-diagonal elements.
Using the true value of the MI and the estimated value, one
can compute the relative absolute error for each run. The result of the average 
absolute relative error over the $1000$ trial runs is shown in Table \ref{tbl:rmseMI}.


\subsection{Example $1$: Simulation experiment}
\label{subsec:example}

Consider the following nonlinear design problem:
\begin{equation} \label{ex1_model}
d = \theta_1 \xi + \theta_2 \xi^2 + f(\xi)\epsilon
\end{equation}

\begin{figure*}
\begin{center}
\subfigure[Initial response (additive noise)]{\includegraphics[width=2.4in]{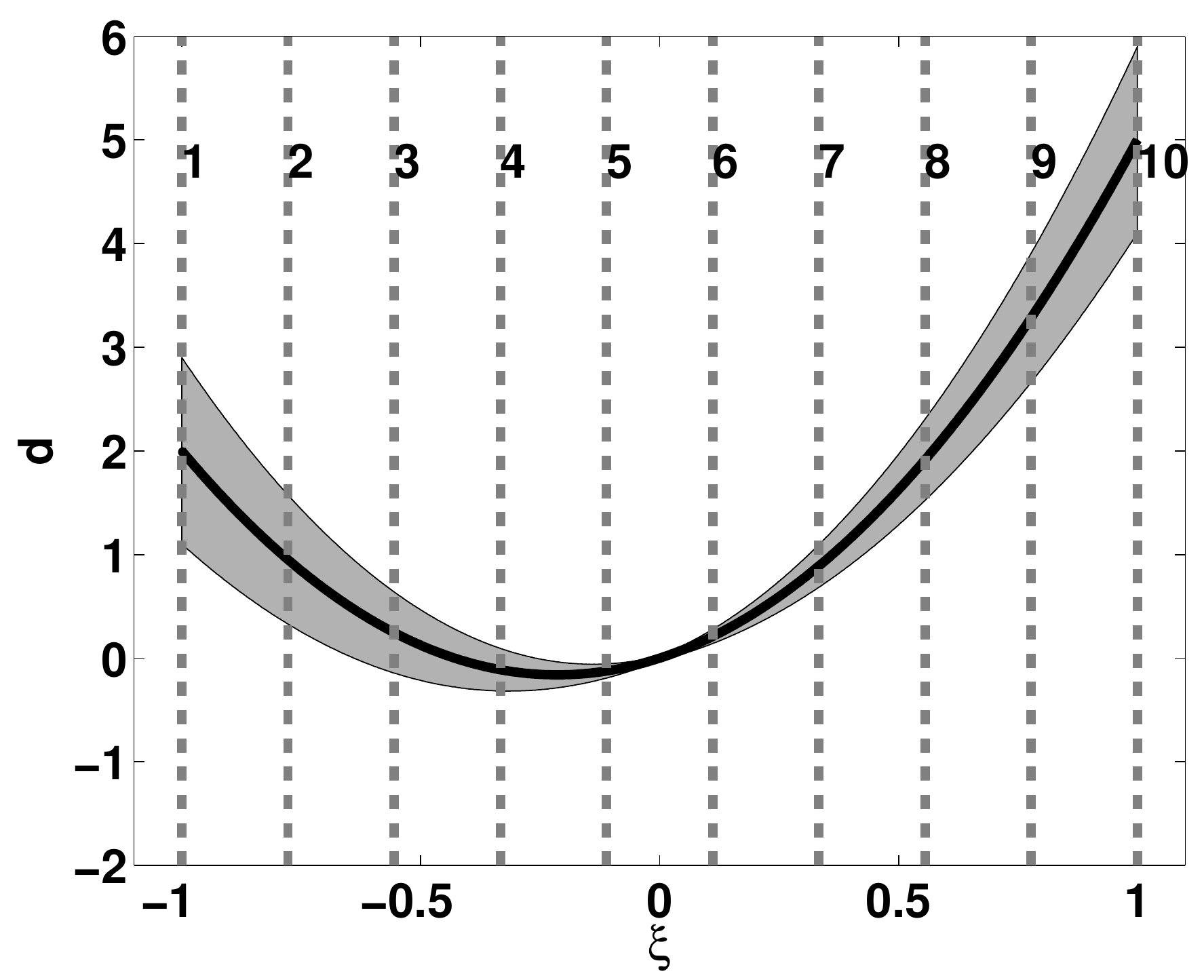}\label{fig:ex1_unc_add}} 
\subfigure[Initial response (multiplicative noise)]{\includegraphics[width=2.4in]{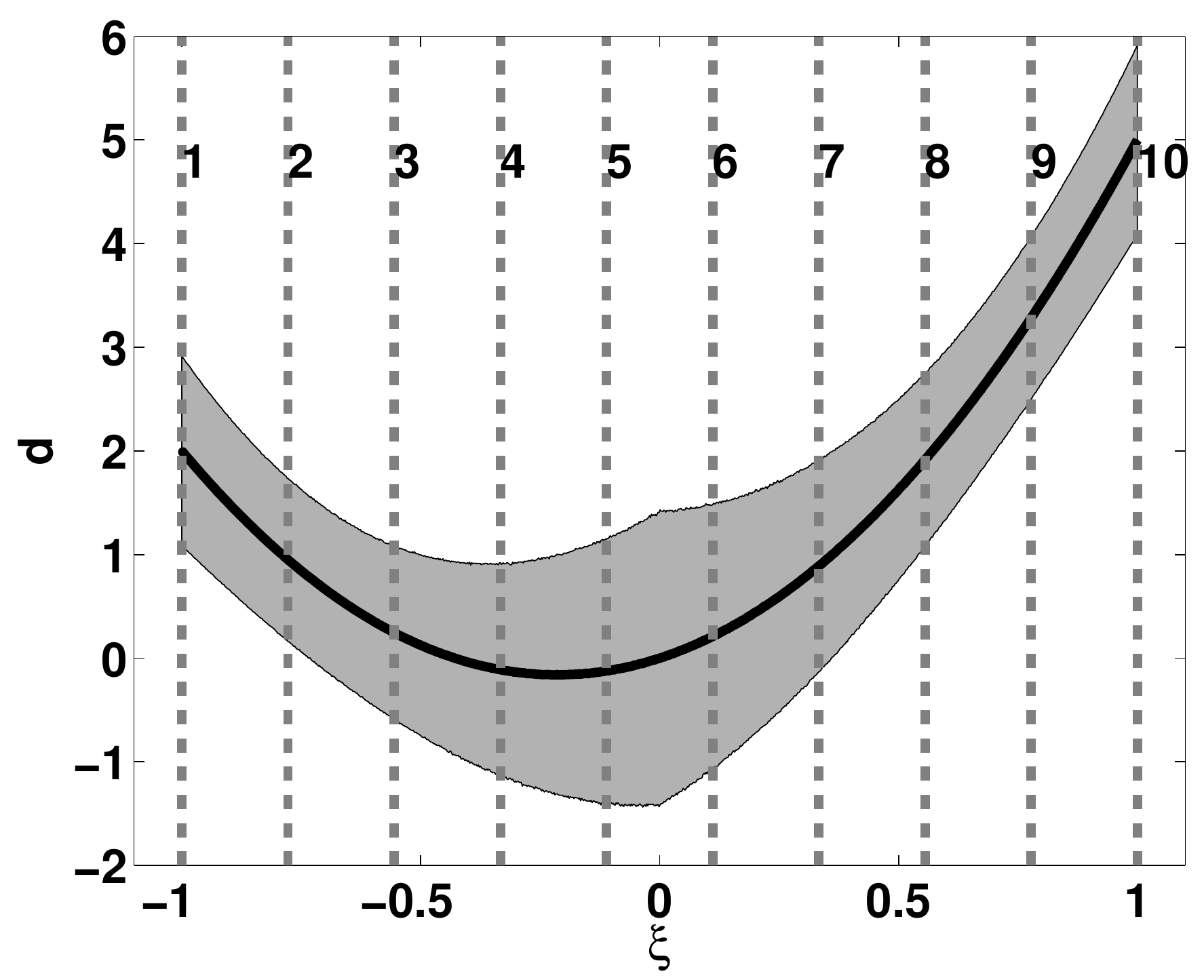}\label{fig:ex1_unc_mult}} 
\caption{Results for example $1$. Model response given initial discrepancy and parametric uncertainty.}
\end{center}
\end{figure*}

The goal of the experiment design in this example is to efficiently learn the model parameters,
$\theta_1$ and $\theta_2$, when the design space is given by $\Xi = [-1,1]$. The discrepancy model
is given by $f(\xi)\epsilon$, where $\epsilon$ is a standard normal random variable, $\mathcal{N}(0,1)$. 
The two parameters are uniformly distributed according to $\theta_1 \sim \mathcal{U}(1,2)$ and 
$\theta_2 \sim \mathcal{U}(3,4)$. Two cases are analyzed: 
additive noise when $f(\xi) = 0.01$ and multiplicative noise with respect to the design when 
$f(\xi) = 0.5(1.1 - |\xi|)$. Given the initial uncertainty of the parameters and the discrepancy function, 
the response of the model it is shown in Figs.\ref{fig:ex1_unc_add}-\ref{fig:ex1_unc_mult}. 
In this respect we want to compare the maximum entropy sampling 
with the information maximization sampling described in this paper. Hence, the two cost functions 
to be maximized in the design stage are,
\begin{itemize}
\item $J(\xi_n) = \mathrm{I}((\theta_1,\theta_2);d_n| D_{n-1}, \xi_n)$ for information maximization (IM) sampling
\item $J(\xi_n) = \mathrm{H}(d_n| D_{n-1}, \xi_n)$ for maximum entropy (ME) sampling 
\end{itemize}

The two sampling approaches, IM and ME, are also compared with a strategy based on random sampling (RND), 
which at each stage it randomly chooses a design. All the designs being equally likely at every stage. 

In this example we consider that up to $10$ experiments can be performed with scenario values
equally distributed in the design space as indicated by the dashed lines in Figs.\ref{fig:ex1_unc_add}-\ref{fig:ex1_unc_mult}. 
For each case corresponding to the discrepancy function, additive noise or multiplicative noise, two 
other cases are analyzed with respect to experimental sampling restriction. In the first case 
no repeated measurements are allowed. Thus, once an experimental design has been chosen from the
set of $10$ designs, it can no longer be used in the subsequent stages. The second case corresponds
to allowing repeated measurements in our calibration. 
Synthetic measurements
are created using Eq.\eqref{ex1_model}, where the true value of the parameters are set to
$\theta_1 = 1.5$ and $\theta_2 = 3.5$. A different noise sample has been generated for each
of the $10$ scenarios. For this example we have used $2000$ samples in computing the MI 
and the $kNN$ has been set to $6$.

For the two cases considered in this example (additive and multiplicative noise) 
each strategy will yield a sequence of designs. At each stage of the design, the performance of 
each strategy is given by the reduction in uncertainty of the parameters which is quantified by the entropy
of the parameter distribution. The entropy is computed here using Eq.\eqref{entropy} and for
comparison purposes, since we are generating synthetic measurements and use a random sampling
approach, the entropy is averaged over $100$ Monte Carlo runs.

\begin{figure*}
\begin{center}
\subfigure[Average entropy (additive noise - repeated measurements not allowed)]{\includegraphics[width=2.4in]{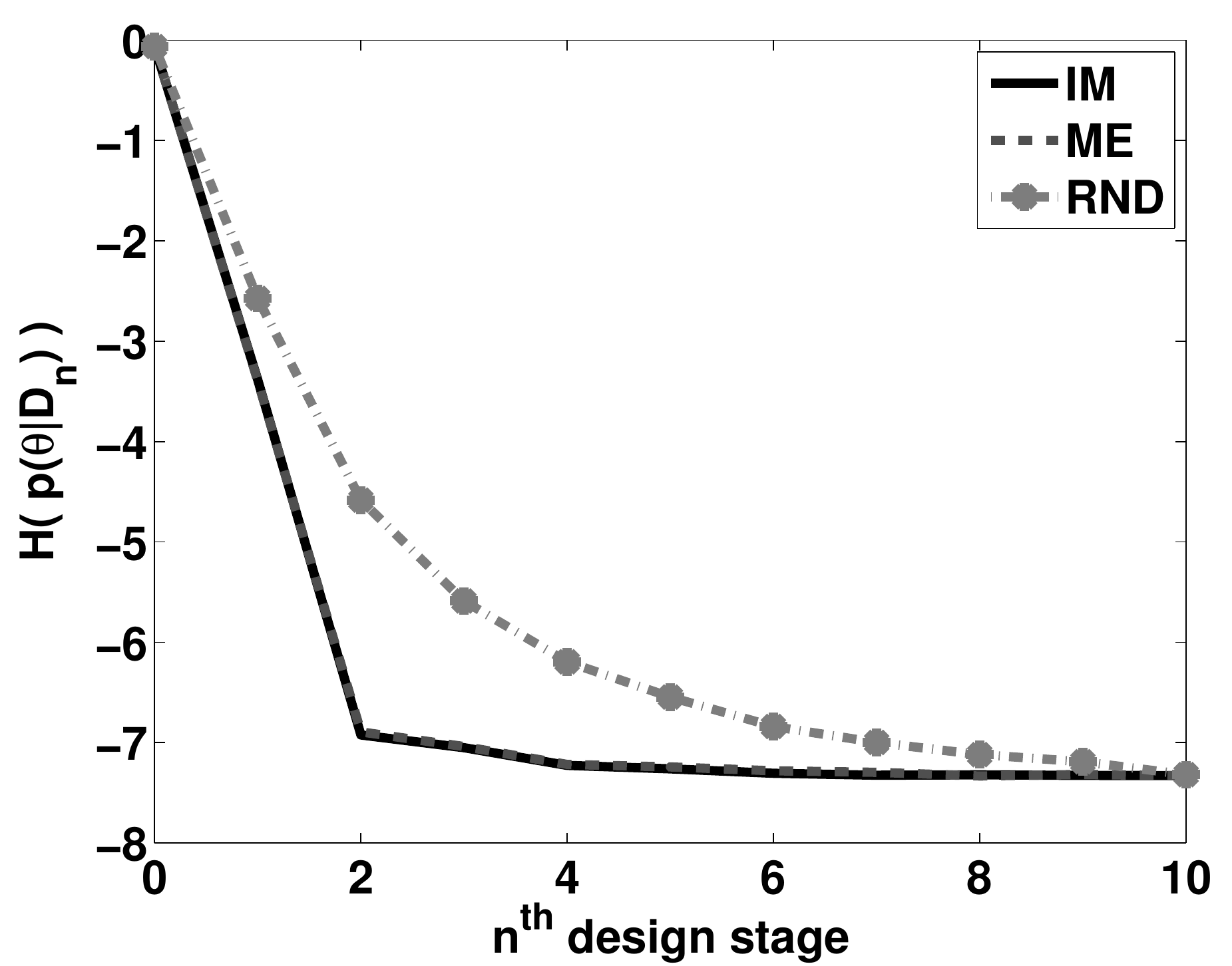}\label{fig:ex1_ent_add_no_rep}} 
\subfigure[Average entropy (multiplicative noise - repeated measurements not allowed)]{\includegraphics[width=2.4in]{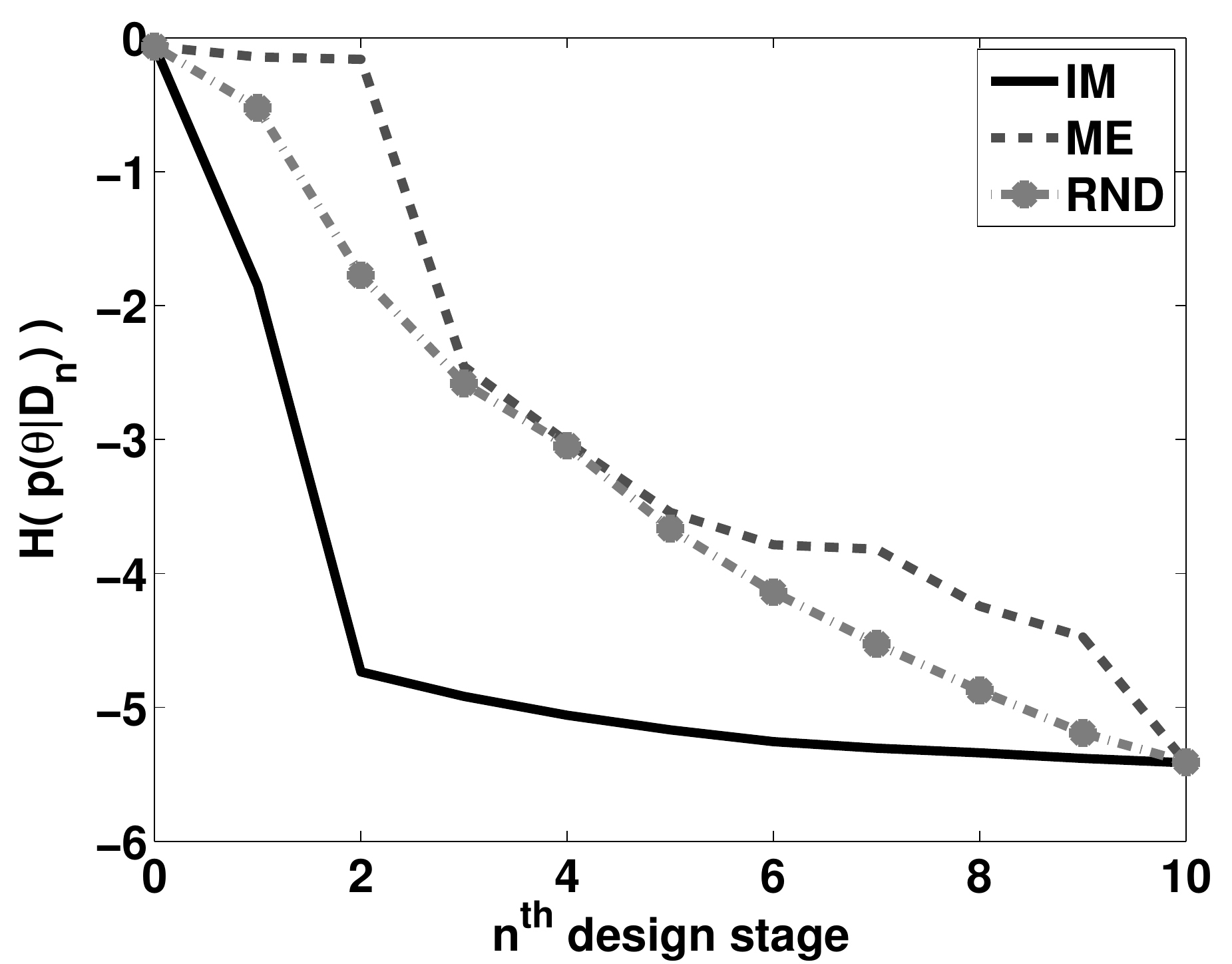}\label{fig:ex1_ent_mult_no_rep}} 
\subfigure[Average entropy (additive noise - repeated measurements allowed)]{\includegraphics[width=2.4in]{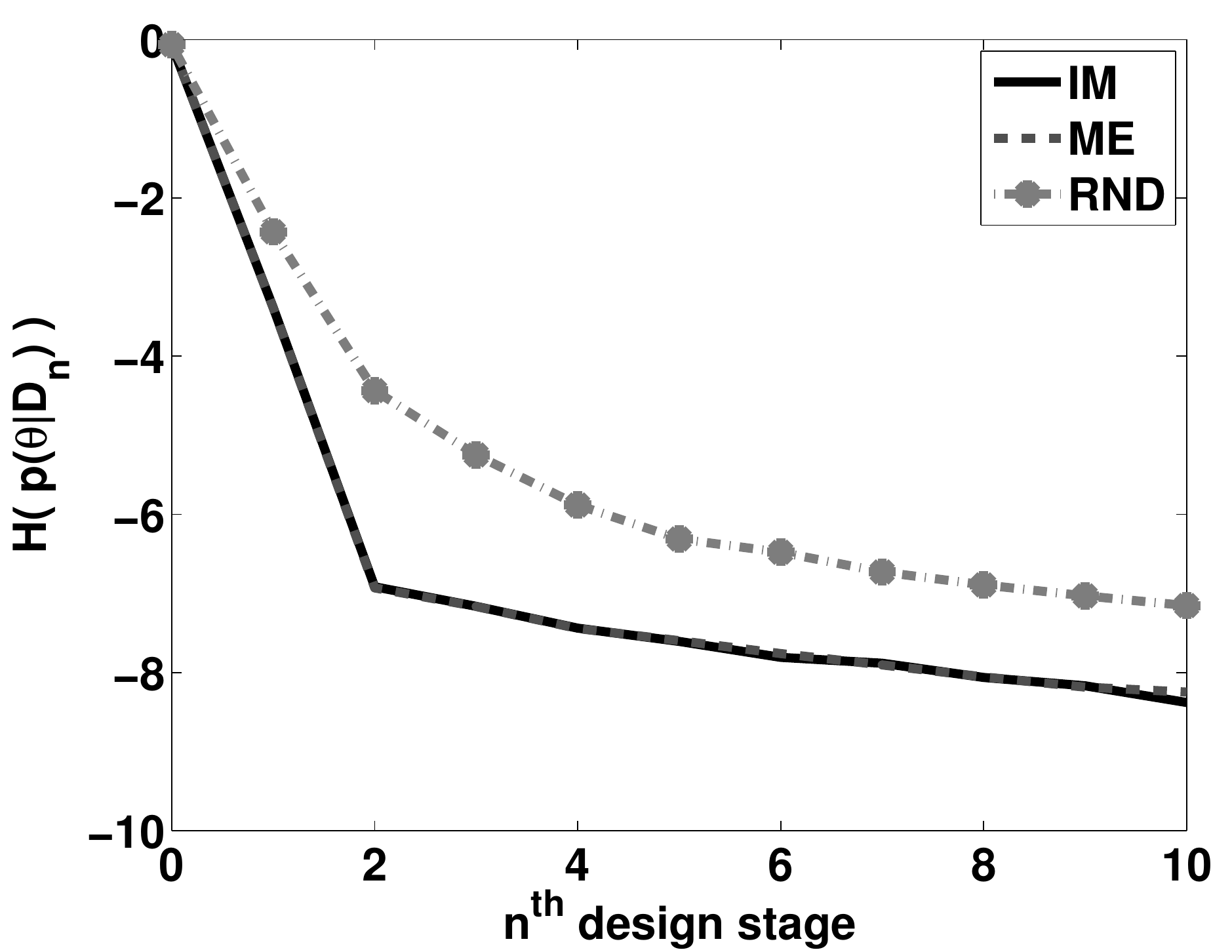}\label{fig:ex1_ent_add}} 
\subfigure[Average entropy (multiplicative noise - repeated measurements allowed)]{\includegraphics[width=2.4in]{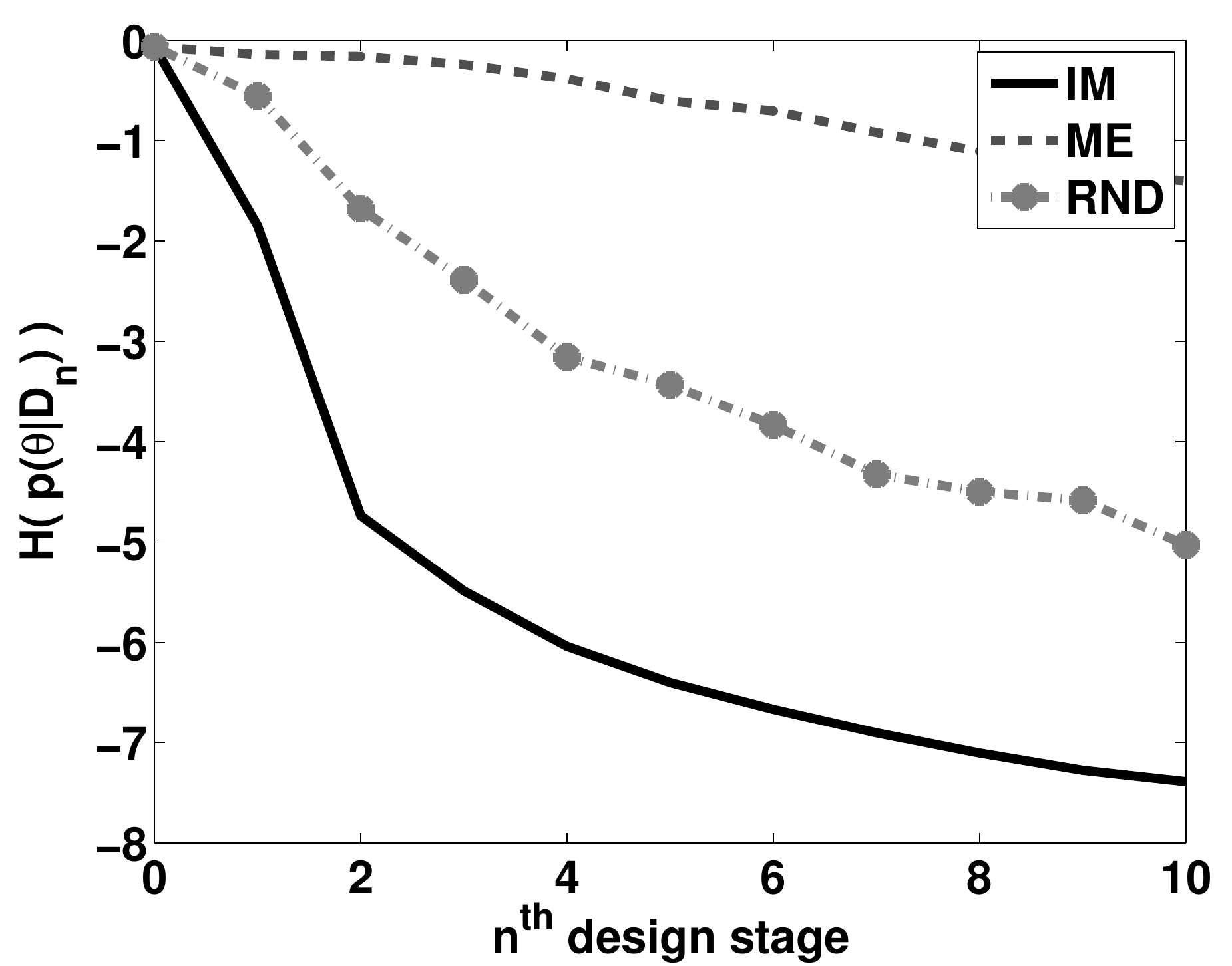}\label{fig:ex1_ent_mult}} 
\caption{The evolution of the entropy of parameter distribution averaged over $100$ Monte Carlo runs for the three strategies: 
IM - Information Maximization Sampling, ME - Maximum Entropy Sampling, RND - Random Sampling, and the two case: no repeated measurements
and repeated measurements.}\label{fig:ex11}
\end{center}
\end{figure*}

As expected, for the additive noise the maximum entropy sampling is equivalent to the information maximization sampling as it is shown in Fig.\ref{fig:ex1_ent_add_no_rep} and Fig.\ref{fig:ex1_ent_add}. Table \ref{tab:ex1_add_seq} shows that in this case the highest initial reduction in uncertainty is obtained by starting the sampling at the boundary of the design domain. For multiplicative noise however, due to the discrepancy model used, the maximum entropy is forced to start the sampling using the design points in the middle of the domain, where the uncertainty is the highest, see Table \ref{tab:ex1_mult_seq} and Fig.\ref{fig:ex1_unc_mult}. Because of the distribution of the noise, the performance of the maximum entropy strategy is worse on average even compared with the random sampling.

\begin{figure*}
\begin{center}
\subfigure[$p(\theta_1,D_2)$ (additive noise)]{\includegraphics[width=2.4in]{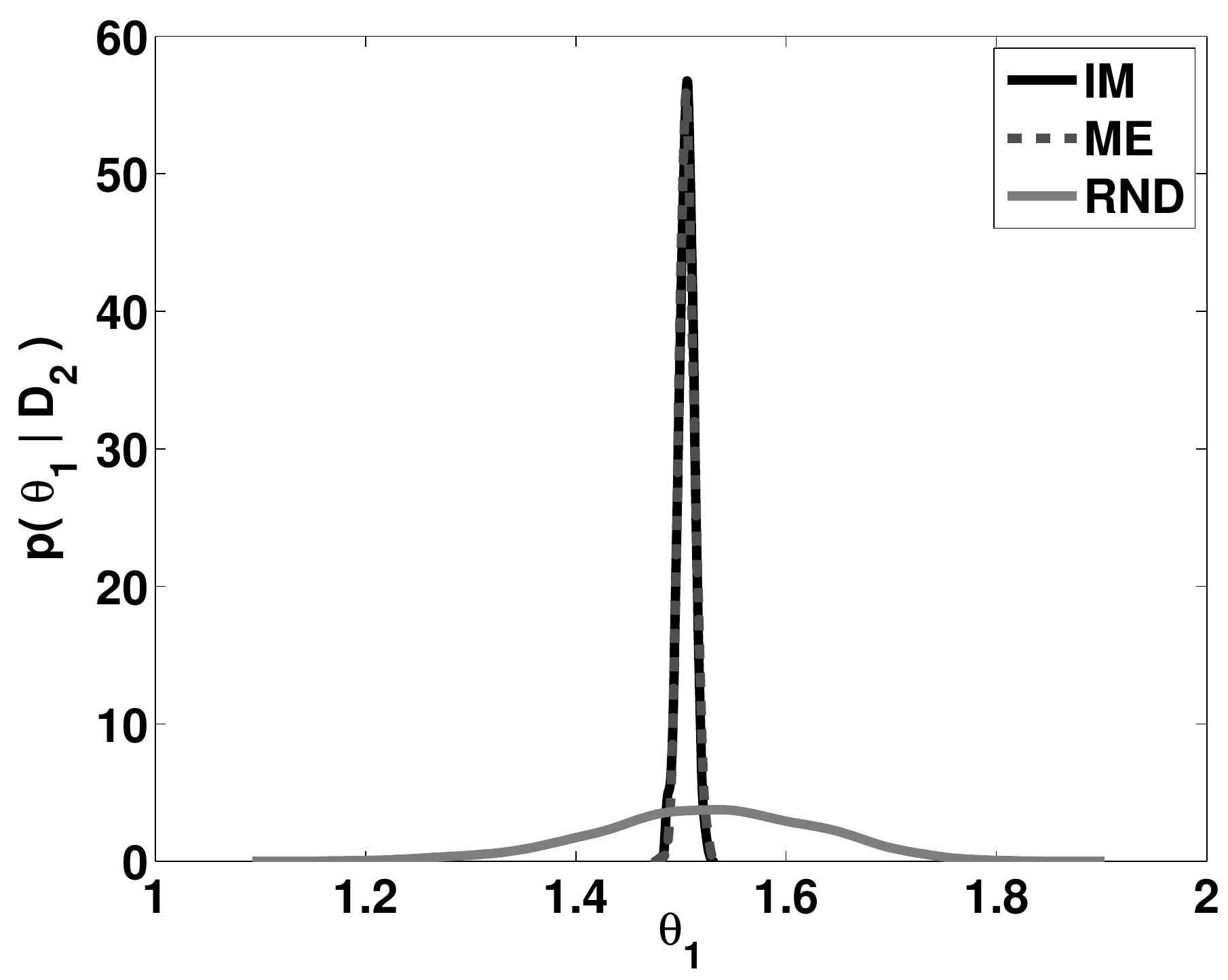}\label{fig:ex1_add_theta1}} 
\subfigure[$p(\theta_1,D_2)$ (multiplicative noise)]{\includegraphics[width=2.4in]{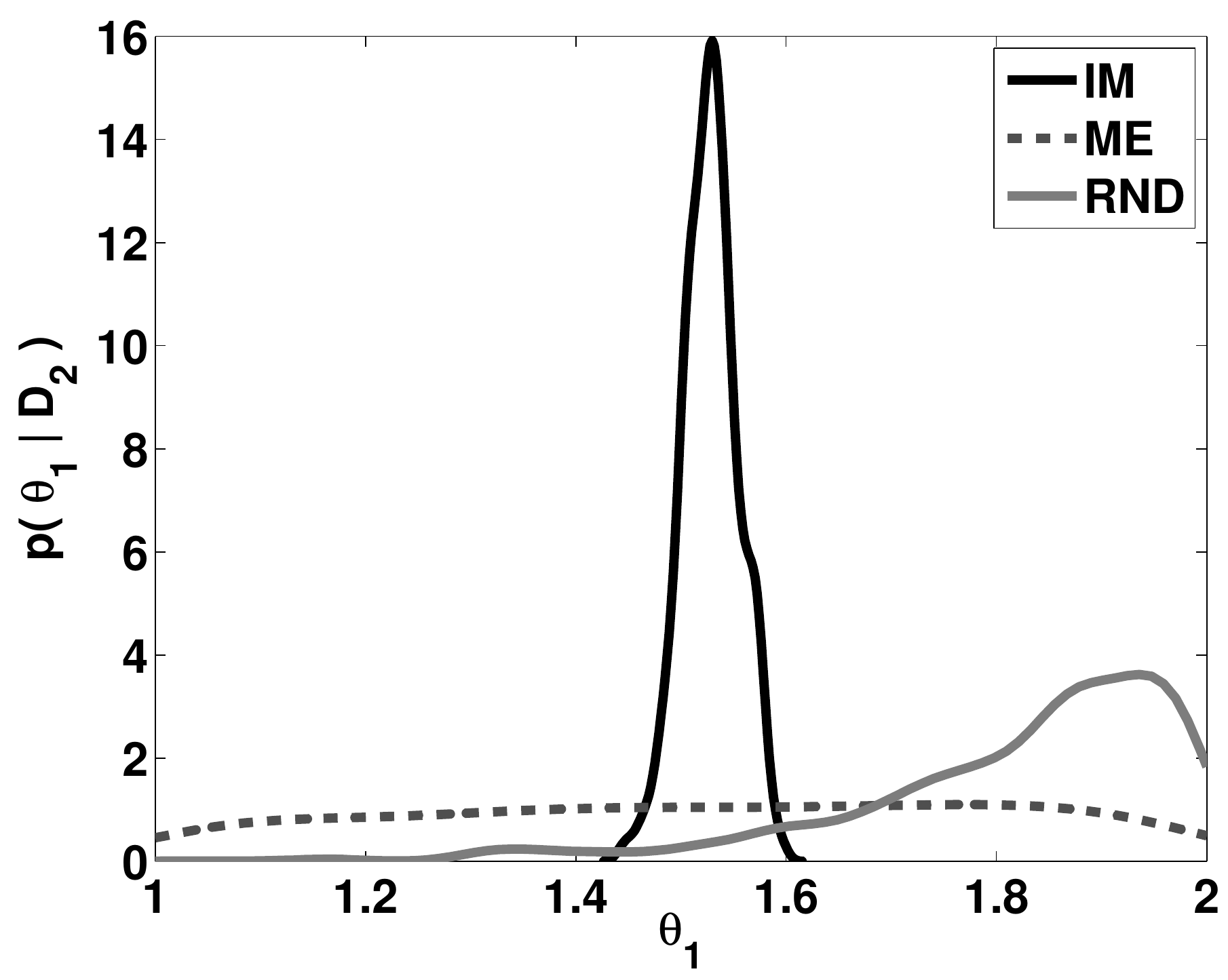}\label{fig:ex1_mult_theta1}} 
\subfigure[$p(\theta_2,D_2)$ (additive noise)]{\includegraphics[width=2.4in]{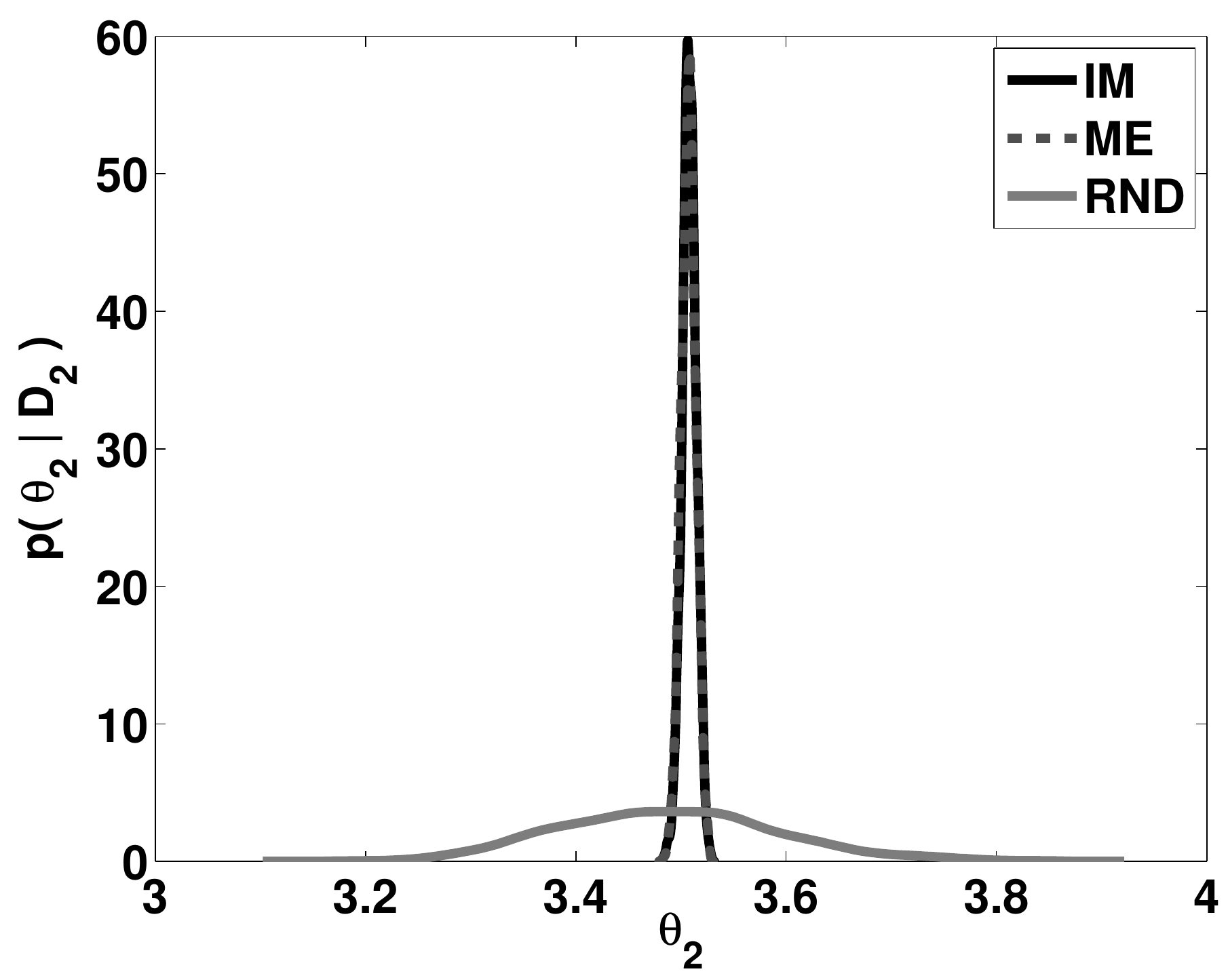}\label{fig:ex1_add_theta2}} 
\subfigure[$p(\theta_2,D_2)$ (multiplicative noise)]{\includegraphics[width=2.4in]{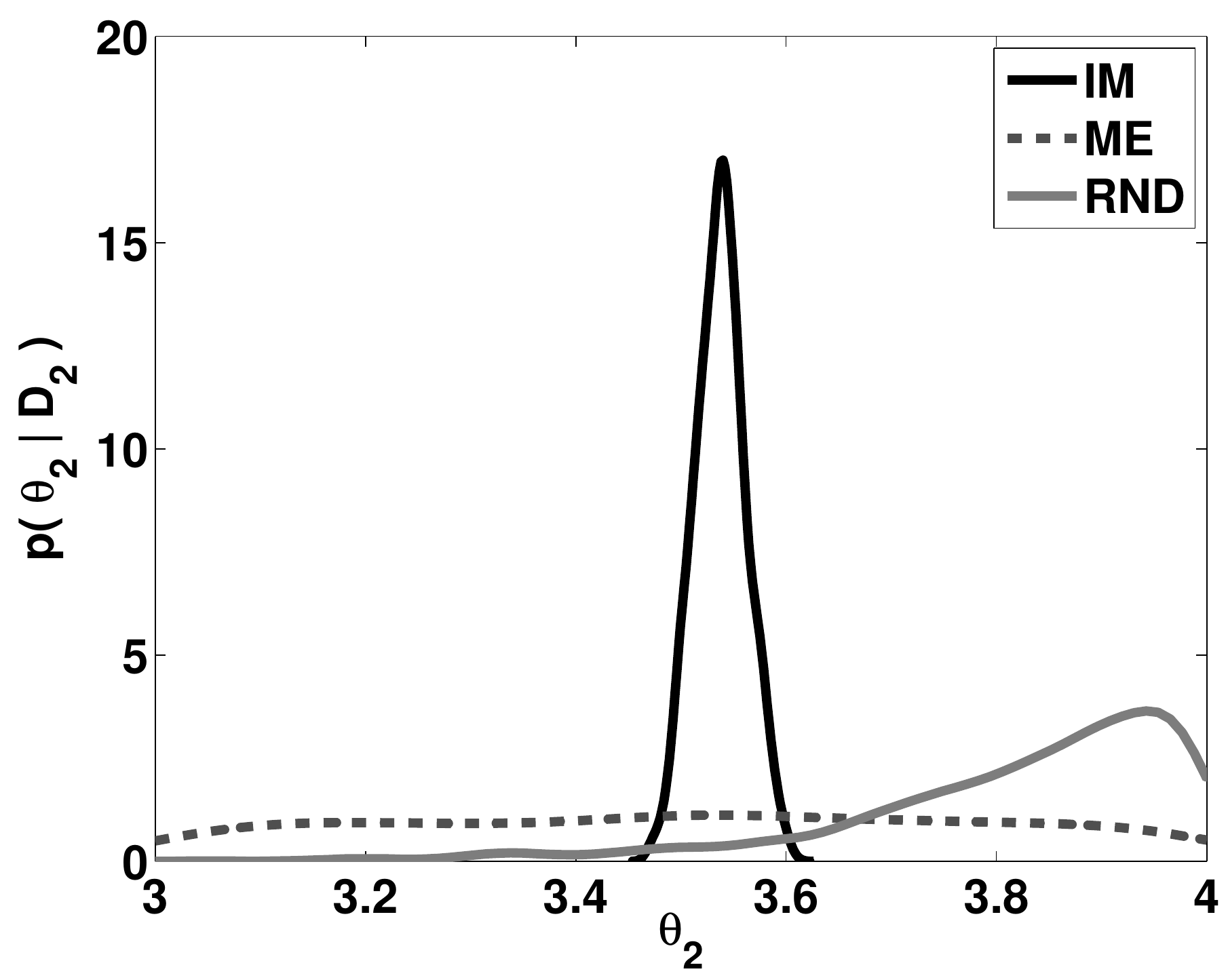}\label{fig:ex1_mult_theta2}} 
\caption{Results for example $1$. Marginal pdfs of model parameters given by the three strategies after the second experimental design stage}\label{fig:ex12}
\end{center}
\end{figure*}

A faster reduction in uncertainty is obtained in this case using the information maximization sampling as it is shown in Fig.\ref{fig:ex1_ent_mult_no_rep} and Fig.\ref{fig:ex1_ent_mult}. For both additive and multiplicative noise cases, an accurate approximation to the true value of the parameters is given by the information maximization sampling using only two experiments as it is shown in Figs.\ref{fig:ex1_add_theta1}-\ref{fig:ex1_mult_theta2}. Therefore, an efficient learning of model parameters in the general nonlinear design problems can be achieved using information maximization sampling.

Regarding the experimental sampling restriction, in both cases when we allow and do not allow repetitive measurements the performance of IM is superior when compared with ME and RND, see Fig.\ref{fig:ex11}. When repeated measurements are allowed and we have additive noise, both IM and ME consistently select the same boundary design points, see Table \ref{tab:ex1_add_seq}. For multiplicative noise and repeated measurements, we have the same consistency in choosing the designs, IM selecting the boundary design points while ME selecting the middle design points, which in this case are the least informative designs. This explains the poor performance of ME compared with IM and RND, see Fig.\ref{fig:ex1_ent_mult}.


\section{A description of experimental setup and data}
\label{Expdata}

\begin{figure*}
  \centering
   \includegraphics[width=1.0\textwidth]{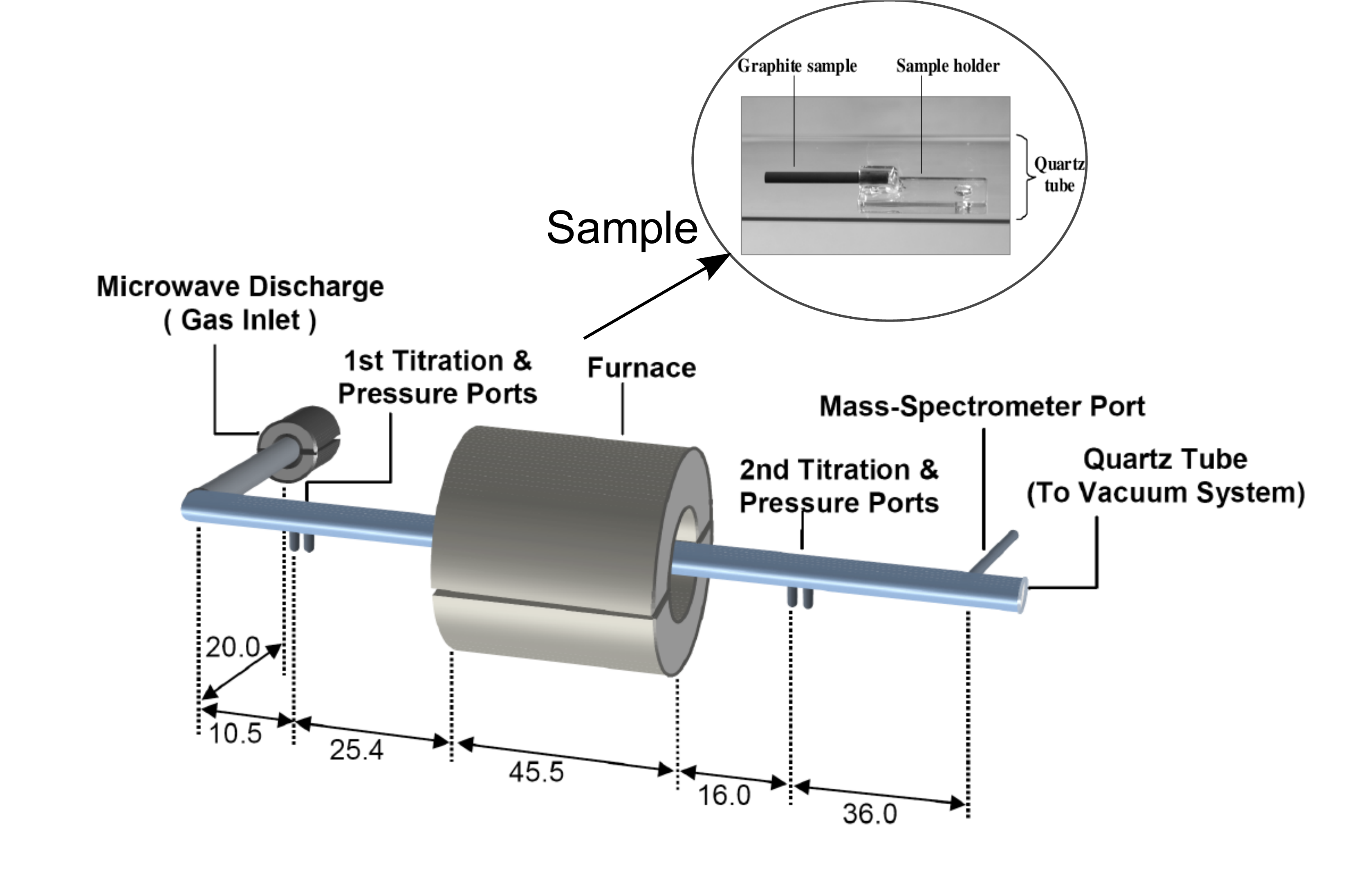}
   \caption{ Schematic of the experimental setup used in \cite{Marschall_nitridation_AIAA}. Also shown in the inset is the sample and sample holder }
   \label{fig:expt_setup}
\end{figure*}

The experimental scenario is described in much detail in \cite{Marschall_nitridation_AIAA}. 
For completeness, we briefly describe the experimental set-up, the measurements and their 
associated uncertainties. The experiments were performed to measure the reaction probability 
of the graphite nitridation reaction, $ C_{\text{Gr}} + \text{N} \rightarrow \text{CN} $. 
The high-purity graphite (grade DFP-2) in the form of 3.175 mm diameter rods was used as 
the sample. The sample was placed in a furnace-heated quartz tube shown in Fig. \ref{fig:expt_setup}. 
A quartz flow tube (22 mm inner diameter) extended through a high-temperature tube furnace 
with a hot region of approximately 45.5 cm. Atomic nitrogen was generated by flowing 
nitrogen through a microwave discharge located upstream of the furnace. The microwave discharge 
was operated at a constant power of around 90-95 W under steady-state temperature and gas 
flow conditions. The pressure was measured before and after the furnace using a 10 torr 
Baratron capacitance manometer. The data sheet from  the instrument reports uncertainty 
as $\pm 1 \%$ of the recorded value. The temperatures of the flow tube were measured using 
Omega Òstick-onÓ type K thermocouples at the location of both pressure ports and near the 
entrance and exit of the furnace. Temperatures were monitored periodically to ensure steady 
state operation. The gas pressure and bulk velocity in the flow tube were varied by simultaneous 
adjustment of the incoming N$_2$ flow rate with a throttling valve located in the downstream 
pumping manifold. The uncertainty in the flow rate is quoted as being no more than $1 \% $. 
The mole fraction of atomic nitrogen  entering and exiting the furnace were measured by gas 
phase titration with nitric oxide. In this method, controlled quantities of nitric oxide 
($\text{NO}$) is introduced into the titration region. The $\text{NO}$ reacts extremely 
rapidly with atomic $\text{N}$ and the disappearance of the $\text{NO}$ signals the titration 
end point. 

The nitrogen atoms that strike the graphite surface react with the graphite sample and cause 
mass loss. Each sample was weighed immediately before it was placed into the 
flow tube and immediately after it was removed from the flow tube, using a Mettler Toledo 
XP105 analytic balance with a 0.01 mg resolution. Control experiments were run at various 
conditions without the microwave discharge to measure the extra mass loss that occurs due 
to de-volatilization of some residual hydrocarbons in the sample or reaction with oxygen 
leaking into the flow system. The measured mass loss was adjusted by subtracting the 
control mass loss. The mass loss rate was obtained by dividing by the time span of the 
experiment. The most extensive set of measurements have been performed at furnace temperature 
of 1273 K. The measurements that are required for the calibration of various parameters 
described in Section \ref{sec:model} are pressure exiting the furnace, and the mass loss 
of the sample.

\section{Model description and uncertain parameters}
\label{sec:model}

In this section, we present the model that is used to compute the measured data as a function of model and control parameters, the system defined abstractly in Eq. (\ref{eq:model_system}) in Section \ref{sec:BayesExpDes}. 
The experiment (\cite{Marschall_nitridation_AIAA}) is briefly described in Section \ref{Expdata}. Since the focus of this work is not in the detailed physical modeling but in the investigation of most informed experimental design strategy, we use an extremely simplified model with several assumptions proposed by \cite{Marschall_nitridation_AIAA}. The flow is assumed to 
be one dimensional and given by the Hagen-Poisseuille flow model with temperature varying 
density given by the ideal gas law. The bulk flow consists entirely of nitrogen as the 
concentrations of atomic nitrogen and cyanates are negligibly small. 

Temperature is assumed to not vary with tube radius. The temperature along the cross section 
of a tube is the same as the wall temperature. The wall temperature is linearly interpolated 
between known temperature values along the tube length. The extra pressure drop due to the 
sample holder is taken into account by using an effective tube diameter in the Hagen-Poisseuille 
flow model. With these assumptions the pressure, mean velocity and density profiles along 
the tube are obtained by solving:
\begin{eqnarray}
P\frac{dP}{dz} &=&- \frac{128 \dot{m}_{N_2} \mu_{N_2} R T(z) }{\pi d_{eff}^4 W_{N_2}} \label{eq:Poisseuille_flow_model}  \\
v(z) &=& \frac{4.0 \dot{m}_{N_2}}{\rho(z) \pi d_{eff}^2} \label{eq:velocity} \\
\rho(z) &=& \frac{P(z) W_{N_2}}{R T} \label{eq:ideal_gas_density} \\
P(z=0) &=& P_{1}; \quad v(z=0) = U_{1}; \quad T(z) = \Psi(z) \label{eq:boundary_conditions}   
\end{eqnarray}
In the above $P(z)$ denotes the temperature as a function of length along the tube $z$, 
$v(z)$ is the bulk flow velocity, $ \dot{m}_{N_2}$ is the mass flow rate of N$_2$, $\rho(z)$ 
is the density of N$_2$, $T(z)$ is the temperature of the wall and also the fluid along the 
tube given by interpolating measured temperatures ($\Psi(z)$ is the linear interpolant), 
$d_{eff}$ is the effective diameter, $W_{N_2}$ is the molecular weight of N$_2$, $P_{in}$ 
and $U_{in}$ are the inlet pressure and velocity respectively and $R$ is the universal 
gas constant. In Eq. (\ref{eq:Poisseuille_flow_model}), Sutherland's model is used for the 
viscosity. Thus, $\mu_{N_2} = \mu_{ref,N_2} \left( \frac{T_{ref}+S_{N_2}}{T + S_{N_2}} \right) 
\left( \frac{T}{T_{ref}} \right)^{3/2}$. The constants appearing in Sutherland's model are 
also assumed to be known precisely. They are: $ \mu_{ref,N_2} = 0.01781e-3 $ Pa s, 
$S_{N_2} = 111.0$ and $T_{ref} = 300.55$ K.

We also need the values of the nitrogen atom concentration as a function of z. The nitrogen 
atom concentration is also assumed to not vary with the tube radius. This means that diffusive 
processes for the N-atom as well as the temperature are assumed to be instantaneous. The 
assumption is justified due to the small radius, reasonably high speed flow and relatively 
slow wall recombination reactions. With these assumptions, the $N$-atom concentration profile 
along the tube is obtained by solving:
\begin{eqnarray}
\frac{d (v(z) C_N) }{dz} &=& - \frac{\gamma_N (T) \bar{v}_N (T) C_N}{d_{eff}} - 2 k (T) C_{N_2} C_N^2 \label{eq:N-atom_conc} \\
C_N(z=0) &=& C_{N,1} \label{eq:inlet_Natom_conc}
\end{eqnarray}
$C_N$ is the molar concentration of atomic N, $\gamma_N$ is the reaction coefficient for 
reactions with the quartz wall defined below in Eq. (\ref{eq:recombination_coeff}), 
$k(T)$ is the reaction rate for gas phase recombination of N atoms defined in Eq.
(\ref{eq:Natom_reaction_rate}), $ \bar{v}_N (T)$ is the thermal velocity of N atoms given 
by kinetic theory defined below in Eq. (\ref{eq:atomic_velocity_N}), $C_{N_2}$ is the 
concentration of N$_2$ gas. The first term in the right hand side models the loss of $N$ 
atoms as a result of recombination at the wall, the second term accounts for the loss 
of $N$ atoms due to reaction in the gas phase to produce molecular nitrogen $N_2$. 
The models/definitions for the varies quantities introduced are:
\begin{eqnarray}
k(T) &=& A_{NN} exp \left[-\frac{E_{a,NN}}{RT} \right] \label{eq:Natom_reaction_rate}  \\
\gamma_N (T) &=& \gamma_N(T_{ref}) \left( \frac{T}{T_{ref}} \right) ^ {\alpha} \label{eq:recombination_coeff} \\
\bar{v}_N (T) &=& \sqrt {\frac{R T}{2 \pi W_N} }  \label{eq:atomic_velocity_N}
\end{eqnarray}  
We assume that sublimation is negligible. The only reaction is a global first order reaction 
between the solid graphite and nitrogen atom to give CN gas: $ C(Gr) + N \rightarrow CN $. 
In the chosen model, called the Gas Kinetic (GK-) model, the transport of the nitrogen atoms to 
the surface is assumed to be instantaneous. The backward reaction is ignored. Hence for the 
GK - model, the mass loss of carbon is given by the following:
\begin{equation}
\Delta M_C = \left( \Delta t \pi d_{sample} W_C \right) \left( \int_{z = z_s}^{z_s + L_s} C_N(z) \sqrt{T(z)} \beta_{N} dz \right) \sqrt {\frac{R}{2 \pi W_N}}  \label{eq:ablation_rate_GK}
\end{equation}
The main quantity of interest for this experiment is the nitridation coefficient $ \beta_{N}$. 
$\Delta t$ is the time interval of the test, $d_{sample}$ is the diameter of the sample, 
$W_C$ is the molecular weight of carbon, $z_s$ is the sample location, $L_s$ is the length 
of the sample and $C_N(z)$ is the molar concentration of N along the sample obtained by 
solution of Eq. (\ref{eq:N-atom_conc}). In Eq. (\ref{eq:ablation_rate_GK}), only the forward 
equation is considered and the wall concentration of $N$ is the same as the free stream 
concentration. 

In the following, the mass loss and the output pressure $\mathbf{d}_n = \{ \Delta m_C, P_2 \}$, are used as the observables. They can be obtained as a solution of equations (\ref{eq:Poisseuille_flow_model}) and (\ref{eq:ablation_rate_GK}) along with the supplementary equations (\ref{eq:velocity}) to (\ref{eq:atomic_velocity_N}). From a review of the literature, we choose the most uncertain parameters to be $\boldsymbol{\theta} = \{d_{eff},\gamma_N,\beta_N\}$. The stochastic forcing term $\boldsymbol{\epsilon}_s$ in Eq. (\ref{eq:model_system})  is neglected. The discrepancy term $\boldsymbol{\epsilon}_m$ in Eq. (\ref{eq:discrepancy_model}) is a Gaussian random variable whose 95$\%$ confidence interval is taken to be the estimated error quoted in \cite{Marschall_nitridation_AIAA}, and it is used here to define the likelihood function in the Bayesian inversion. The control variables are the volumetric flow rate, inlet pressure and the inlet N atom mole fraction : $\boldsymbol{\xi}_n = \{$ $F_{N_2,in}$, $P_1$, $\mathcal{X}_{N,1}$ $\} $ from which the mass flow rate, $\dot{m}_{N_2}$ and inlet nitrogen atom concentration $C_{N,1}$ defined in Eqs. (\ref{eq:velocity}) and (\ref{eq:inlet_Natom_conc}) can be computed.

\section{Results}
\label{sec:result}

To recapitulate, the most uncertain parameters appearing in the model formulation described in 
Section \ref{sec:model} are the effective diameter $d_{eff}$, the reaction efficiency 
$\gamma_{N}$ and, the nitridation coefficient $\beta_N$. To learn the model parameters
$\boldsymbol{\theta} = \{d_{eff},\gamma_N,\beta_N\}$ the mass loss and the output pressure 
are used as observables, $\mathbf{d}_n = \{ \Delta m_C, P_2\}$. In the followings
we present the experimental design analysis on both simulated data and real experimental data as given in \cite{Marschall_nitridation_AIAA}. For these examples, we have used 
$1000$ samples per level in our adaptive MCMC and $2000$ samples at the last level. 
These $2000$ samples have also been used in the forward uncertainty propagation and in
the MI calculation. We have set the $k$NN to $6$ in the following results.

\subsection{Example $2$: Simulated measurements}

In this example we use the model presented in Section \ref{sec:model} to generate
synthetic measurements for pressure and mass loss by adding noise to model predictions using
$1 \%$ uncertainty for pressure and $5 \%$ uncertainty for mass loss. To generate 
the measurements the following values have been used for the normalized parameters: 
$\frac{d_{eff}}{d_{eff0}} = 0.95$, $\log\bigg(\frac{\gamma_N}{\gamma_{N0}}\bigg) = 1.0$ and, 
$\frac{\beta_N}{\beta_{N0}} = 2.0$, where the nominal values are given by: 
$d_{eff0} = 0.022$, $\gamma_{N0} = 10^{-3}$, and $\beta_{N0} = 3.29 \times 10^-3$. 

The values for the experimental scenarios ($F_{N_2,in}$, $P_1$, $\mathcal{X}_{N,1}$, $\Delta t$, $L_S$) are the same as in \cite{Marschall_nitridation_AIAA}, and presented in Fig.\ref{fig:ex2_inflow}.
Similar with the previous example three strategies are used to select the 
optimal sequence of designs from the set of $28$ designs: maximum entropy sampling, information 
maximization sampling and, ascending sampling which starts with the first tabulated design in \cite{Marschall_nitridation_AIAA} (enumerated in Fig.\ref{fig:ex2_inflow} from $E0$ to $E27$), and continues providing the next available one at each stage.

\begin{figure*}
  \begin{center}
    \subfigure[IM: Relative mutual information]{\includegraphics[width=3.5in]{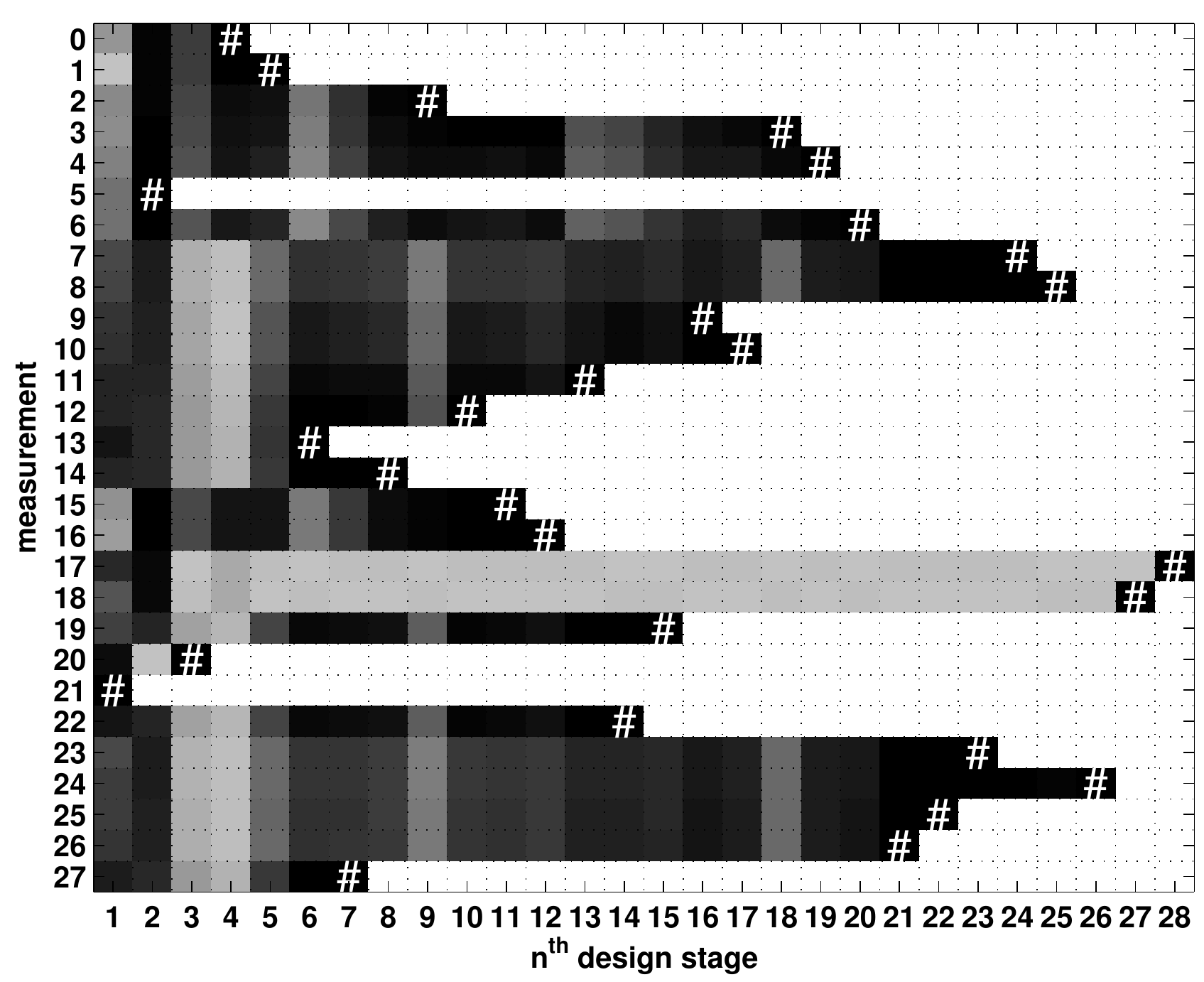}\label{ex2_meas_order_IM}} 
    \subfigure[ME: Relative entropy]{\includegraphics[width=3.5in]{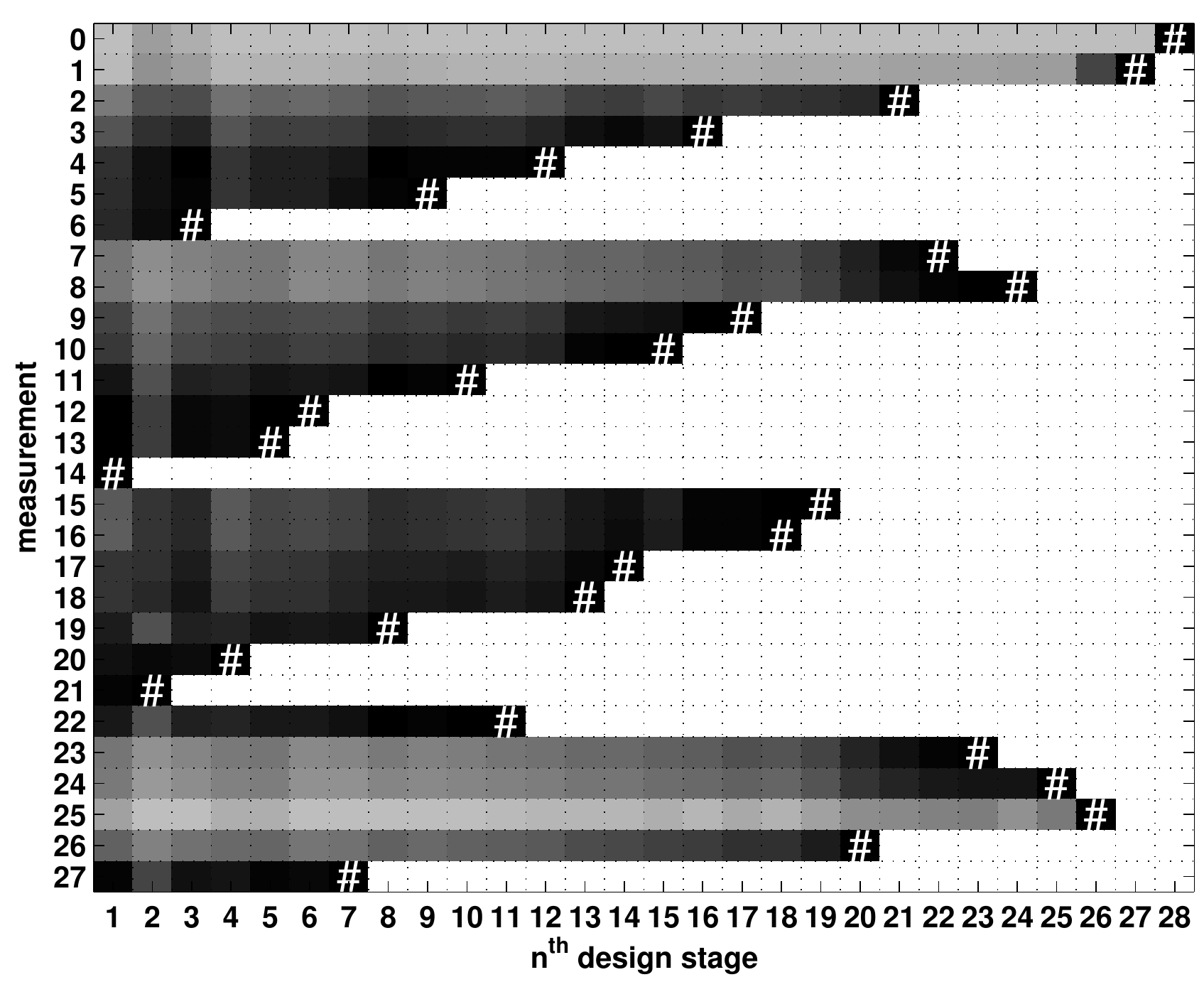}\label{ex2_meas_order_ENT}} 
    \caption{Results for example $2$ (simulated data). Relative expected utility for Information Maximization Sampling and 
      Maximum Entropy Sampling. Darker shadings indicate large expected utilities (mutual information for IM or
      entropy for ME) relative to all available scenarios within a design stage. The $\#$ sign indicates which
      scenario has been selected at a given stage. Previously selected scenarios are excluded from subsequent
      analysis which is indicated here by white cells.}\label{fig:ex2_meas_order}
  \end{center}
\end{figure*}

\begin{figure*}
  \begin{center}
    \subfigure[Entropy]{\includegraphics[width=3.5in]{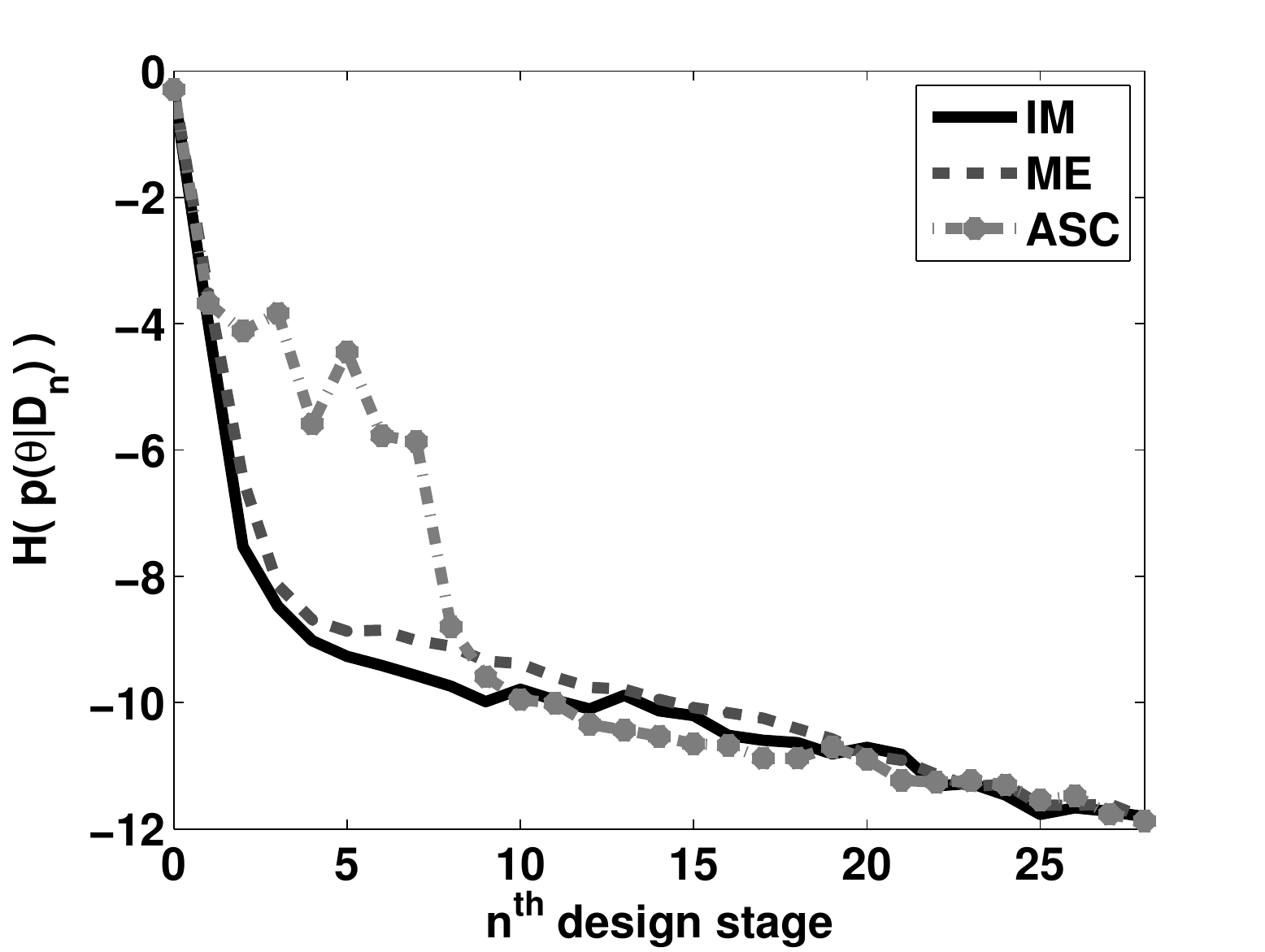}\label{ex2_ent}} 
    \subfigure[Kullback Leibler divergence]{\includegraphics[width=3.5in]{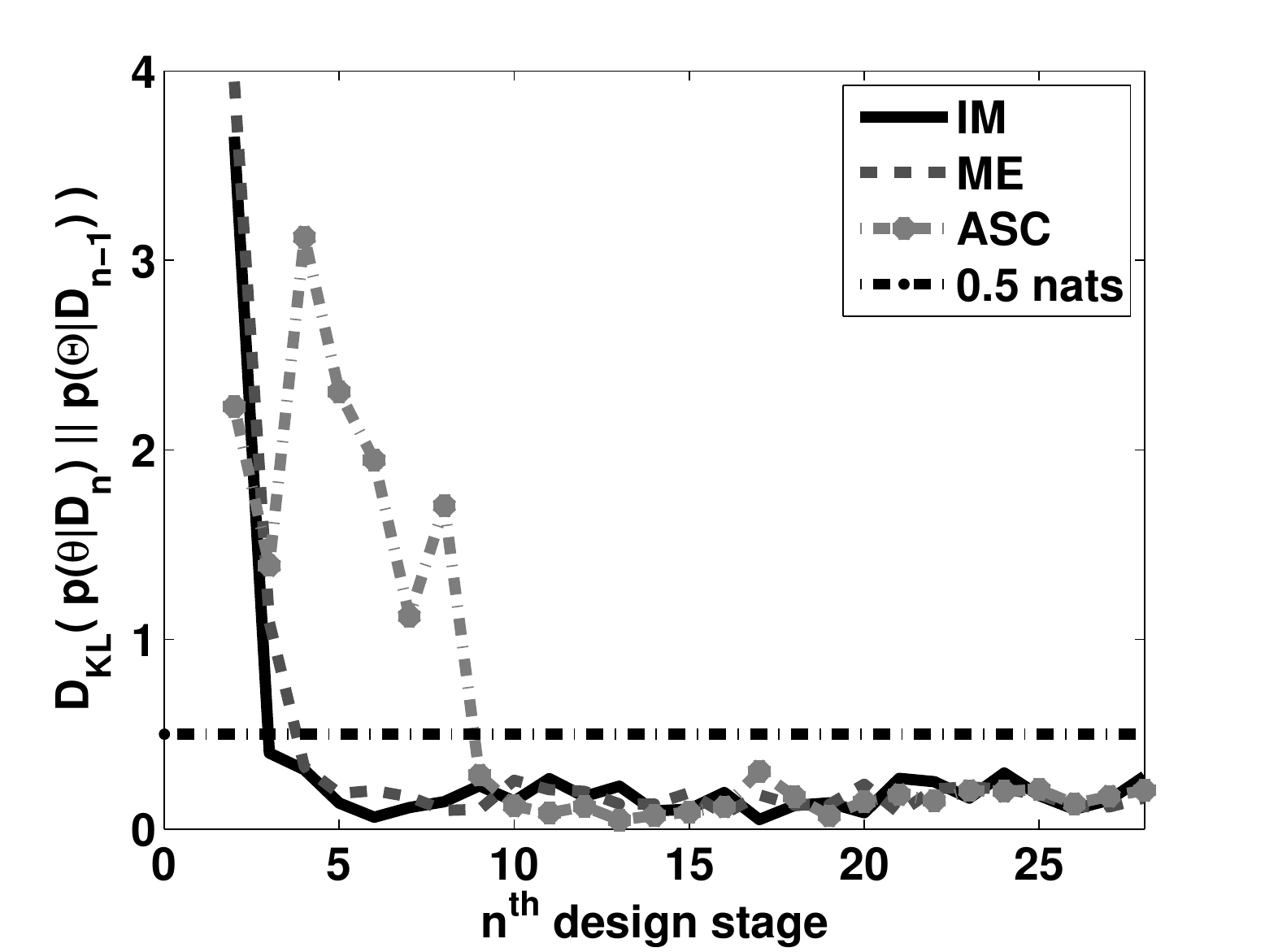}\label{ex2_kl}} 
    \caption{Results for example $2$ (simulated data). The evolution of the entropy and the KL divergence 
      given the posterior distributions yielded by the three strategies for experimental design:
      IM - Information Maximization Sampling, ME - Maximum Entropy Sampling and ASC - Ascending Sampling}\label{fig:ex2}
  \end{center}
\end{figure*}

Fig.\ref{ex2_ent} shows the drop in uncertainty versus the number of design stages. 
The design sequences provided by both maximum entropy sampling and the information maximization 
sampling, see Table \ref{tab:ex2_seq}, give a comparable learning rate, which is superior to 
the ascending sampling. Fig. \ref{fig:ex2_meas_order} gives the relative entropy for ME sampling 
and relative mutual information for IM sampling at each design stage. The difference between the 
information theoretic approaches and the ascending sampling is that the former ones give higher 
priority to the designs with high inflow. 

ME sampling preferably selects the experimental scenarios with large values for inflow conditions, either 
$F_{N_2,in}$, $P_1$, $\mathcal{X}_{N,1}$. For example, first the scenario of 14$^{th}$ measurement has the largest inflow rate 
and pressure among others. It follows the same type of experimental scenario. A physical explanation for this trend could 
be hypothesized as follows. A large value of $F_{N_2,in}$ results in the large pressure drop through the experimental
region. This enhances the sensitivity of pressure on $d_{eff}$ through Eq.\eqref{eq:velocity}. 
Also, a large pressure (e.g., 14$^{th}$, 6$^{th}$ measurements) accelerates chemical reactions especially the gas-phase 
recombination through the three-body reaction: $N+N+N_2\leftrightarrow N_2+N_2$. The significance of 20$^{th}$ and  
21$^{st}$ measurements preferably selected in the early stage is related to the combination of the large volumetric flow 
rate and relatively small pressure, resulting the large velocity.  This reduces the residence time of flow and consequently
has a large influence on the determination of N-atom concentration. On the contrary, the scenarios with the small
values of  inflow conditions (e.g., 0$^{th}$, 1$^{st}$ measurements) are selected in the later stages.

On the other hand, IM sampling prefer alternately selecting the experimental scenarios of the large values and small 
values for inflow conditions after the first 3 scenarios. These difference is clearly seen in the different order of 
selecting  0$^{th}$ and 1$^{st}$ measurements.  
It seems to the authors that IM sampling performs in a more logical manner by mixing a variety of information.
This can be seen in Fig.\ref{fig:ex2_inflow} which provides a graphical representation for
the inflow conditions preferred by both IM and ME.

\begin{figure*}
\begin{center}
\subfigure[Stage $2$ - $d_{eff}$]{\includegraphics[width=1.5in]{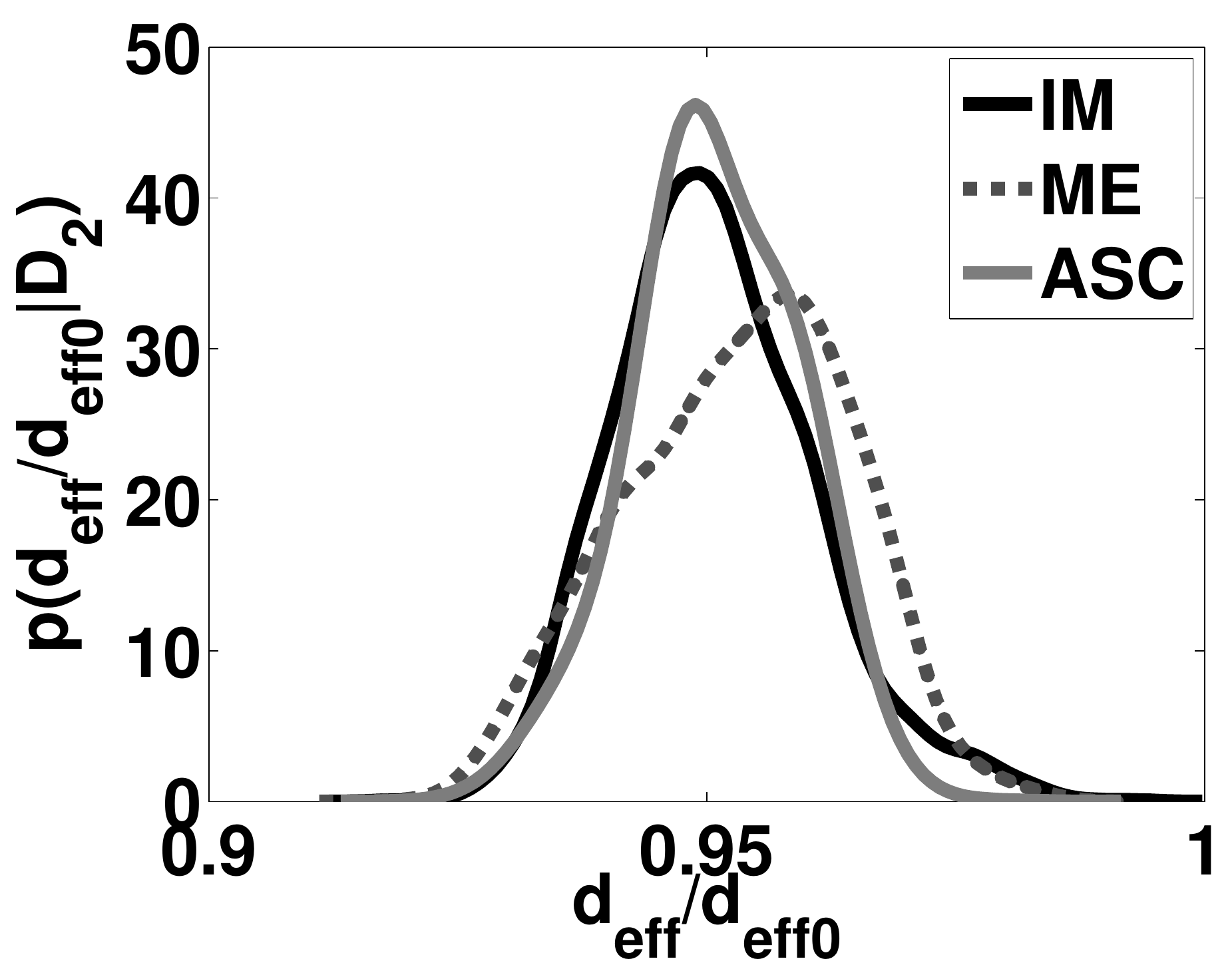}} 
\subfigure[Stage $2$ - $\gamma_N$]{\includegraphics[width=1.5in]{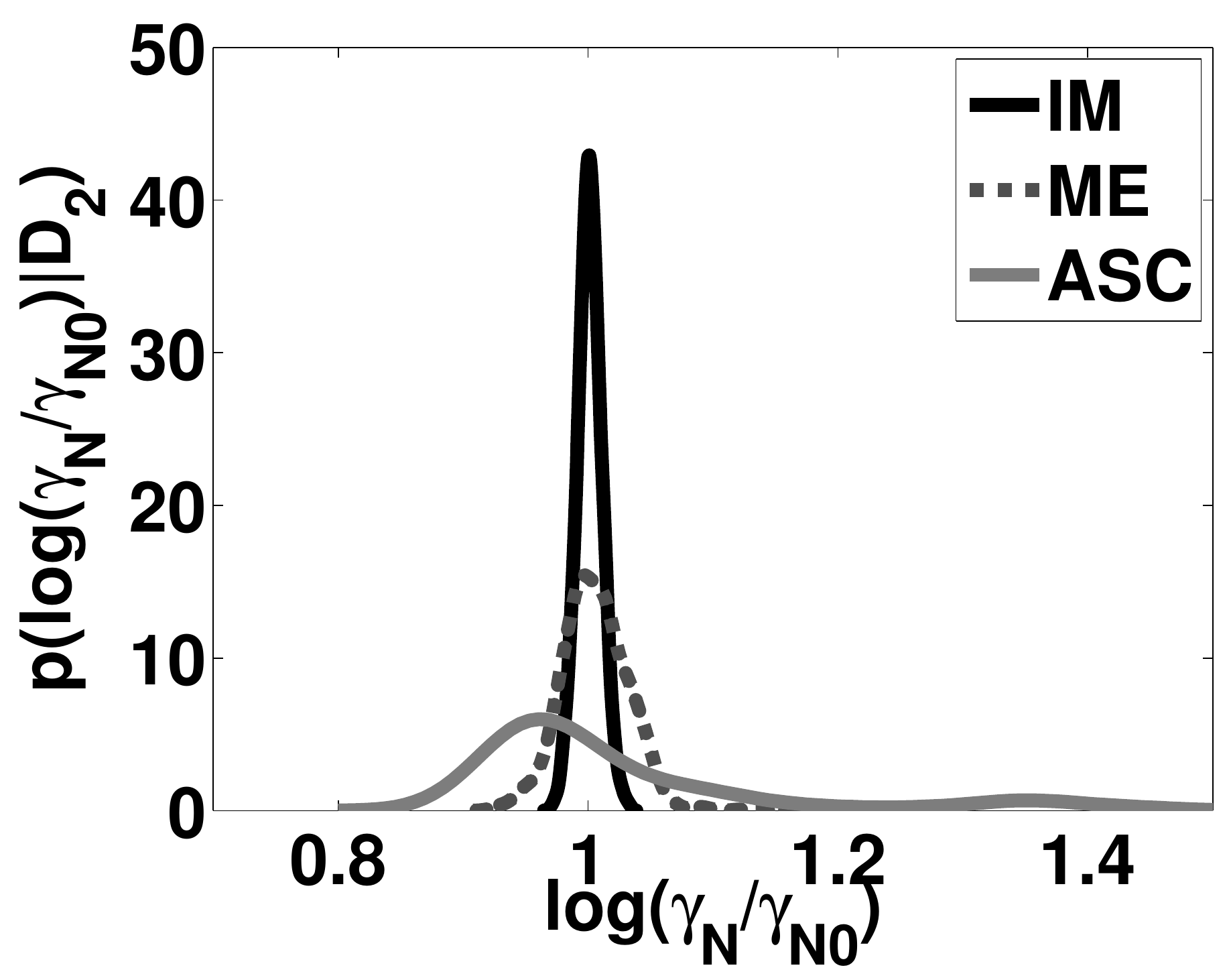}} 
\subfigure[Stage $2$ - $\beta_N$]{\includegraphics[width=1.5in]{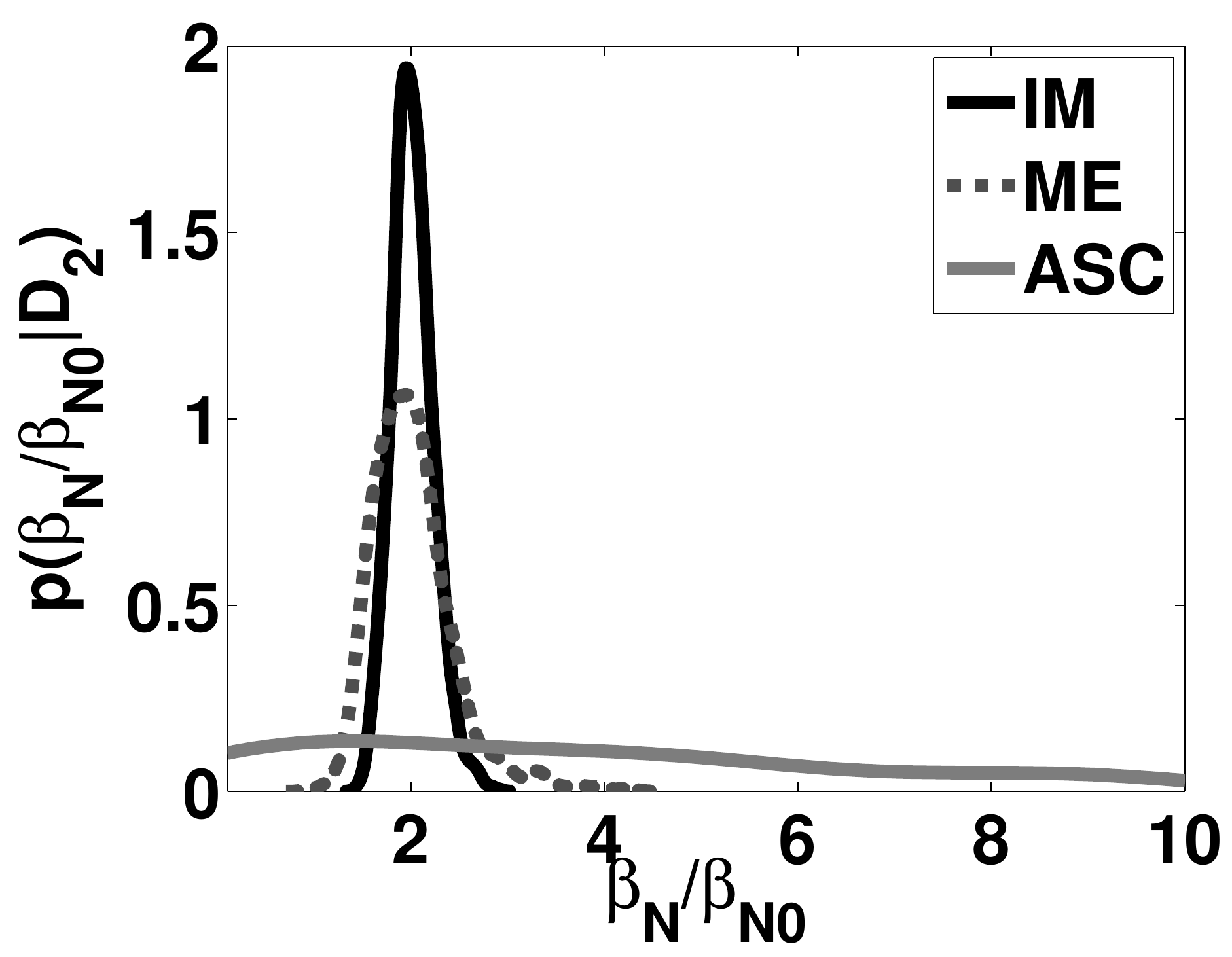}} 
\subfigure[Stage $4$ - $d_{eff}$]{\includegraphics[width=1.5in]{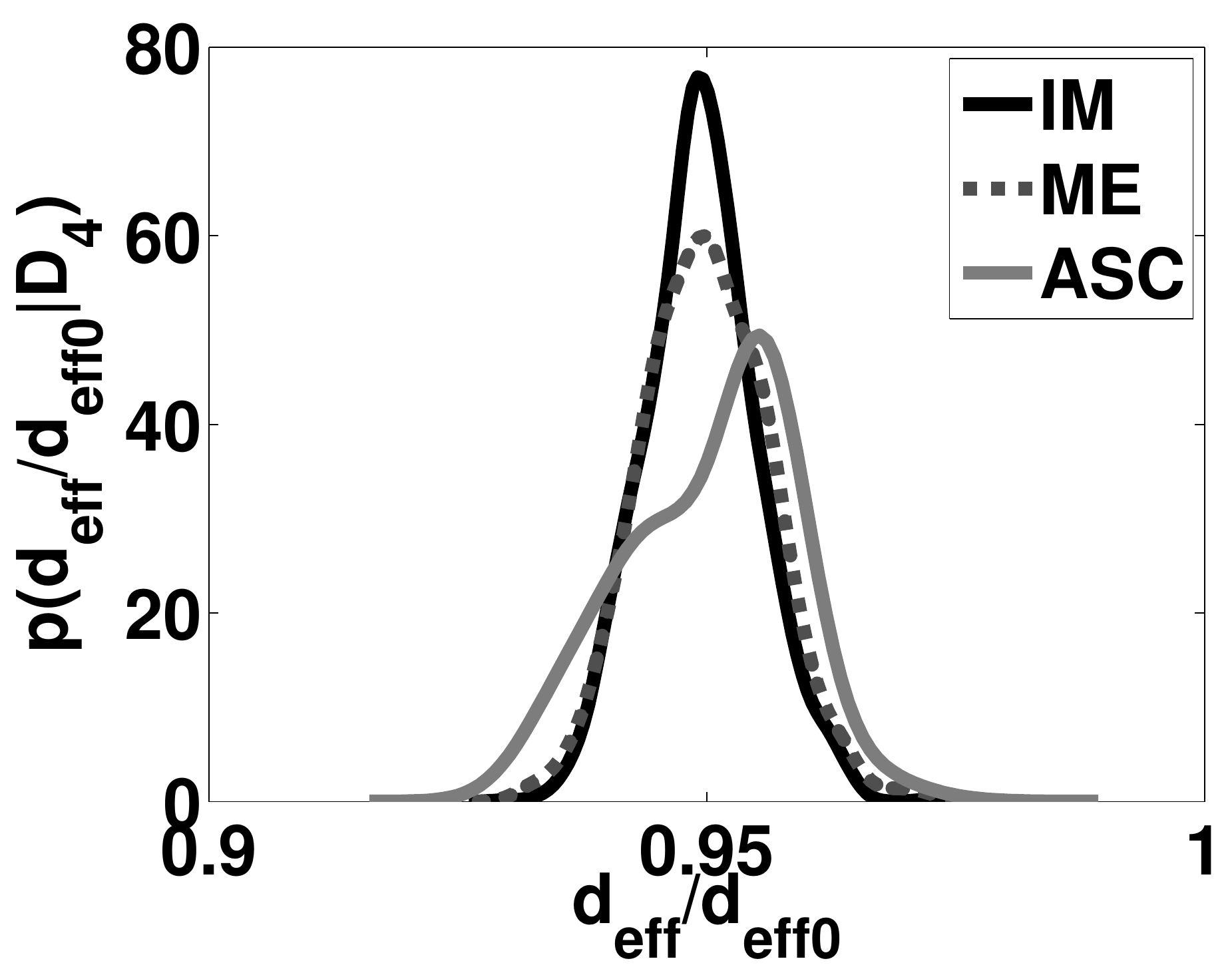}} 
\subfigure[Stage $4$ - $\gamma_N$]{\includegraphics[width=1.5in]{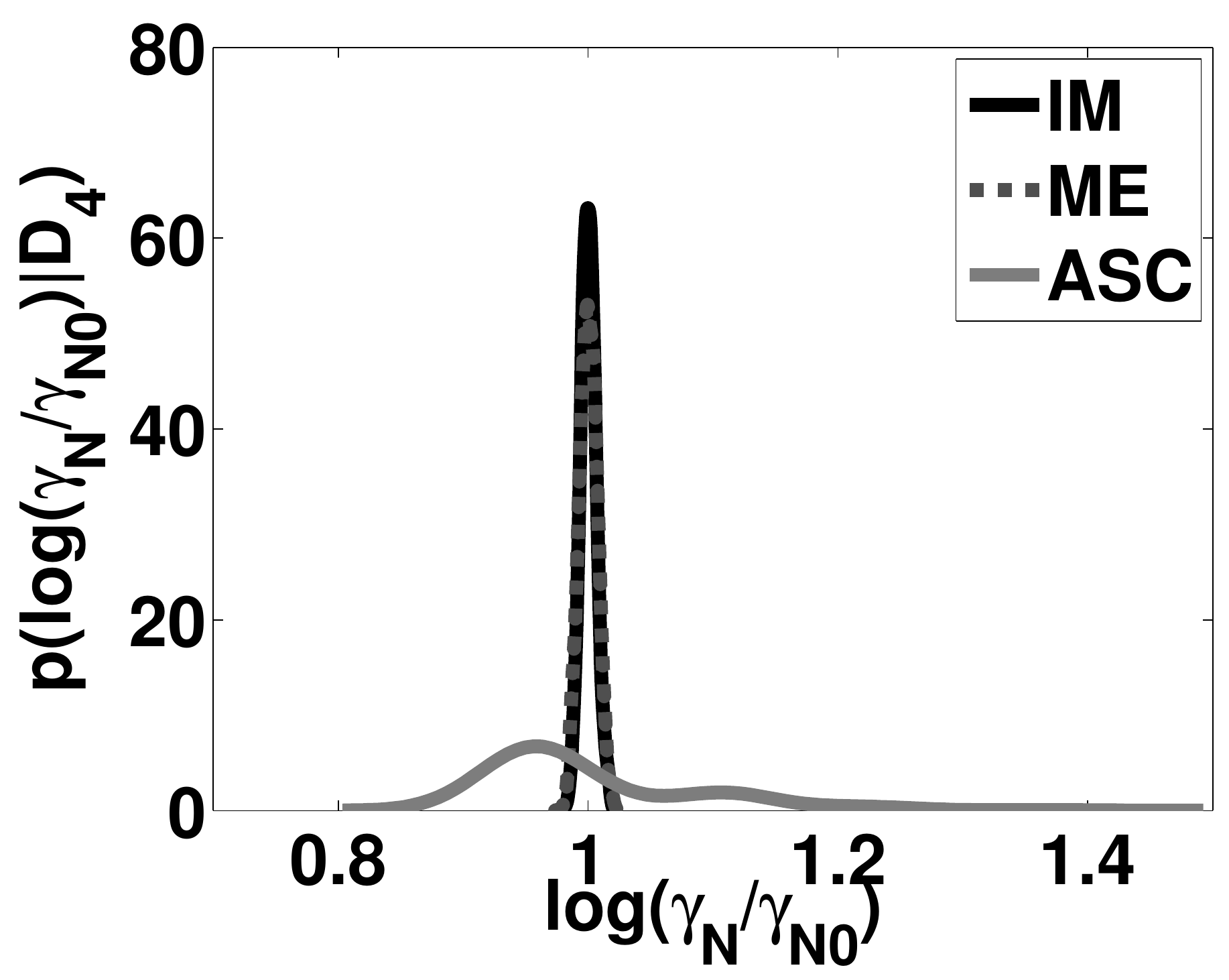}} 
\subfigure[Stage $4$ - $\beta_N$]{\includegraphics[width=1.5in]{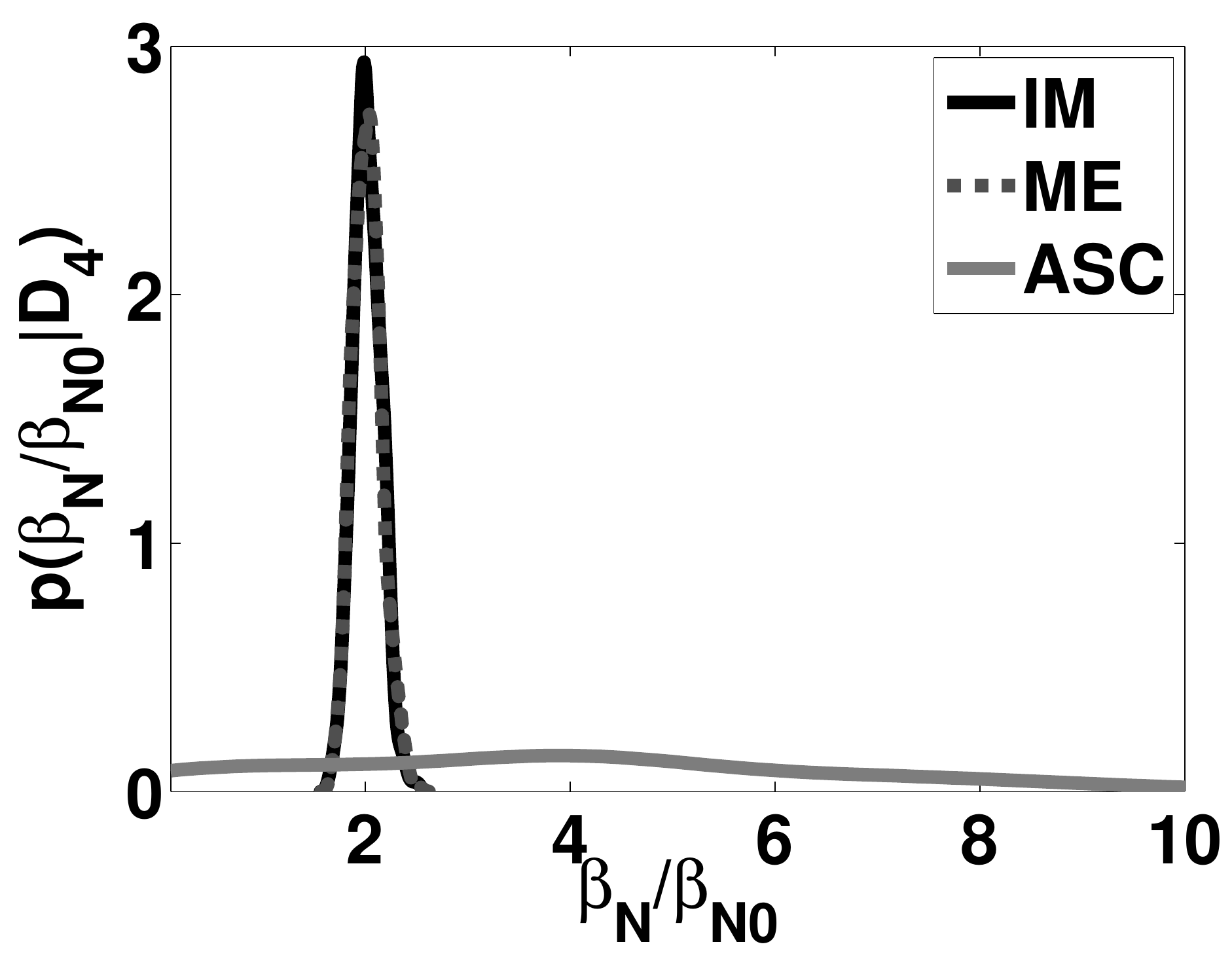}} 
\subfigure[Stage $6$ - $d_{eff}$]{\includegraphics[width=1.5in]{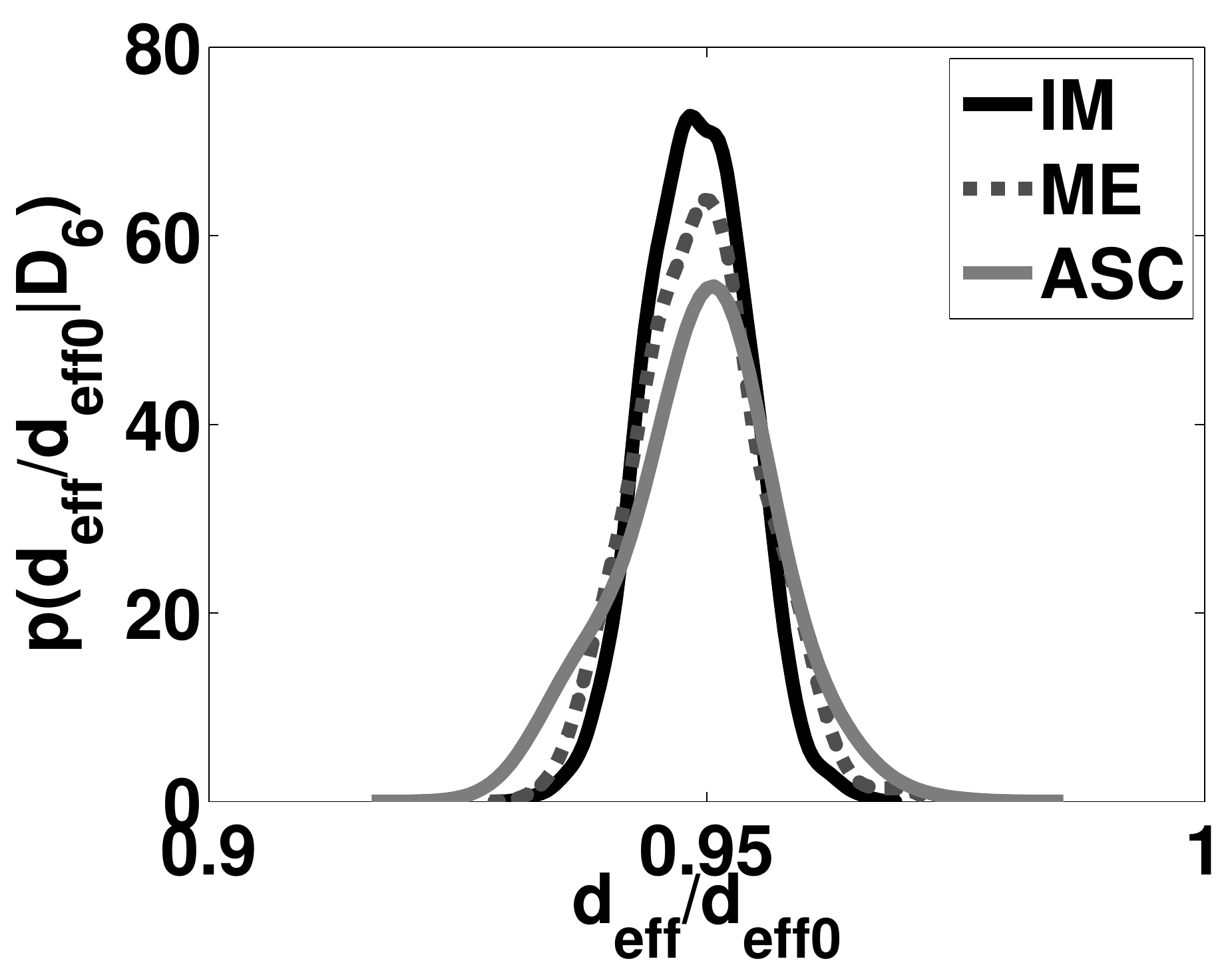}} 
\subfigure[Stage $6$ - $\gamma_N$]{\includegraphics[width=1.5in]{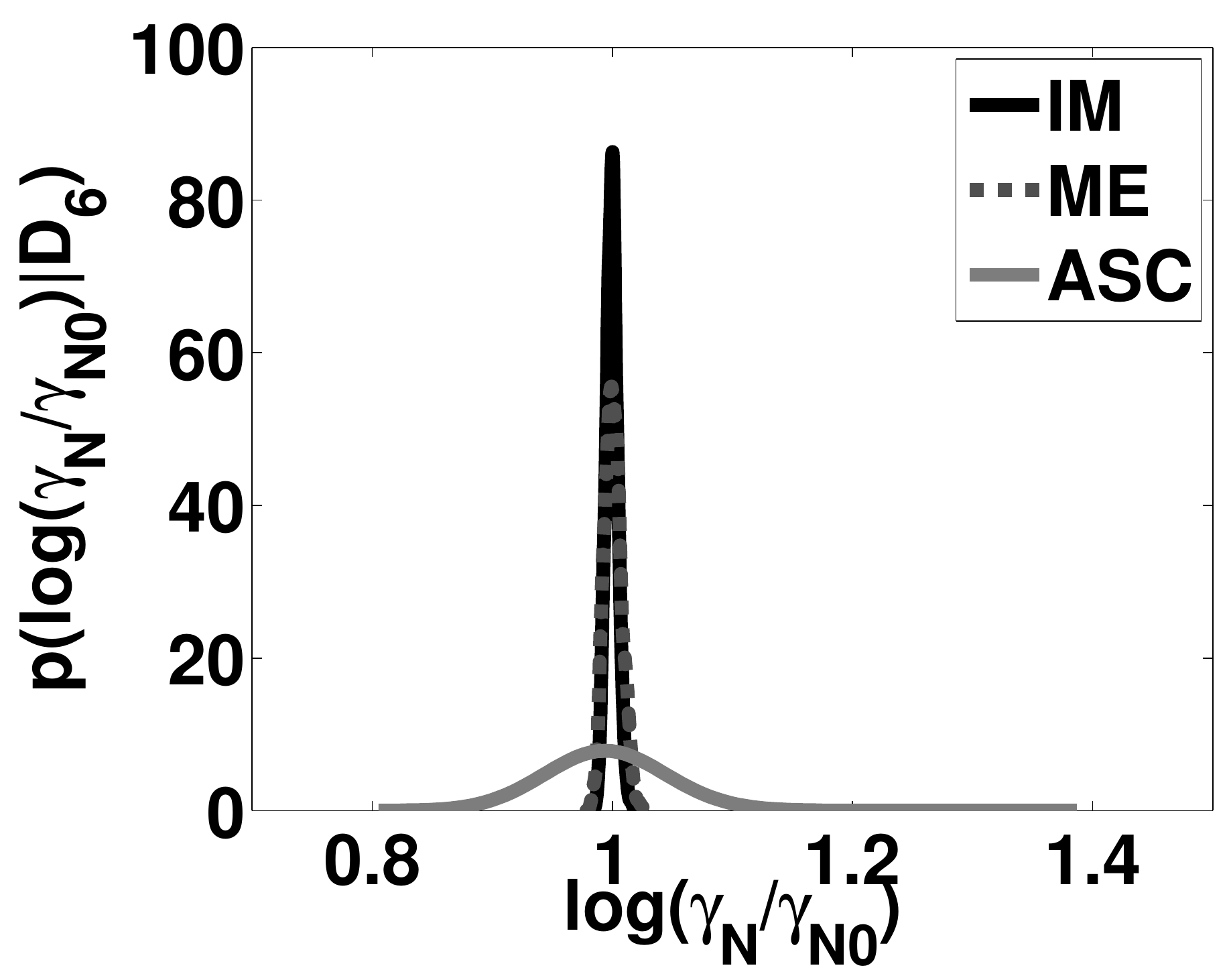}} 
\subfigure[Stage $6$ - $\beta_N$]{\includegraphics[width=1.5in]{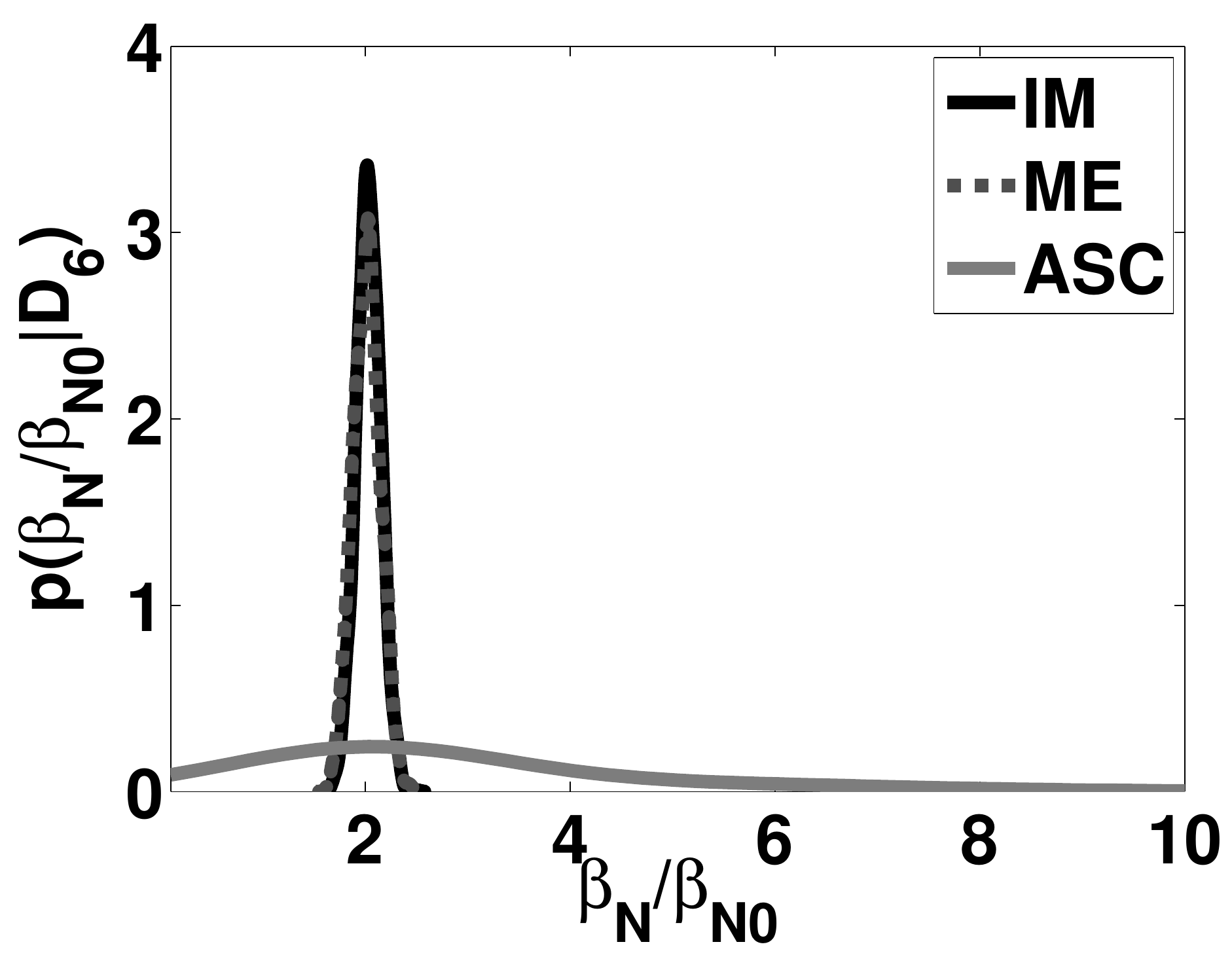}} 
\subfigure[Stage $8$ - $d_{eff}$]{\includegraphics[width=1.5in]{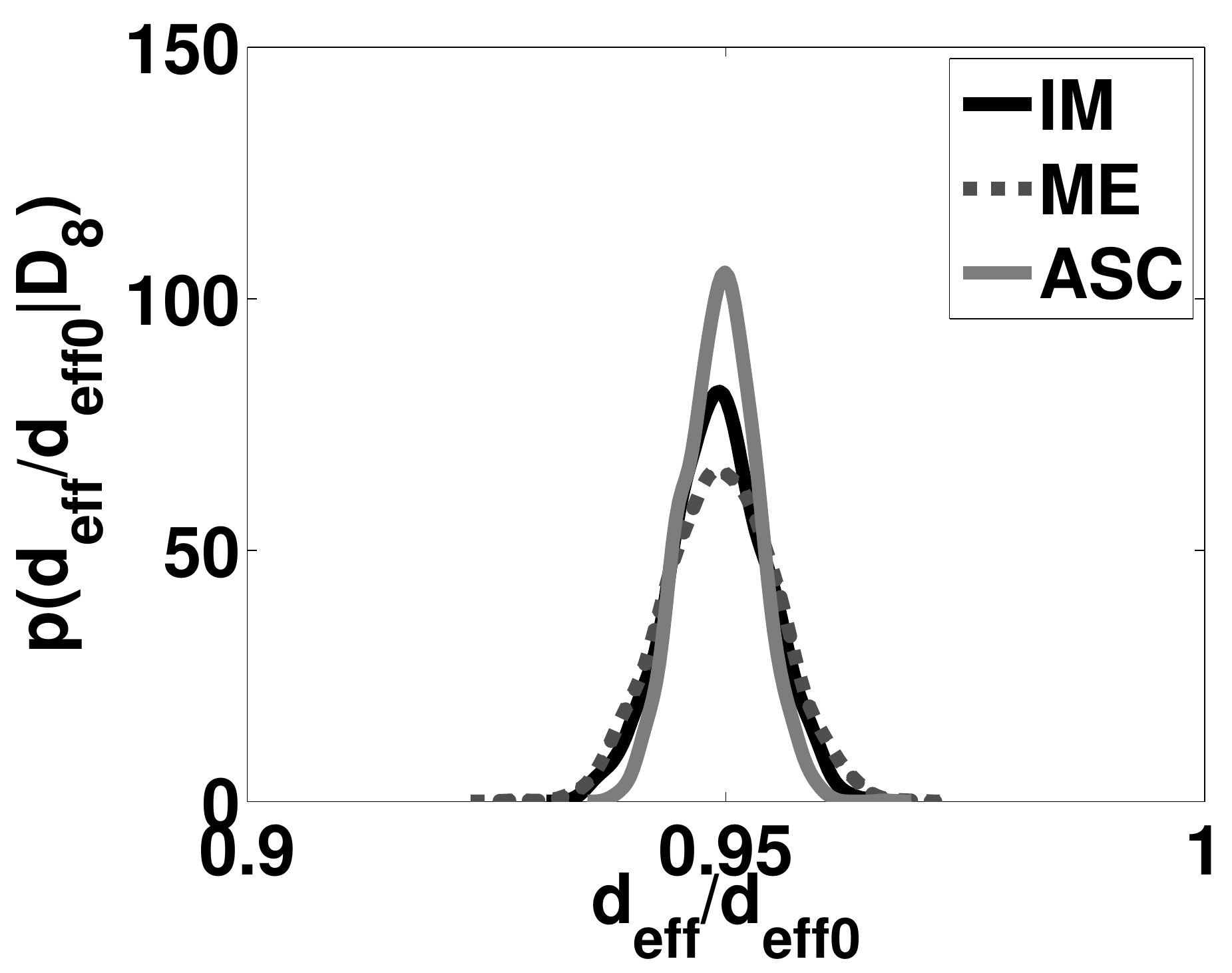}} 
\subfigure[Stage $8$ - $\gamma_N$]{\includegraphics[width=1.5in]{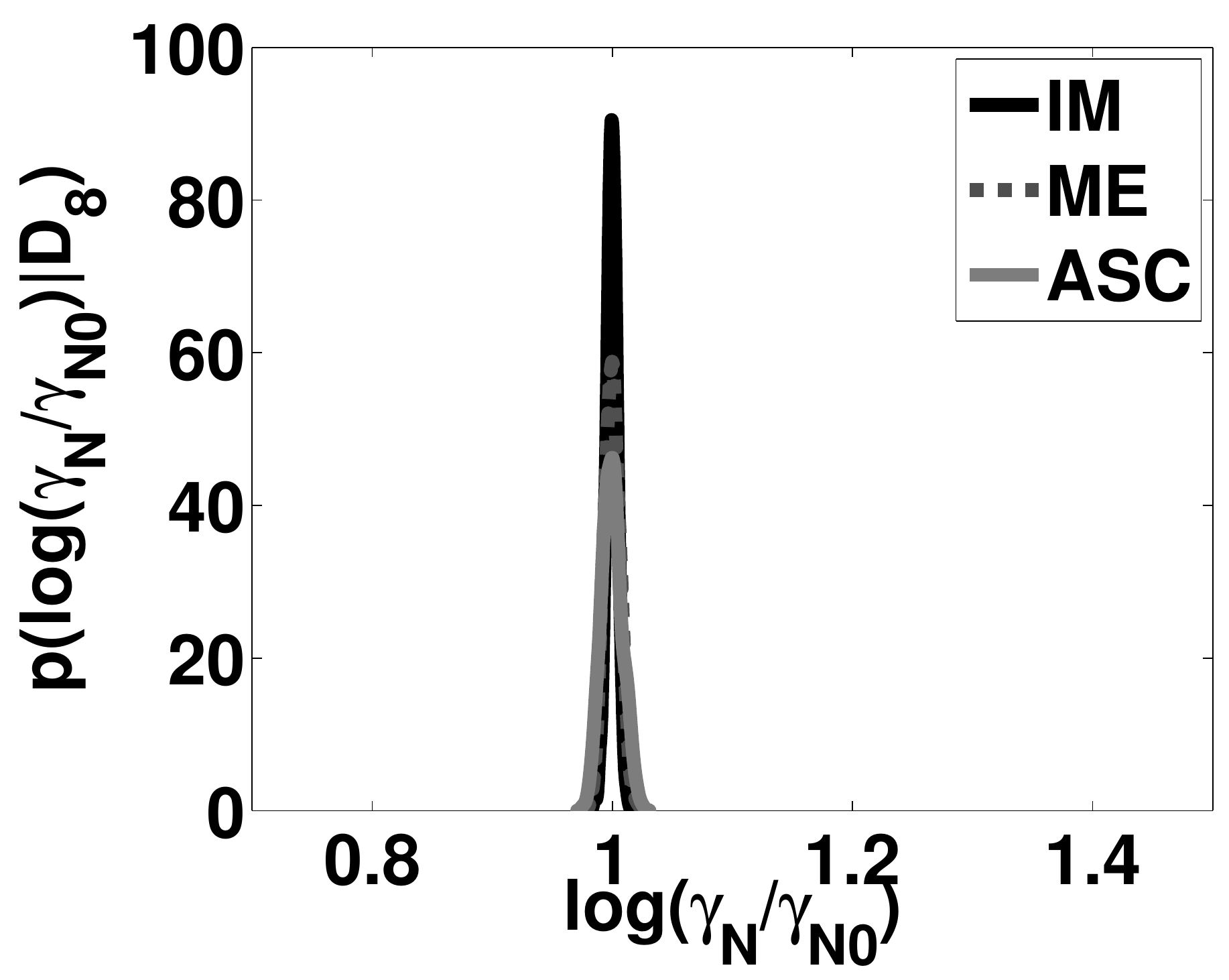}} 
\subfigure[Stage $8$ - $\beta_N$]{\includegraphics[width=1.5in]{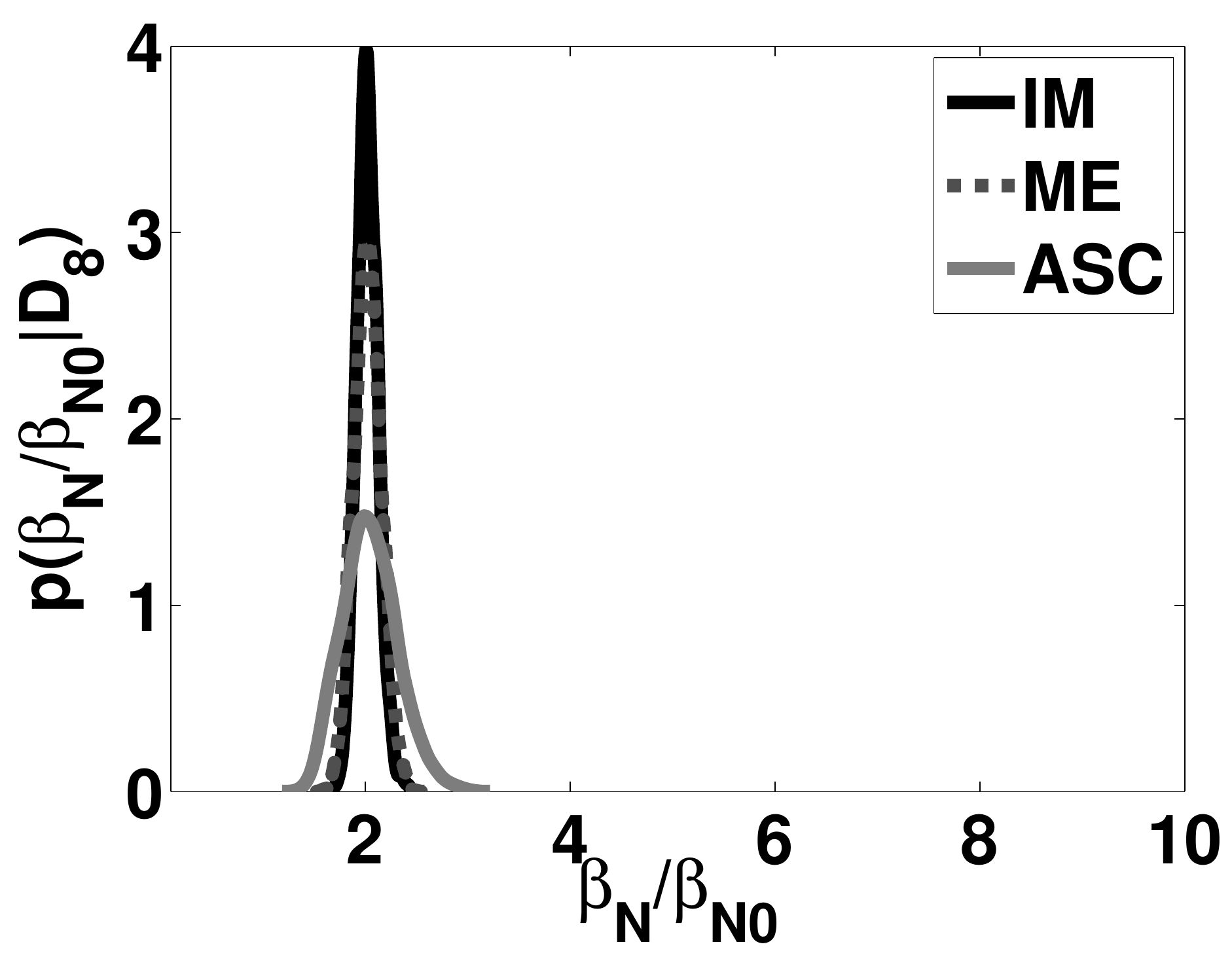}} 
\subfigure[Stage $28$ - $d_{eff}$]{\includegraphics[width=1.5in]{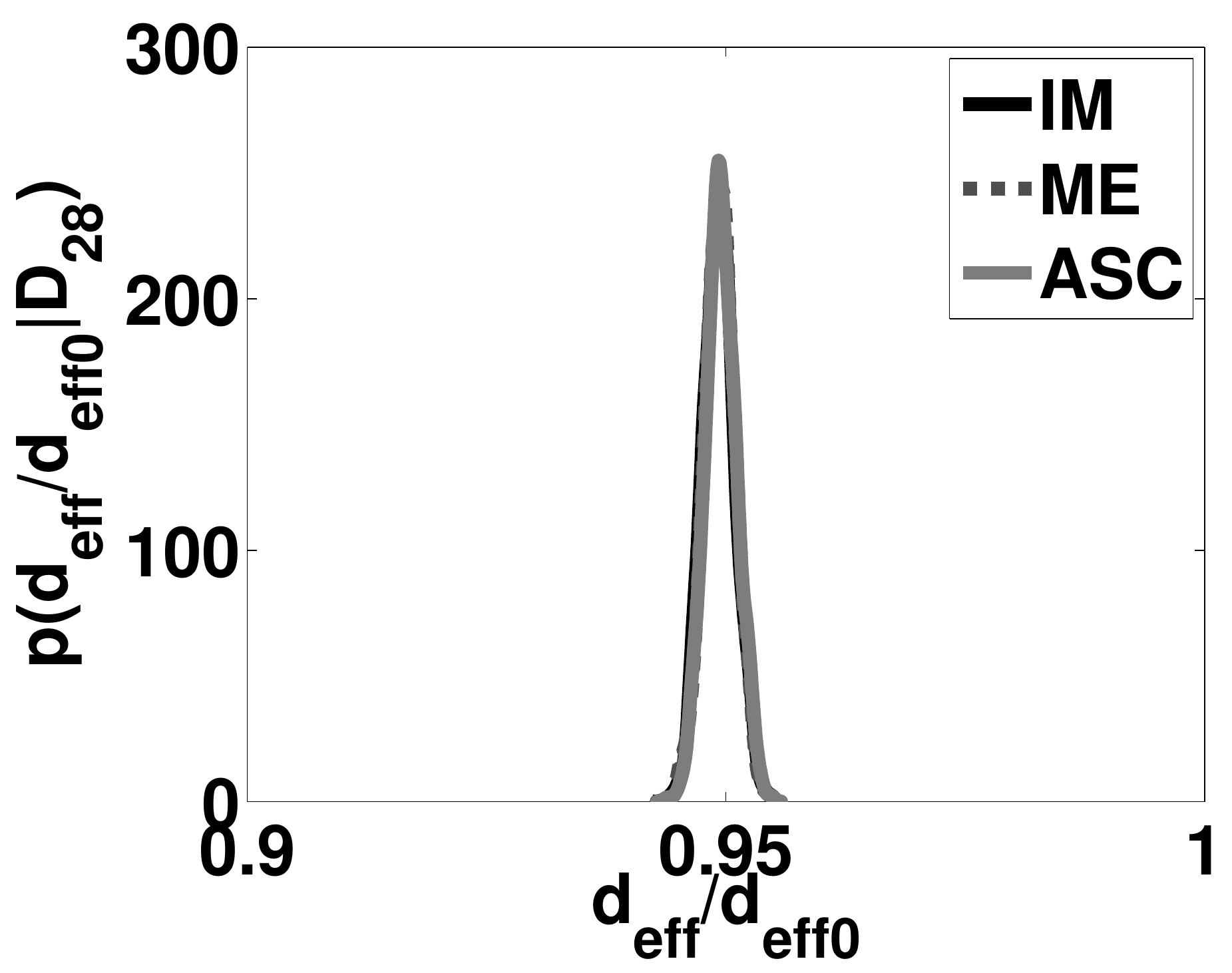}} 
\subfigure[Stage $28$ - $\gamma_N$]{\includegraphics[width=1.5in]{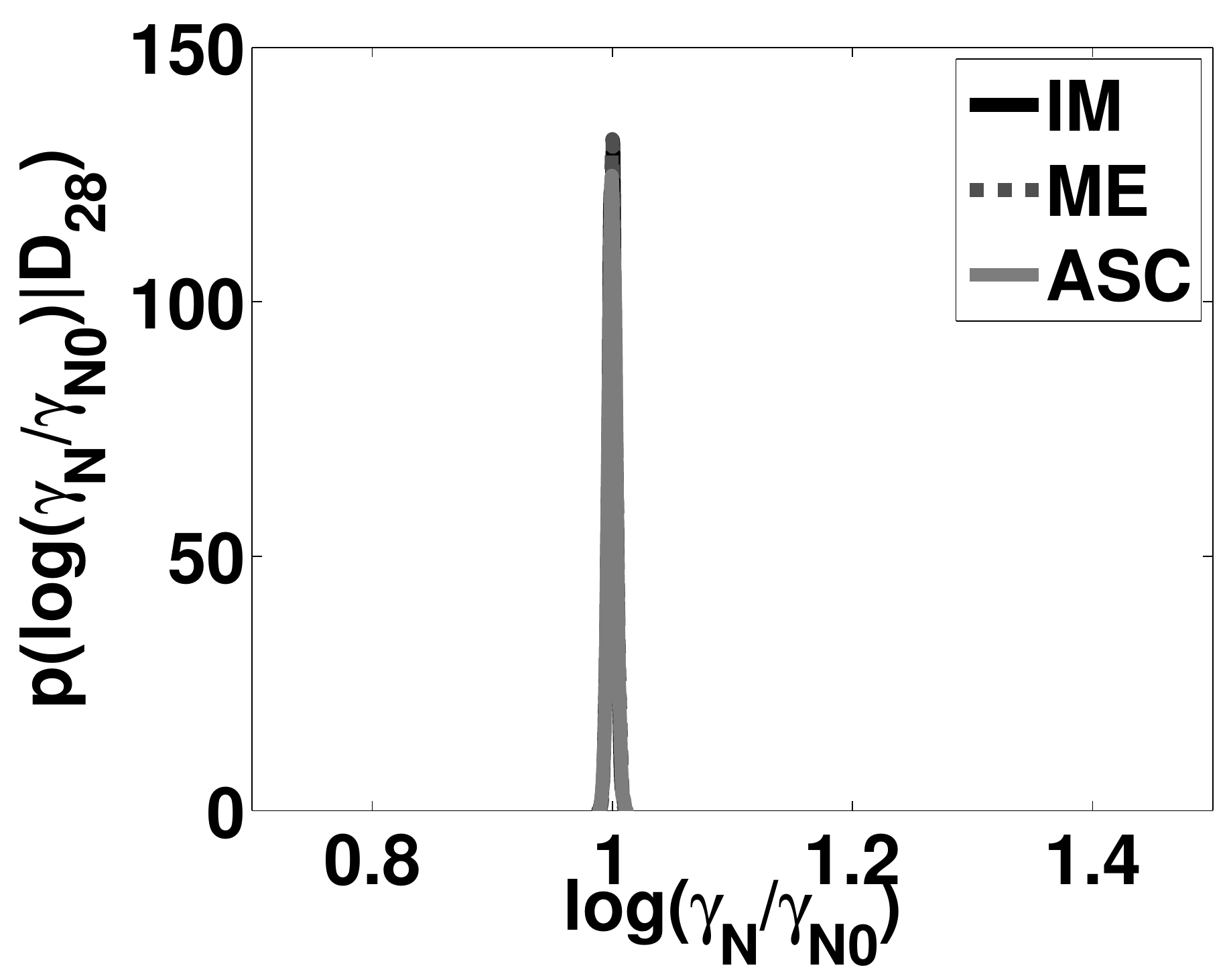}} 
\subfigure[Stage $28$ - $\beta_N$]{\includegraphics[width=1.5in]{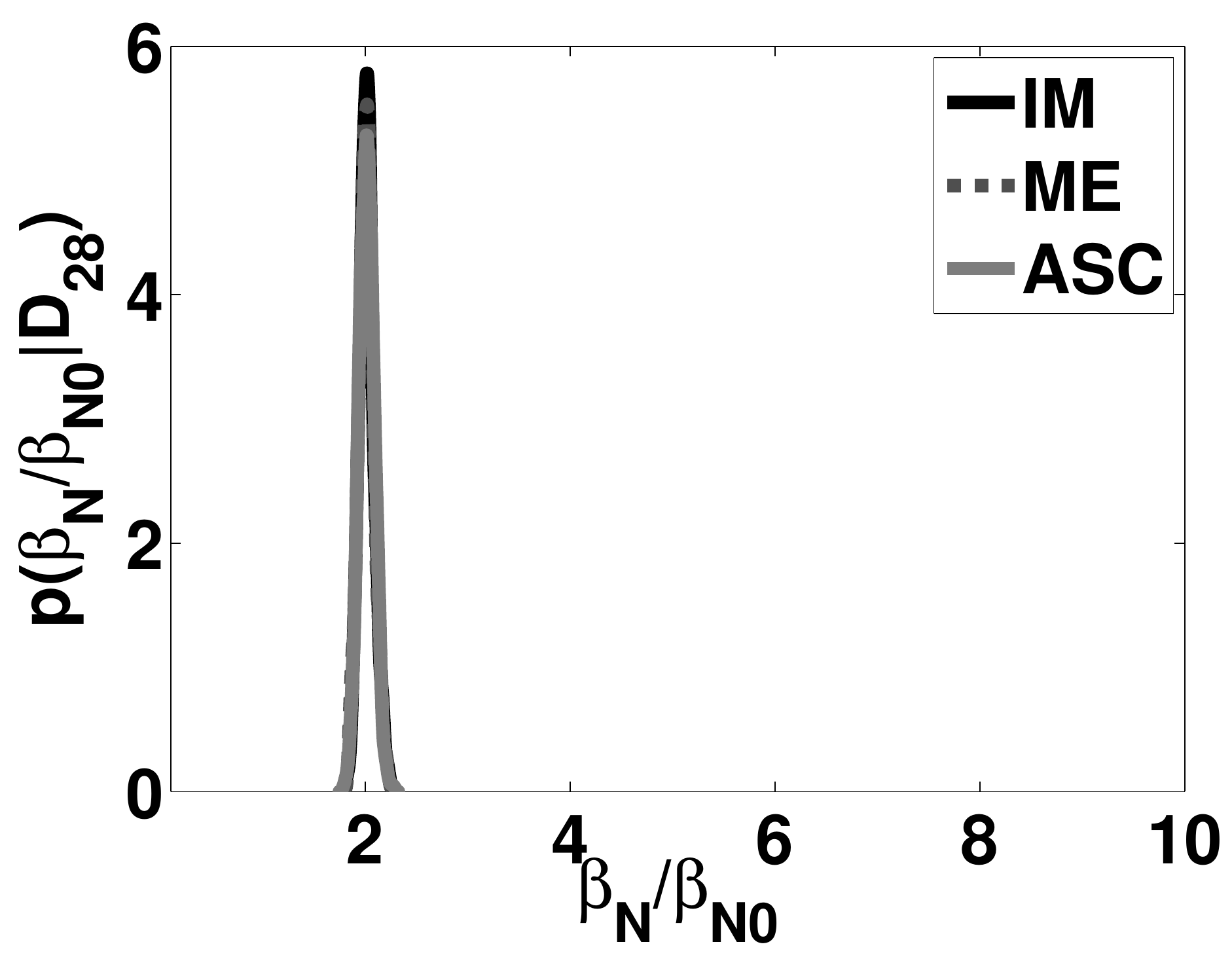}} 
\caption{Results for example $2$ (simulated data). Marginal pdfs of model parameters given by the three strategies after a selected number of experimental design stages}\label{fig:ex2_pdfs}
\end{center}
\end{figure*}

The marginal pdfs of the parameters at different stages are plotted in Fig.\ref{fig:ex2_pdfs}. 
Note that both information theoretic approaches provide a better determination of the model 
parameters than the ascending sampling approach. While including all the $28$ measurements we 
are able to recover with good accuracy the model parameters, information maximization can provide 
an adequate estimate after the first two stages, followed closely by the maximum entropy sampling. 

Kullback-Leibler (KL) divergence between the pdfs obtained at the current stage and the previous one 
is also included in order to monitor the evolution of the inference process, see Fig.\ref{ex2_kl}. 
The KL divergence is calculated using also a $k$NN estimator as described in the Appendix A.
While the entropy quantifies how certain we are about the model parameters, the KL divergence 
quantifies how much information has been gained by adding the current measurement. This includes
the reduction in uncertainty as well as a change in the support of the pdf.
Decreasing entropy and large values of KL are indicative that the model is still learning. 
This is equivalent with obtaining tighter pdfs at each stage however their high probability support 
is changing from one stage to another. A sequence of low entropies as well as low KL 
divergence indicates that the learning process has slowed down and one can stop the experimental 
process. In this example, by thresholding the KL at $0.5$ nats, depending on the level of 
uncertainty, given here by the entropy, one can stop the experimental process at any stage 
greater than $5$. Thus for simulated data not all the $28$ measurements are needed to recover 
the model parameters with a given level of accuracy.

\subsection{Example $3$: Study case on existing experimental measurements}

\begin{figure*}
  \begin{center}
    \subfigure[IM: Relative mutual information]{\includegraphics[width=3.5in]{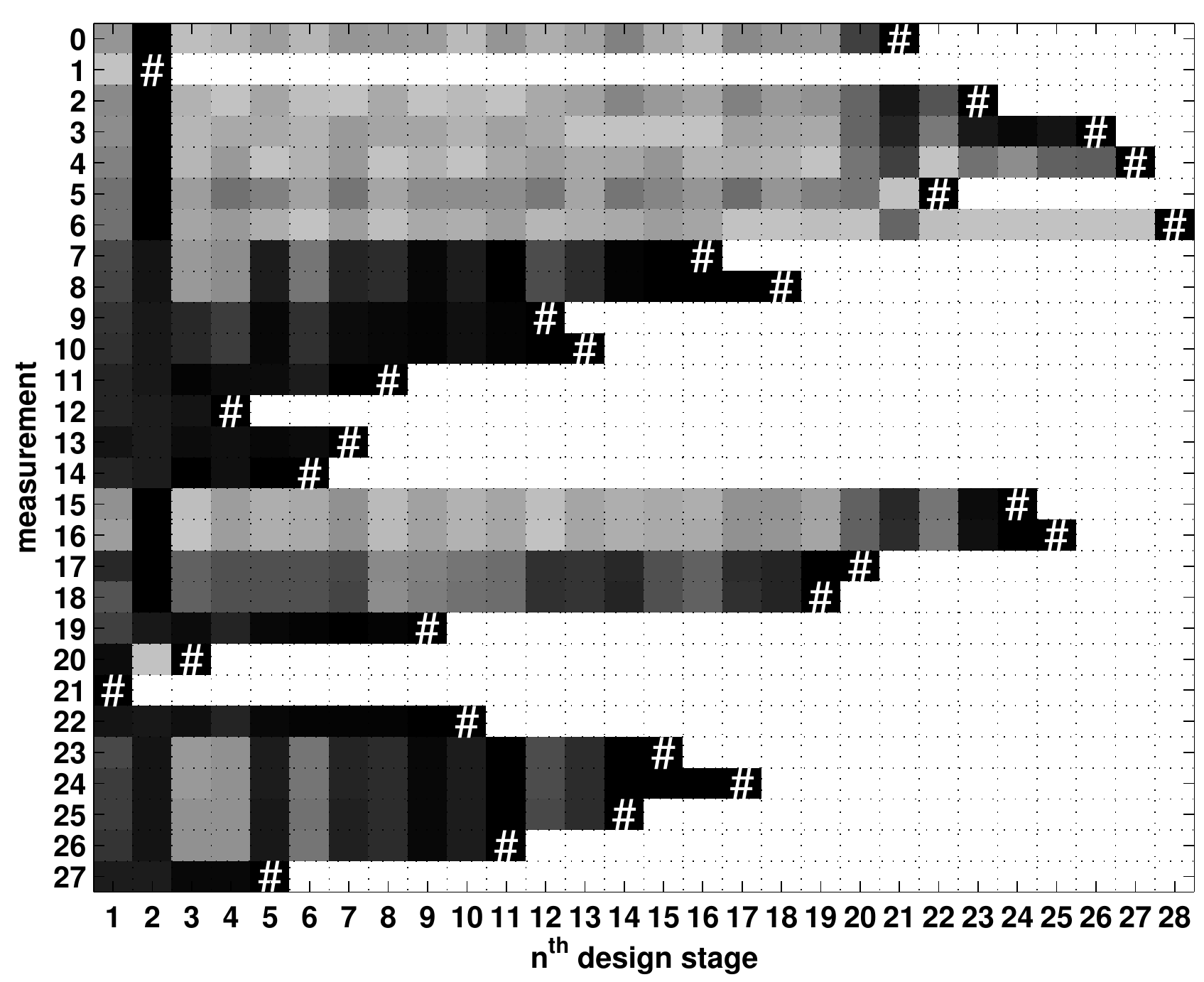}\label{ex3_meas_order_IM}} 
    \subfigure[ME: Relative entropy]{\includegraphics[width=3.5in]{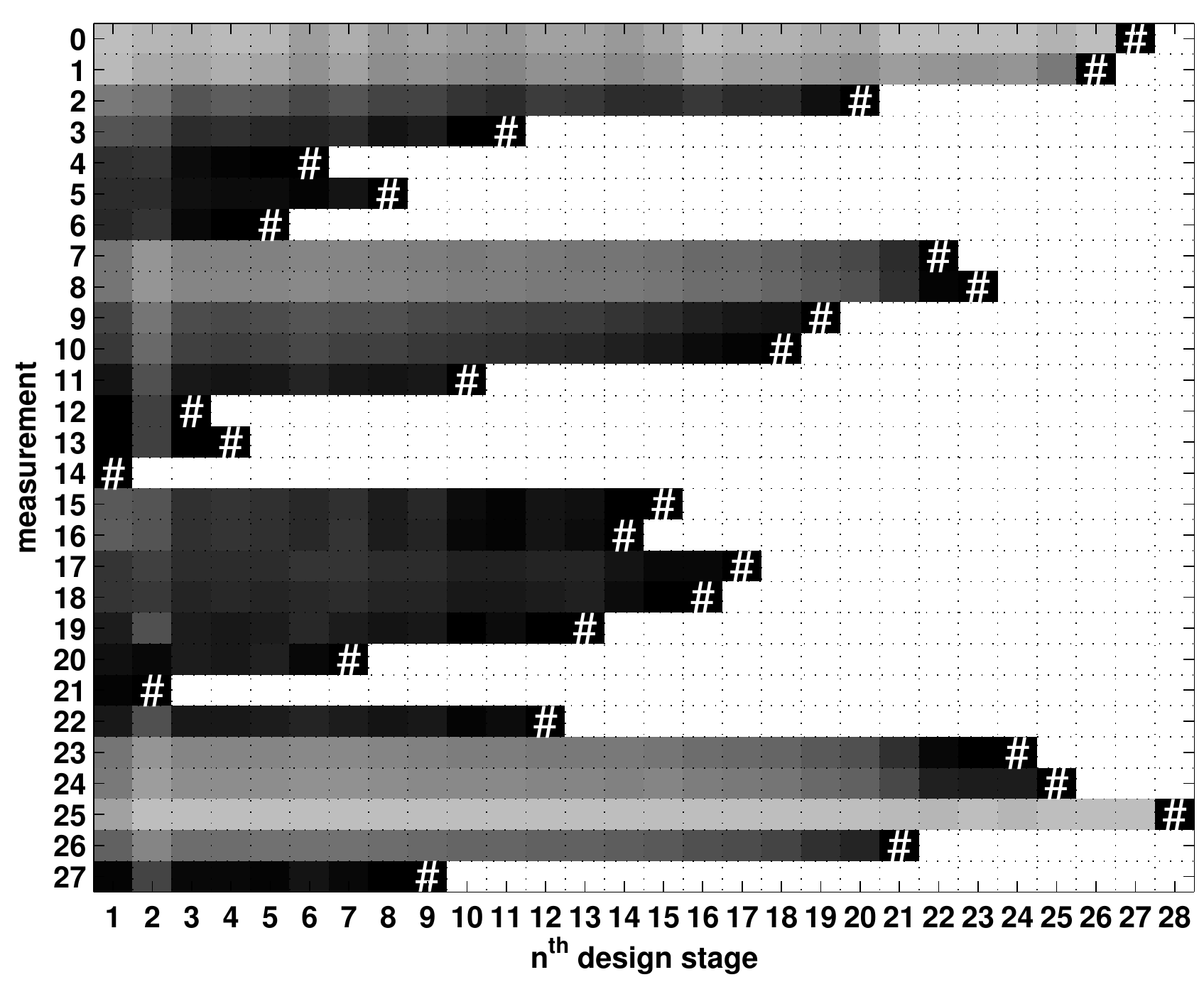}\label{ex3_meas_order_ENT}} 
    \caption{Results for example $3$ (real data). Relative expected utility for Information Maximization Sampling and 
      Maximum Entropy Sampling. Darker shadings indicate large expected utilities (mutual information for IM or
      entropy for ME) relative to all available scenarios within a design stage. The $\#$ sign indicates which
      scenario has been selected at a given stage. Previously selected scenarios are excluded from subsequent
      analysis which is indicated here by white cells.}\label{fig:ex3_meas_order}
  \end{center}
\end{figure*}

\begin{figure*}
  \begin{center}
    \subfigure[Entropy]{\includegraphics[width=3.5in]{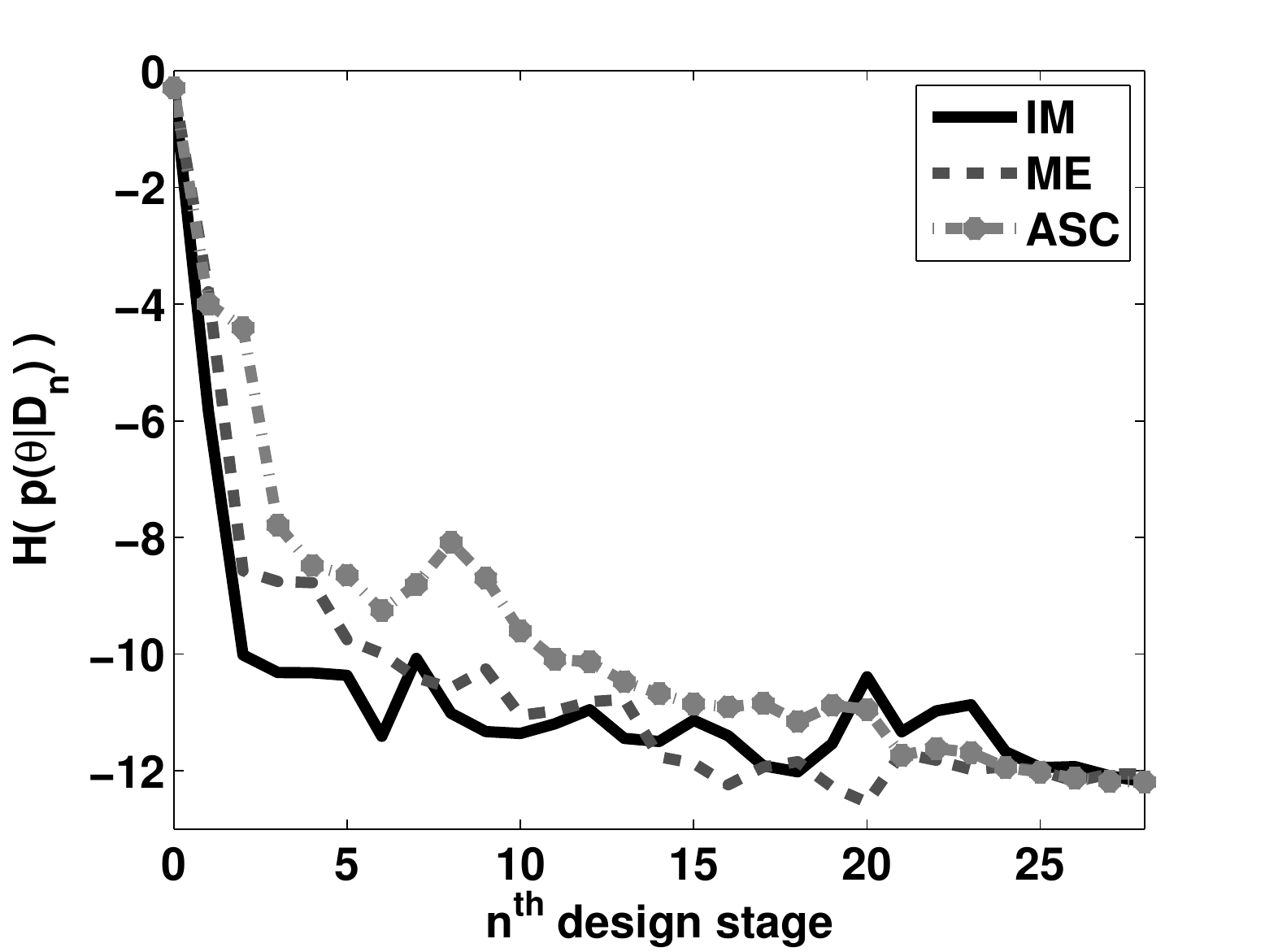}\label{ex3_ent}} 
    \subfigure[Kullback Leibler divergence]{\includegraphics[width=3.5in]{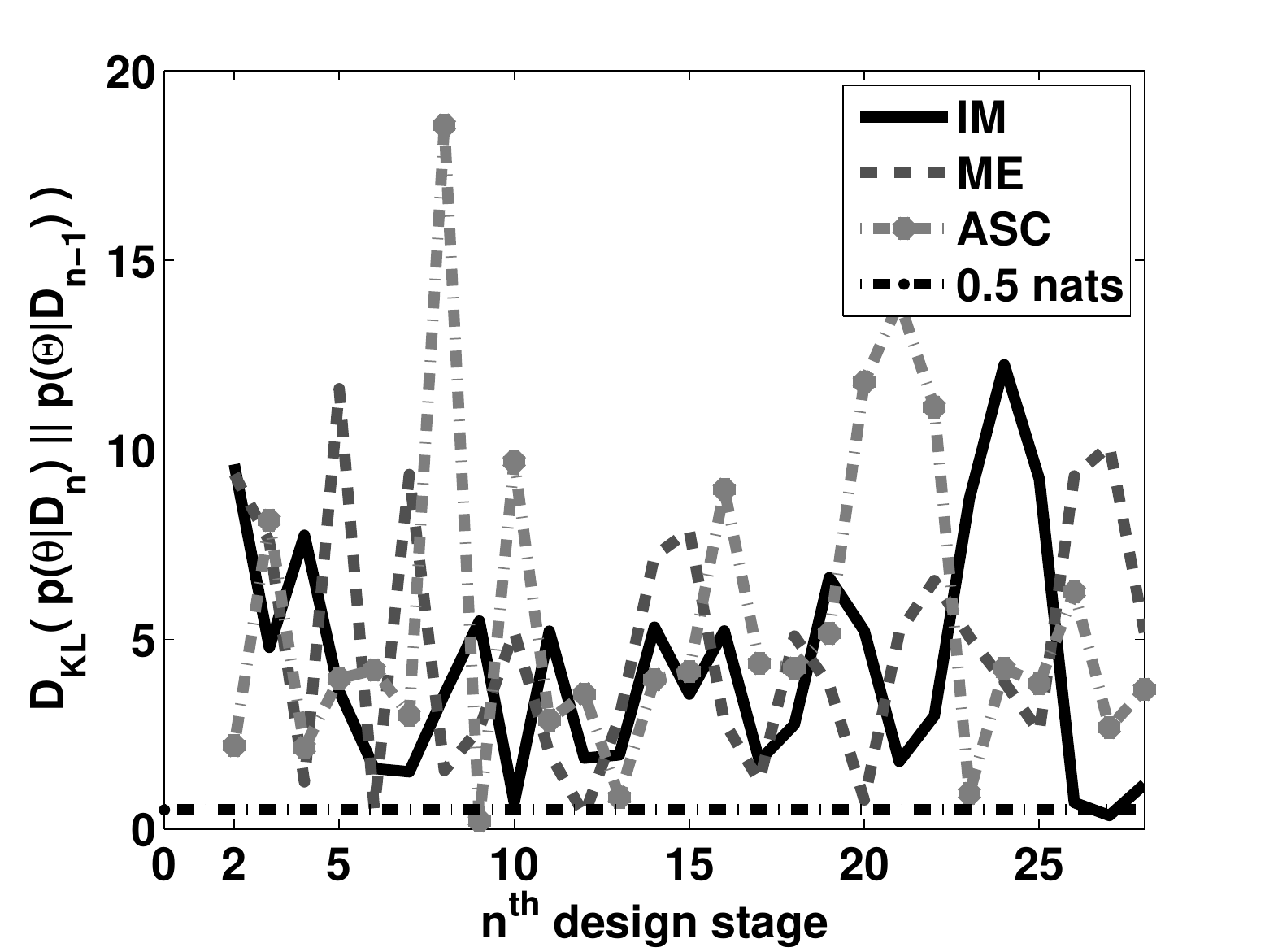}\label{ex3_kl}} 
    \caption{Results for example $3$ (real data). The evolution of the entropy and the KL divergence 
      given the posterior distributions yielded by the three strategies for experimental design:
      IM - Information Maximization Sampling, ME - Maximum Entropy Sampling and ASC - Ascending Sampling}\label{fig:ex3}
  \end{center}
\end{figure*}

\begin{figure*}
\begin{center}
\subfigure[Stage $6$ - $d_{eff}$]{\includegraphics[width=1.5in]{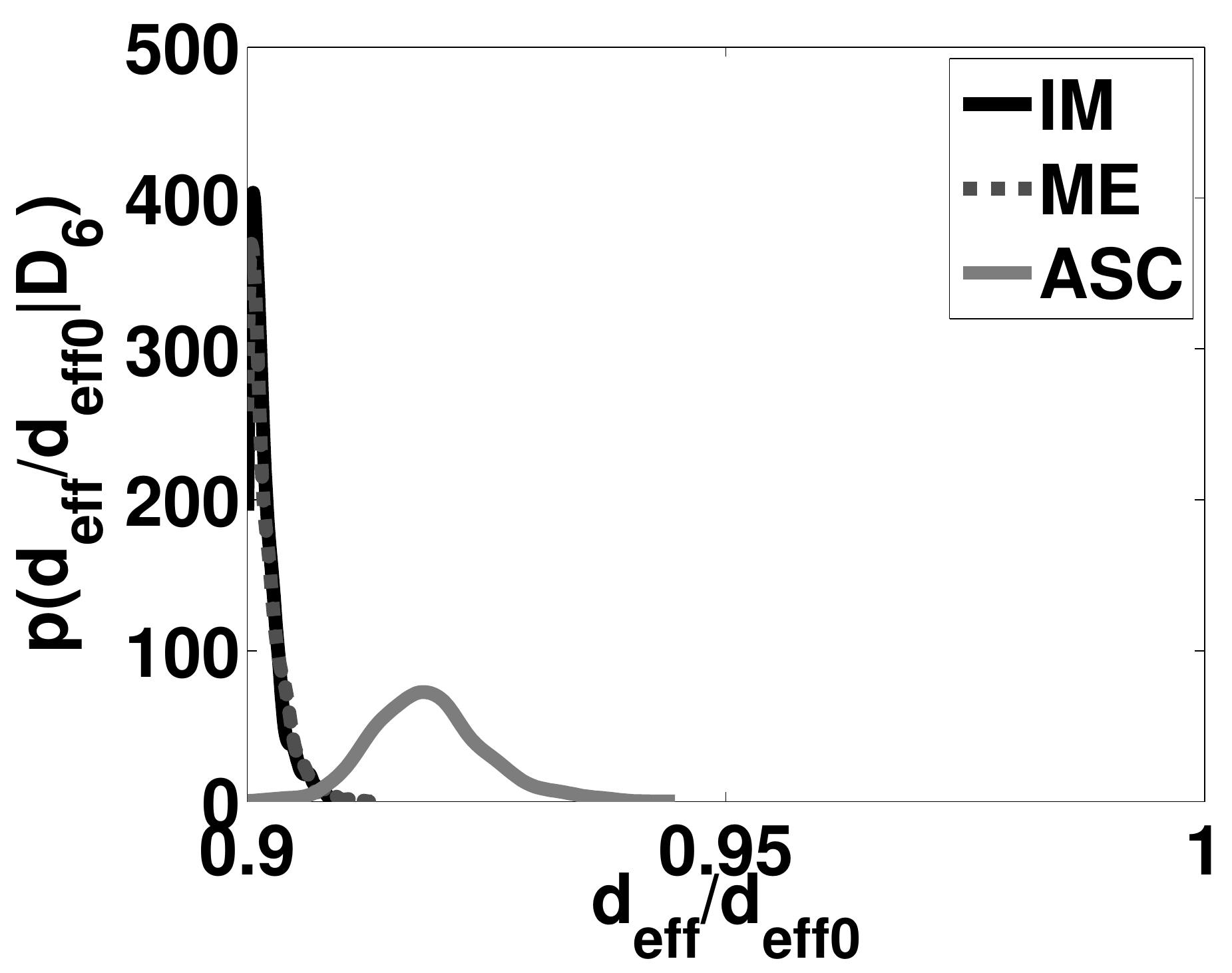}} 
\subfigure[Stage $6$ - $\gamma_N$]{\includegraphics[width=1.5in]{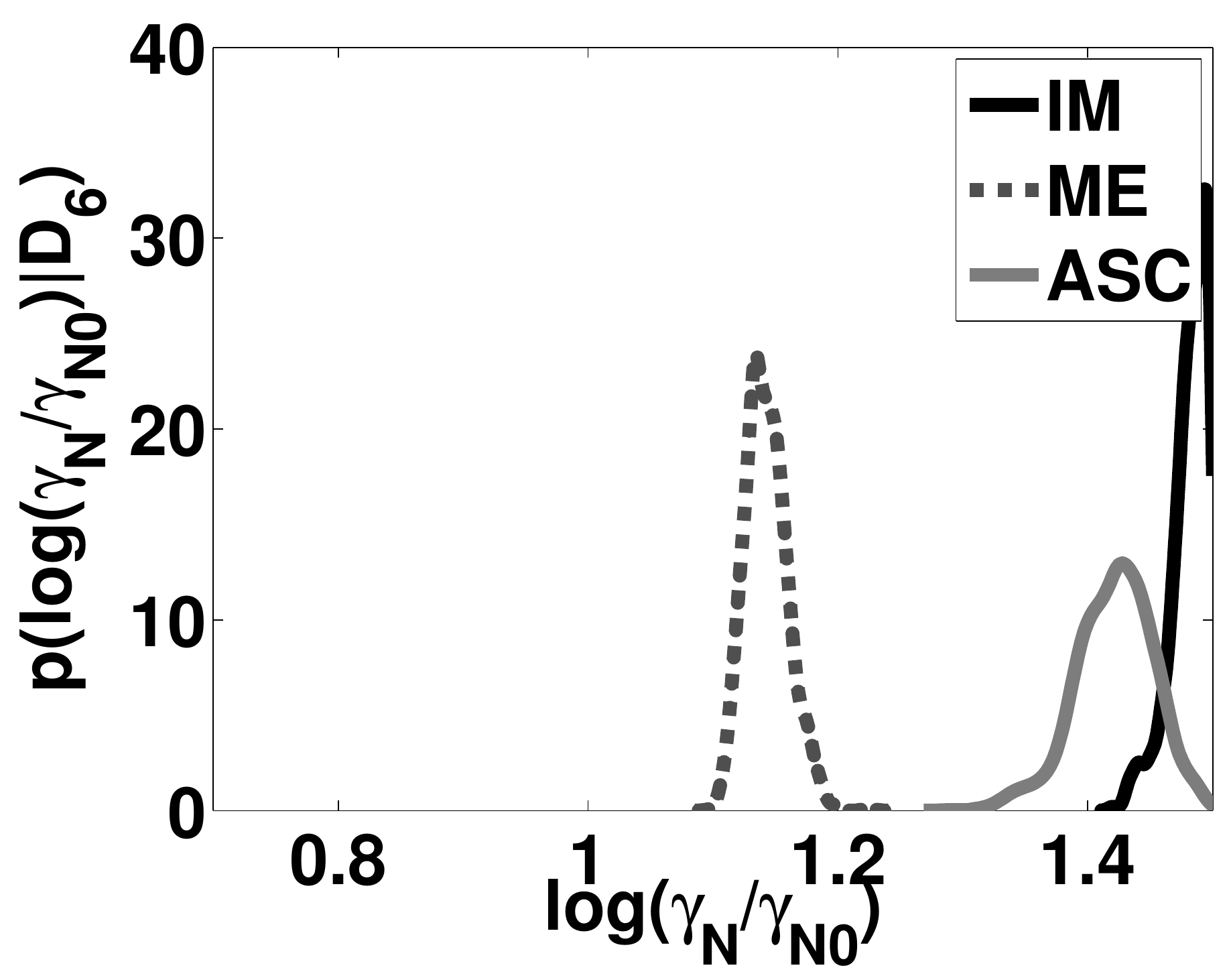}} 
\subfigure[Stage $6$ - $\beta_N$]{\includegraphics[width=1.5in]{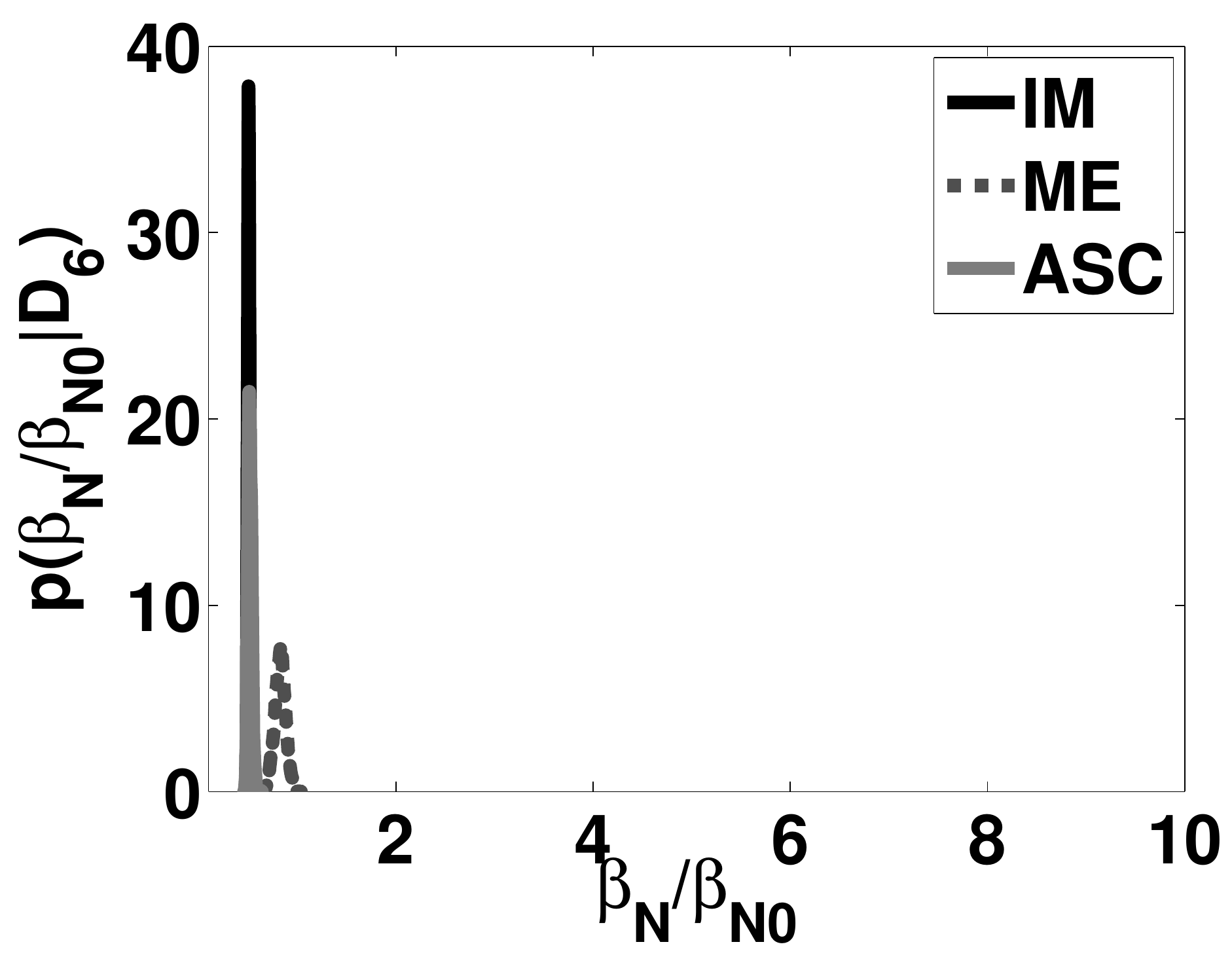}} 
\subfigure[Stage $12$ - $d_{eff}$]{\includegraphics[width=1.5in]{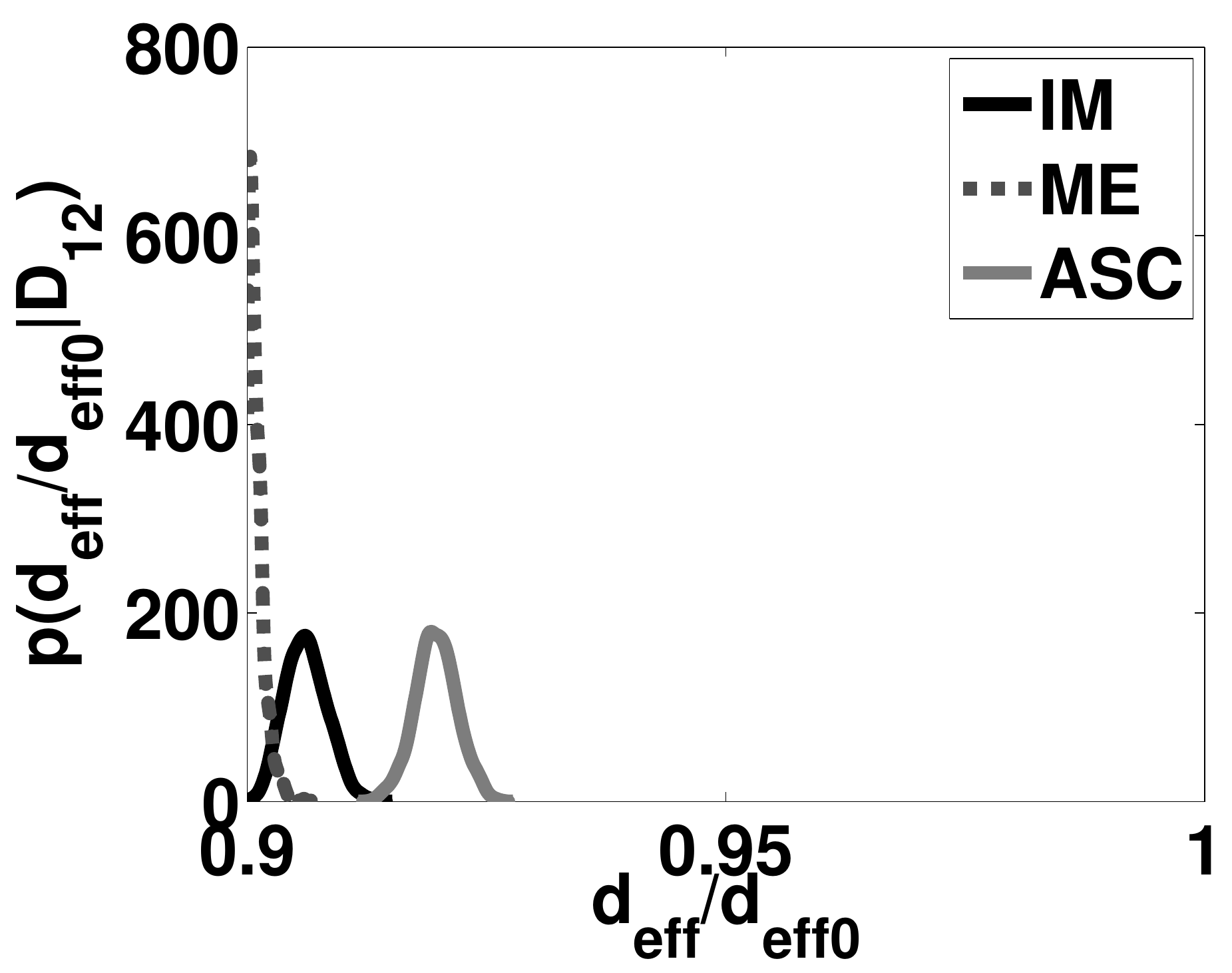}} 
\subfigure[Stage $12$ - $\gamma_N$]{\includegraphics[width=1.5in]{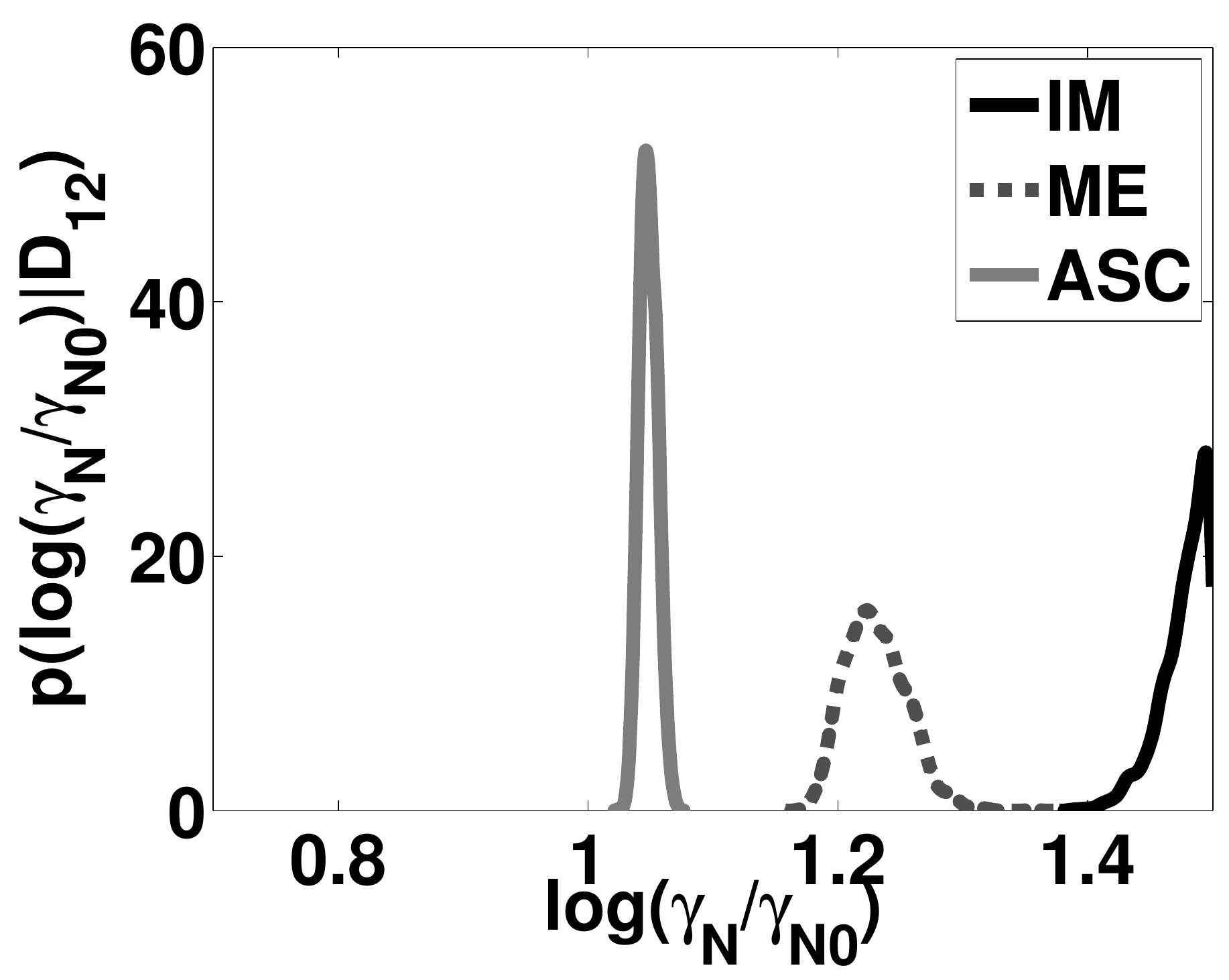}} 
\subfigure[Stage $12$ - $\beta_N$]{\includegraphics[width=1.5in]{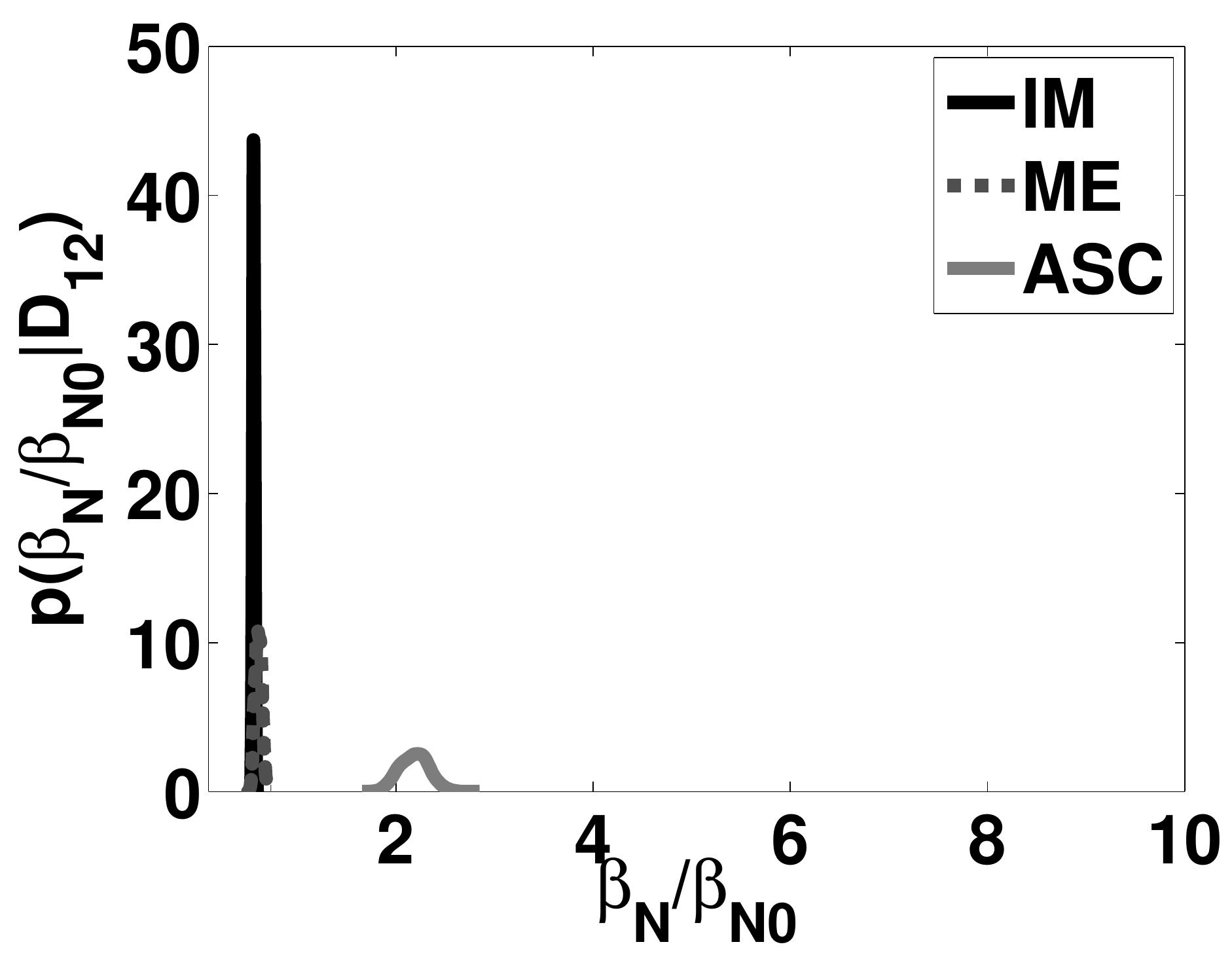}} 
\subfigure[Stage $18$ - $d_{eff}$]{\includegraphics[width=1.5in]{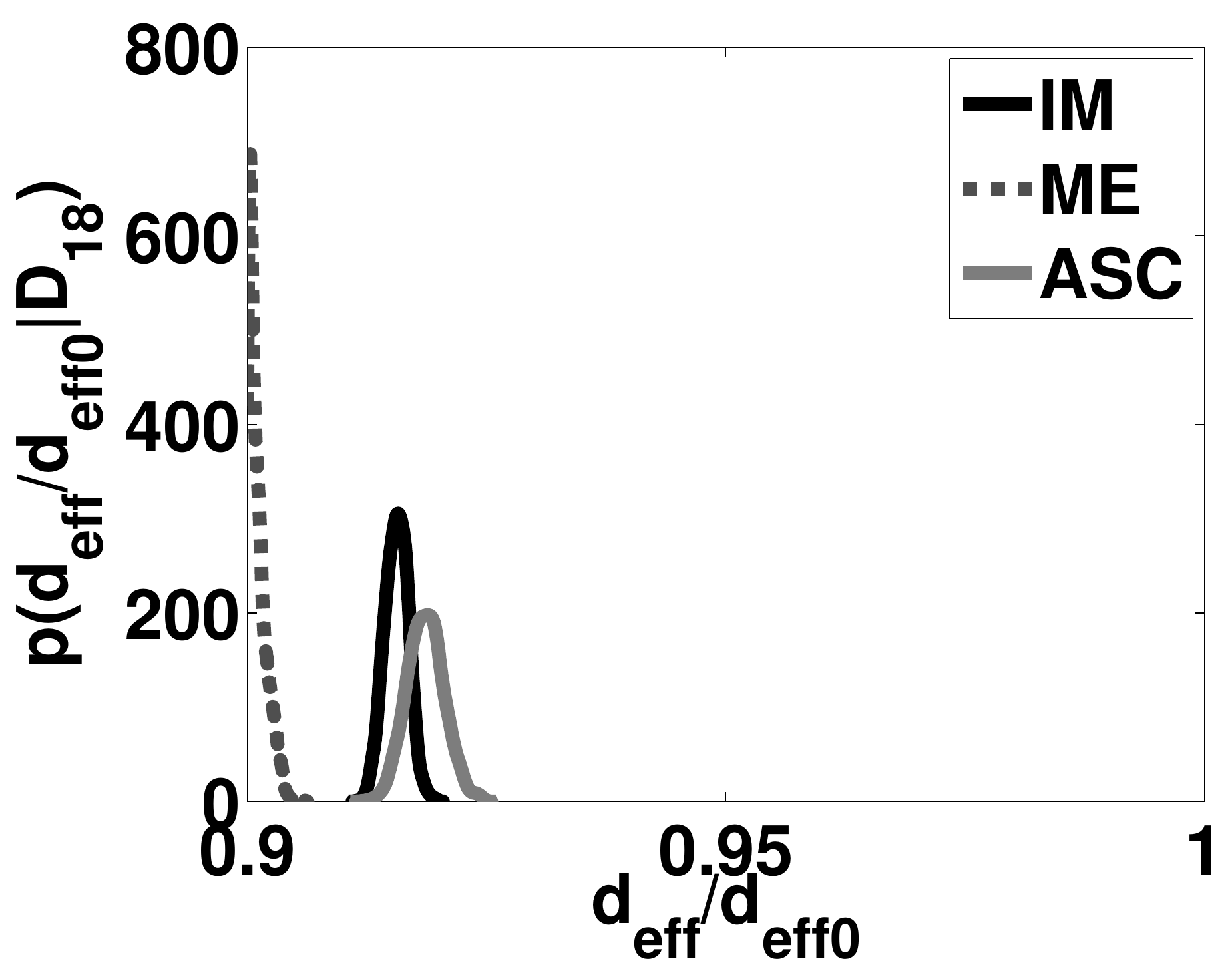}} 
\subfigure[Stage $18$ - $\gamma_N$]{\includegraphics[width=1.5in]{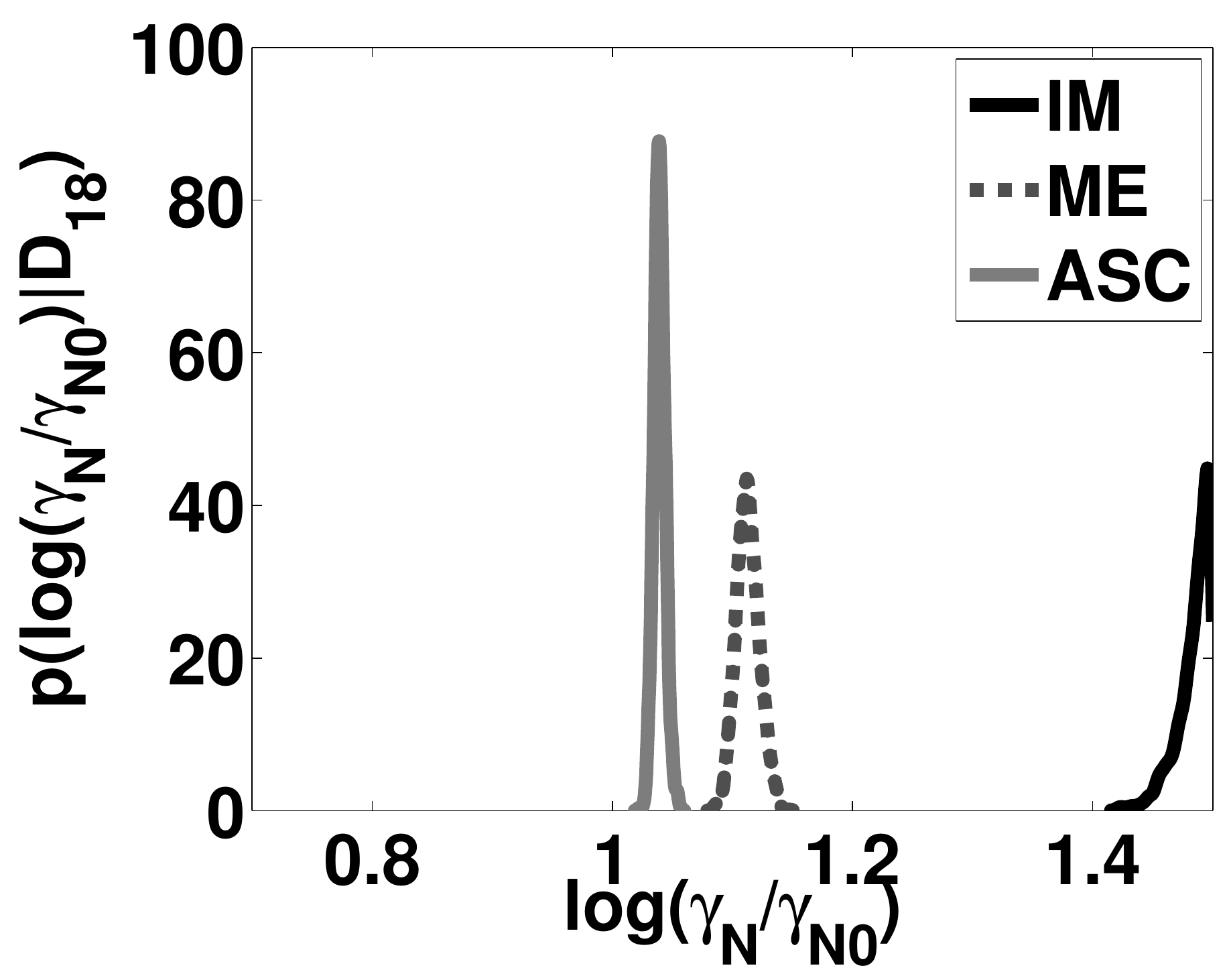}} 
\subfigure[Stage $18$ - $\beta_N$]{\includegraphics[width=1.5in]{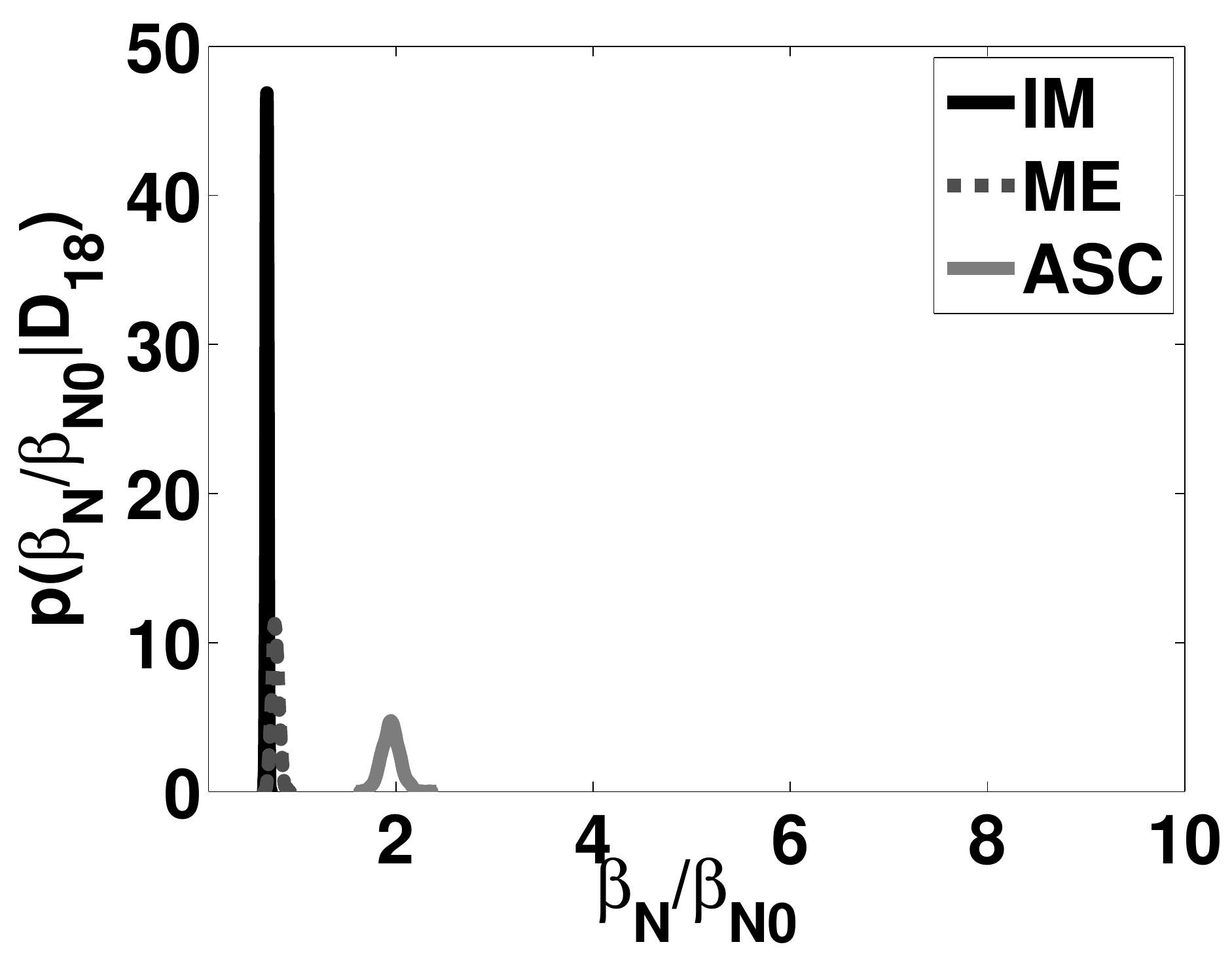}} 
\subfigure[Stage $24$ - $d_{eff}$]{\includegraphics[width=1.5in]{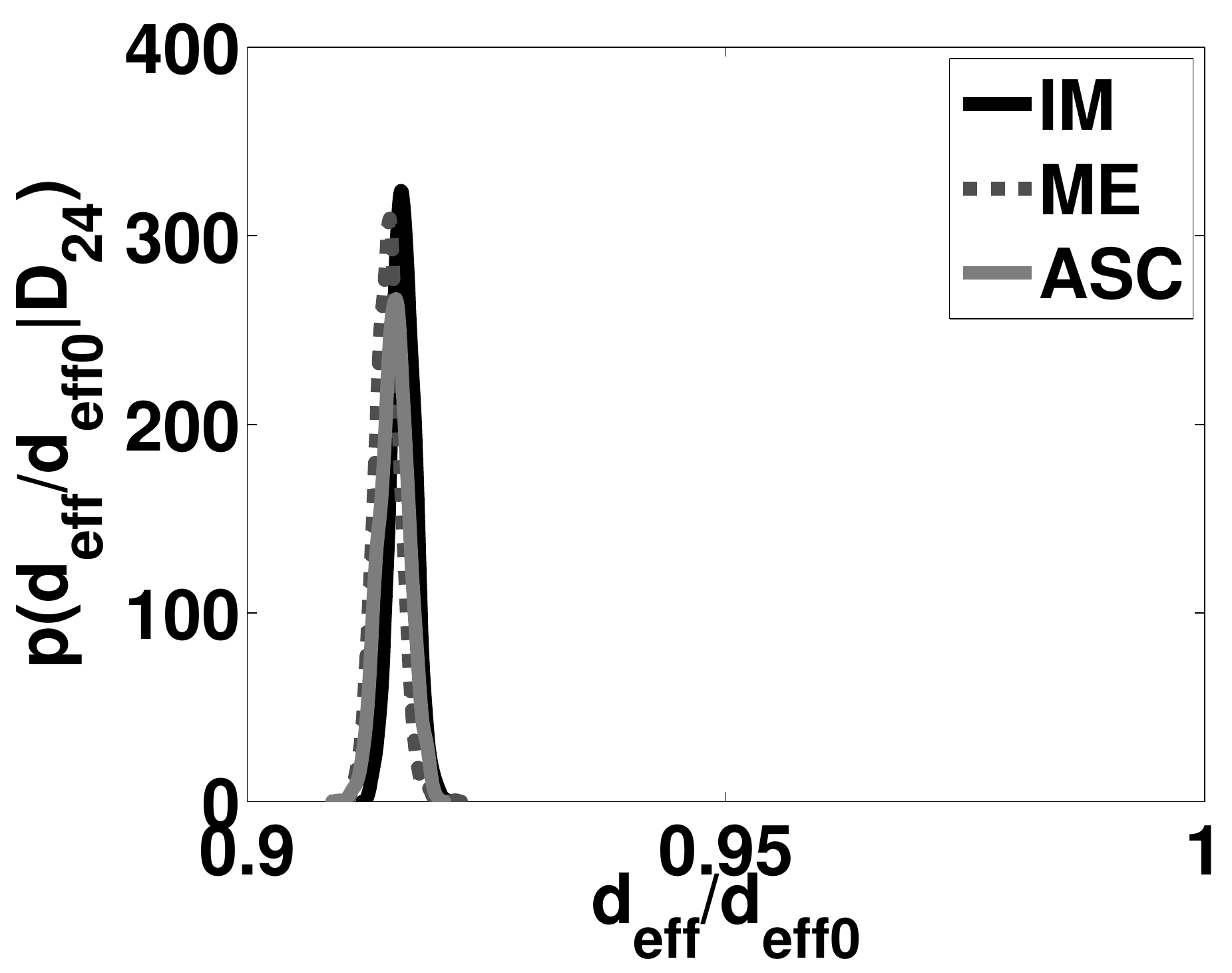}} 
\subfigure[Stage $24$ - $\gamma_N$]{\includegraphics[width=1.5in]{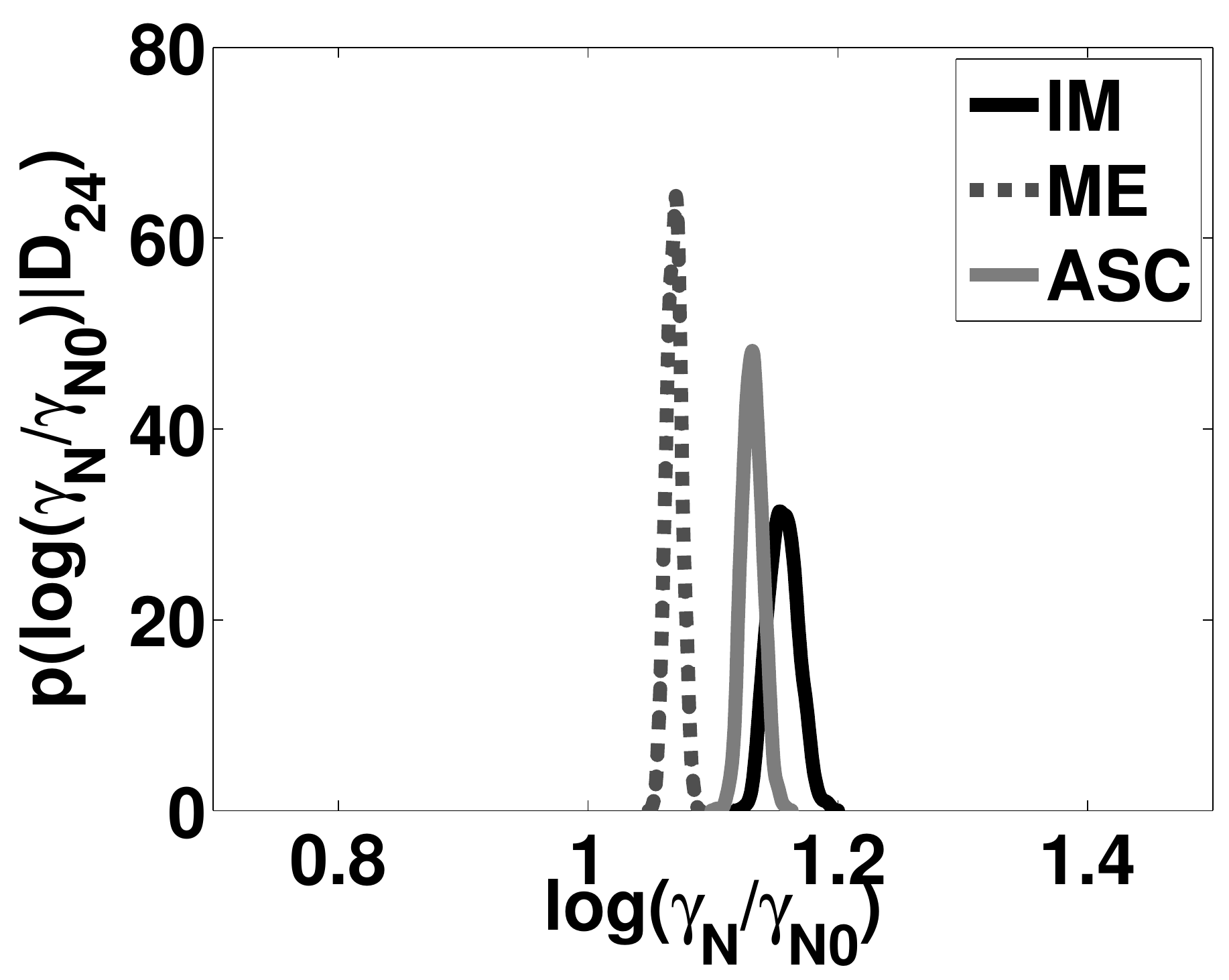}} 
\subfigure[Stage $24$ - $\beta_N$]{\includegraphics[width=1.5in]{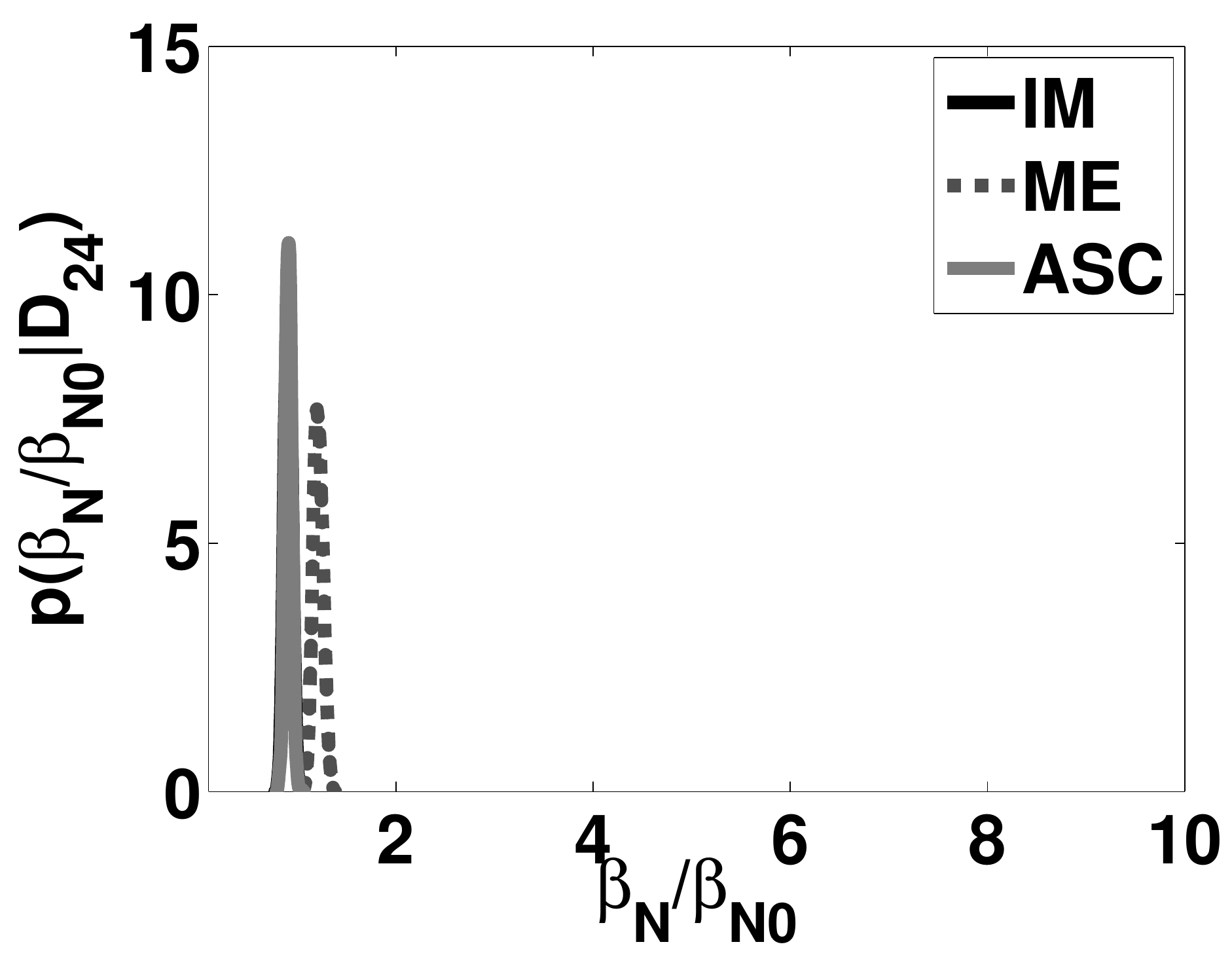}} 
\subfigure[Stage $28$ - $d_{eff}$]{\includegraphics[width=1.5in]{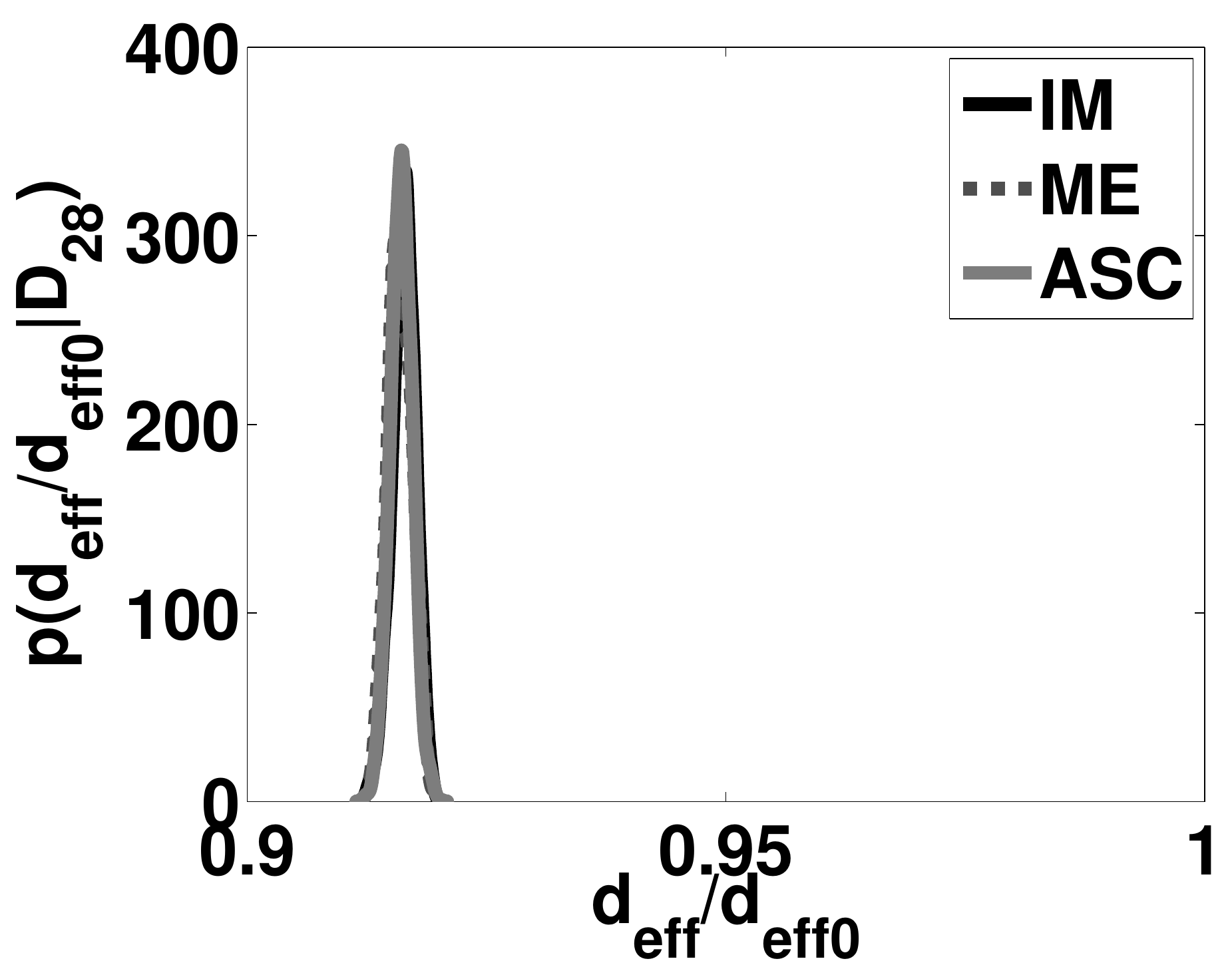}} 
\subfigure[Stage $28$ - $\gamma_N$]{\includegraphics[width=1.5in]{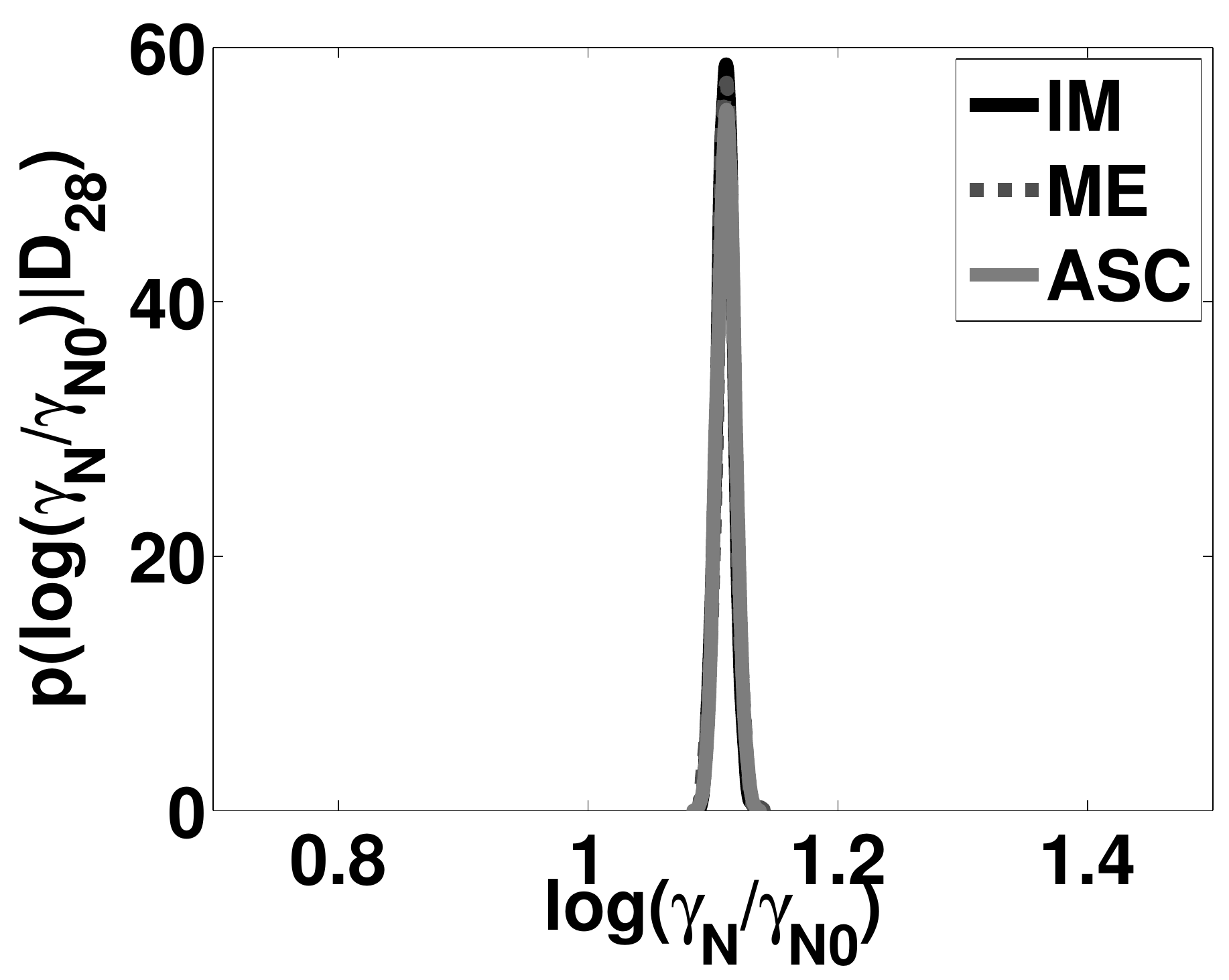}} 
\subfigure[Stage $28$ - $\beta_N$]{\includegraphics[width=1.5in]{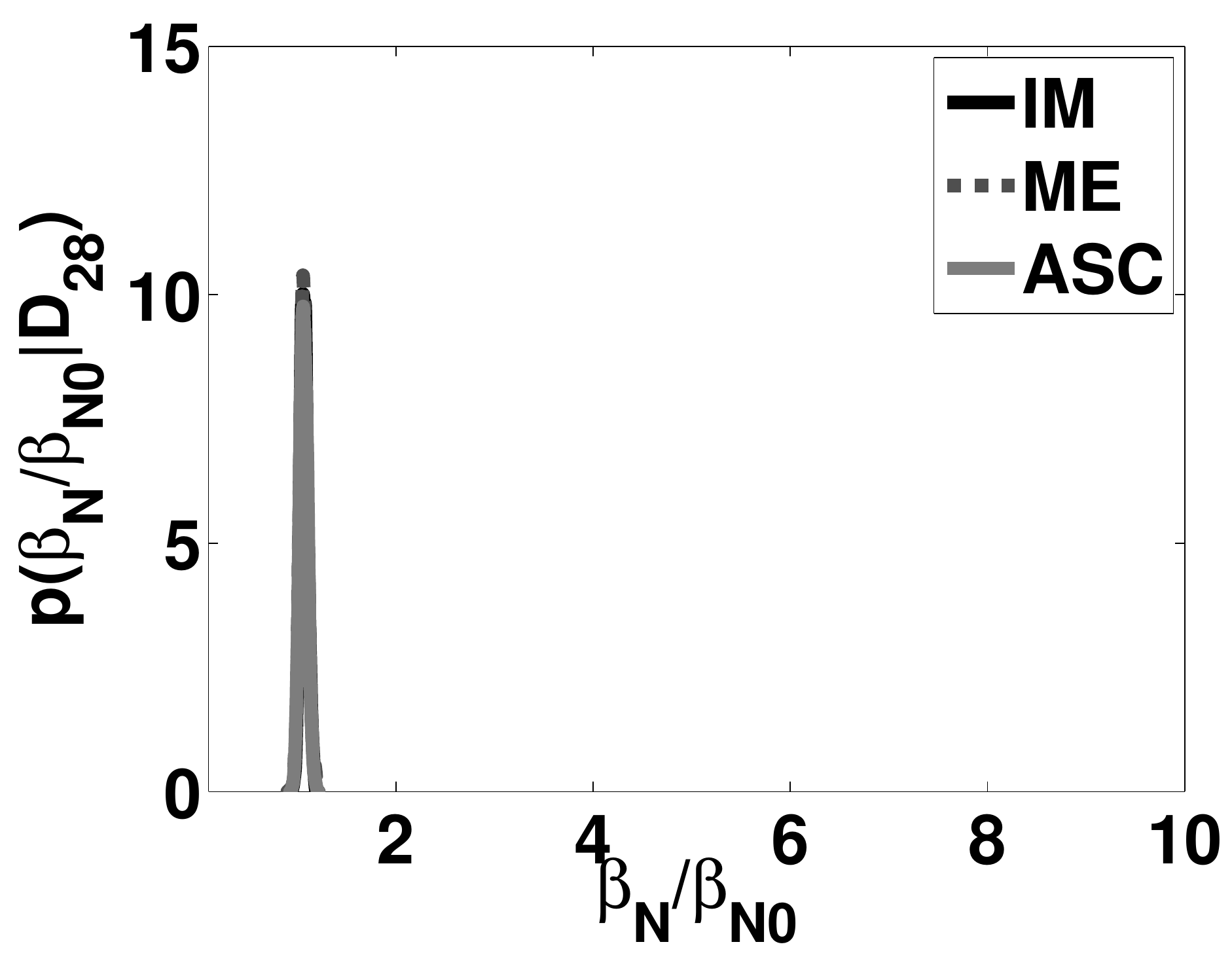}} 
\caption{Results for example $3$ (real data). Marginal pdfs of model parameters given by the three strategies after a selected number of experimental design stages}\label{fig:ex3_pdfs}
\end{center}
\end{figure*}

The same analysis as in the previous example is now carried on real experimental data as provided in 
\cite{Marschall_nitridation_AIAA}. As in the simulated measurement case, both the information theoretic 
approaches give design sequences which start with high flow designs. See Table \ref{tab:ex3_seq} for
the corresponding design sequences or Fig. \ref{fig:ex3_meas_order} for the relative expected utility 
corresponding to the two information theoretic strategies at each design stage. 
The profile of the uncertainty reduction showed in Fig.\ref{ex3_ent} is similar with the one in 
Fig.\ref{ex2_ent}. However, the same cannot be said about the KL divergence plotted in Fig.\ref{ex3_kl}. 
No matter which design strategy is used, each subsequent measurement brings new information which moves the 
high probability support of the pdf from one stage to another, see Fig.\ref{fig:ex3_pdfs} for marginal pdfs 
at different design stages. As mentioned in the previous section, the model discrepancy is assumed to be 
Gaussian and thus in this case the likelihood function has nonzero tails. Given that the prior is 
uniform, the subsequent posterior pdfs will have nonzero tails over the support of the prior. 

When compared with the simulated data, this behavior is indicative of conflicting information 
between experimental measurements and model predictions. In this case the posterior distribution
asserts that the high probability support of both the prior and the data are highly implausible, 
for more information see \cite{OHagan2004}. When dealing with real data this is not unexpected
given that most of the time we have model error or some of the measurements are not self-consistent. 
In other words, either different biases exist in the measurements or the model along with the 
probabilistic assumptions have limited explanatory power of the real phenomenon, or both
model errors and corrupted measurements can exist at the same time. In such situations, 
the experimental process should be stopped after a pre-determined number of experiments
to find the cause of the disagreement. While one can use a more direct way to check for 
this disagreement, the authors believe that using the KL divergence in this settings can provide
another useful tool for model calibration diagnostics.

\begin{figure}[!htp]
  \begin{center}
    \includegraphics[width=5.0in]{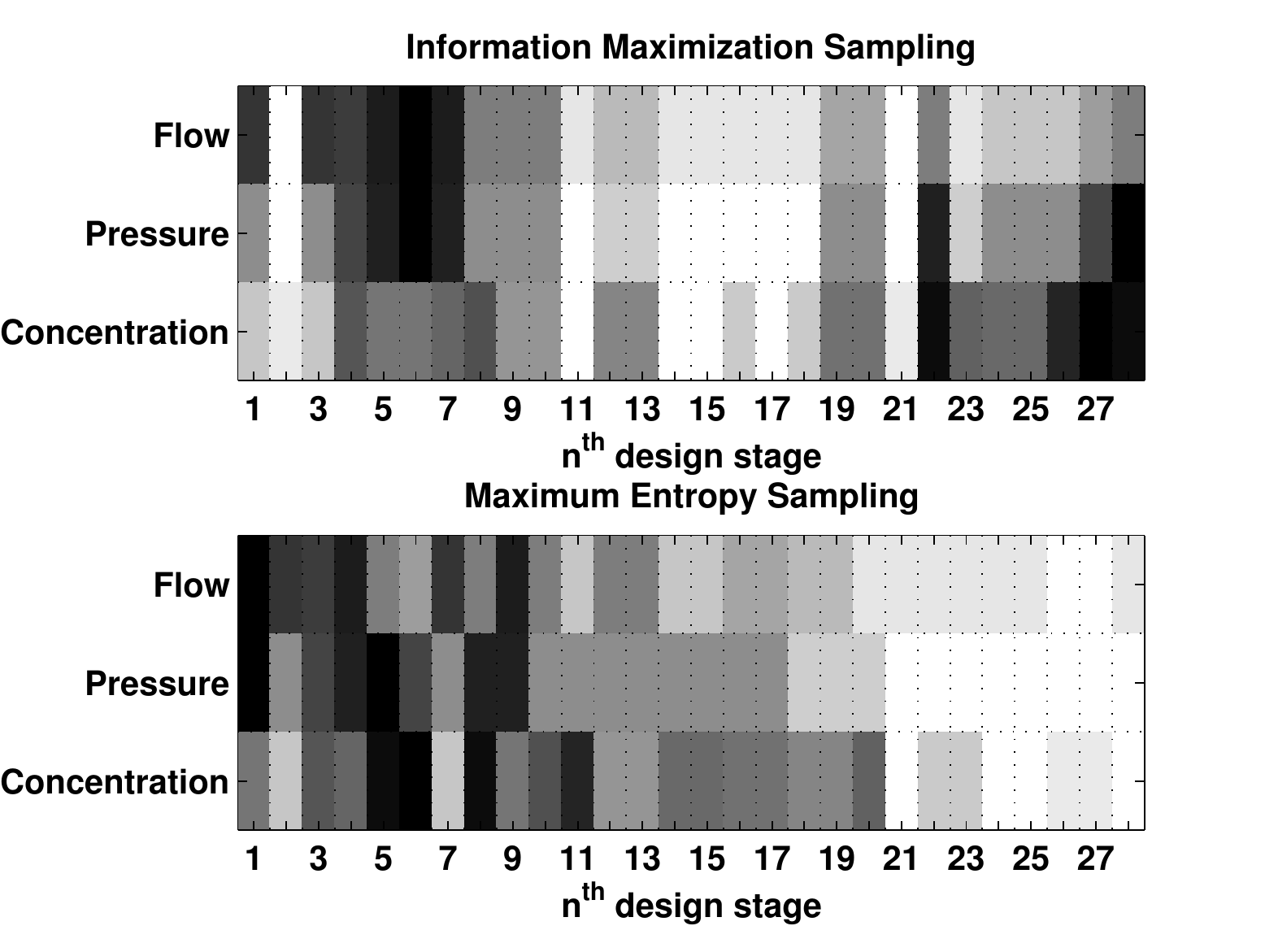}
  \end{center}
  \caption{Results for example $3$ (real data). Graphical representation for the
    inflow conditions selected by the two information theoretic strategies. Darker shadings
    indicate large inflow conditions such as flow, pressure, and concentration.}
  \label{fig:ex3_inflow}
\end{figure}

Generally speaking, the order of measurement sequence picked by ME sampling is very close to one with the 
simulated data case. In other words, it picks the scenario based on the large values of either 
$F_{N_2,in}$, $P_1$, $\mathcal{X}_{N,1}$, see Fig.\ref{fig:ex3_inflow}. However, IM sampling picks the experimental scenarios 
in a slightly different way from ones with the simulated data. For example, with the real data, IM sampling prefers 
the large values of inflow conditions in the early stage and does not alternatively pick different scenarios 
until 10$^{th}$ design stage. Note 1$^{st}$ measurement is selected at the second stage, however the choice 
of 0$^{th}$ measurement is further delayed. 

As mentioned above, the large  values of the inflow scenario parameters are more informative due to the abundance of nitrogen atoms at the surface of the sample. However, when the real
experimental data is used, this kind of data can be problematic because the physical model error and experimental 
uncertainty tend to be magnified in such a condition.

In Fig. \ref{fig:ex3_pdfs}, the pdfs of $d_{eff}$ from ME sampling stay close to the lower bound until 
24$^{th}$ design stage. However, the pdfs from IM sampling start to shift toward the final estimate earlier (say, after 
it considers 26$^{th}$ measurement that has a quite small volumetric flow rate). This explains why the sampling in
ascending order considering 0$^{th}$ and 1$^{st}$ measurements first  does the better job for this parameter.
For the other parameters, $\gamma_N$ and $\beta_N$, the large values of the inflow scenario parameters are preferable for
getting more information due to enhancement of the chemical reactions. 
In the middle column of Fig. \ref{fig:ex3_pdfs}, the posterior pdfs of $\gamma_N$ from IM sampling and ME sampling 
appreciably varies from Stage 6 (Fig. \ref{fig:ex3_pdfs}(b) to Stage 12 (Fig. \ref{fig:ex3_pdfs}(e)).  
 
This can be partially understood by the experimental data uncertainties related to  $13^{th}$ measurement and 
 27$^{th}$ measurement. The authors recognize these measurements are problematic and probably bring 
quite different information on this parameter. The scenarios of these experiments are almost identical.
Nevertheless, the observed N-atom concentration is quite different (i.e., $\chi_{N,13}/\chi_{N,27}\approx 
1.6$). As a result, the measurements of the mass loss that are used for calibration significantly differ each other.  
In fact, the local peaks seen in Fig. \ref{ex3_ent} at 9$^{th}$ design stage from ME sampling and
at 7$^{th}$ design stage from IM sampling corresponds to these experiments.      
Moreover, the knurling feather seen in Fig. \ref{ex3_kl} appears to  be related to some inconsistency in the
experimental measurement. For example, for ME sampling, the first peak around 4$^{th}$ and 5$^{th}$  
design stage are related to 13$^{th}$ and 6$^{th}$ measurements. The problem can be substantiated by a simple
analysis as it follows. From Eq. \eqref{eq:ablation_rate_GK}, we get the following relationship in an 
approximate manner:
\begin{eqnarray} \label{approx_M_C}
\Delta M_C \approx \Delta t \bar{C}_N \approx \Delta t \overline{P\chi}_N,   
\end{eqnarray}
where $\bar{\ast}$ is the averaged value along the sample and $\chi_N$ is the mole fraction ($=C_N R T/P$).
Using the linear interpolation for the pressure and concentration, we can roughly estimate $\bar{\chi}_N$ and 
$\bar{P}$ by averaging the experimental measurements of $P$ and $\chi$ at two measurement locations. 
The estimated ratio of the mass loss for these experiment by Eq. \eqref{approx_M_C} is given by
$\Delta M'_{C,6}/\Delta M"_{C,13}\approx 0.9$. This is significantly different from the actual measurement, $0.7$.
The same kind of analysis explains the first peak appearing at 3$^{rd}$ and 4$^{th}$ design stage,
corresponding to 20$^{th}$ and 12$^{th}$ measurements, for IM sampling.
($\Delta M'_{C,20}/\Delta M"_{C,12}\approx 0.35$ that is much larger than the measurement, $0.18$).    
These simple analysis illustrates different experimental data provide the model with different information. 
As  a consequence, Fig. \ref{ex3_kl} is significantly different from Fig. \ref{ex2_kl} where the simulated 
data is perfectly consistent with the model itself.

\section{Conclusions}
\label{Conclusion}

In this paper the experimental design problem is formulated in the Bayesian framework,
and it is shown to be equivalent with an information theoretic sensitivity analysis
between model parameters and observables. The optimal design selected by the information
maximization sampling is the one which provides the highest statistical dependence
between the two random variables. The statistical dependence is quantified using mutual 
information which is approximated in this paper using a $k$ nearest neighbor estimator. 
Theoretically the information maximization sampling is more general than the maximum
entropy sampling and more efficient than random sampling.

The entropy and Kullback-Leibler information metrics are used to monitor the learning
process. A decreasing trend in entropy is indicative that the uncertainty in model 
parameters is reduced by assimilating each additional measurement. The information gain
brought by each measurement is quantified by the Kullback-Leibler divergence. Large 
values are indicative that the model is still learning. A large divergence between 
subsequent posterior distributions might be explained by a large uncertainty reduction or
the existence of conflicting information between model predictions and measurements. 
When the trend in the Kullback-Leibler divergence is decreasing one can stop the
experimental process given a level of accuracy, otherwise one may still choose to
stop the experimental process to check the formulation of the model, the probabilistic
assumptions and/or analyze the collected measurements to find potential outliers.

\section*{Appendix A: Estimating the Kullback-Leibler divergence using $k$NN}
\label{app:kl_div}

Similar with the entropy and mutual information, the approximation of the Kullback-Leibler 
divergence is based on a $k$ nearest neighbor approach as proposed in \cite{Wang2006}. 
\begin{eqnarray}
  D_{KL}(p(x|D_n) || p(x|D_{n-1}) \approx \frac{d_X}{N_n} \sum_{i=1}^{N_n} 
  \log \frac{\nu_{n-1}(i)}{\rho_n(i)} + \log \frac{N_{n-1}}{N_n-1}
\end{eqnarray}
where $d_X$ is the dimensionality of the random variable $X$, $N_n$ and $N_{n-1}$ give the 
number of samples $\{X_n^i|i=1,\ldots,N_n\} \sim p(x|D_n)$ and 
$\{X_{n-1}^j|j=1,\ldots,N_{n-1}\} \sim p(x|D_{n-1})$ respectively, and the two distances 
$\nu_{n-1}(i)$ and $\rho_n(i)$ are defined as follows:
\begin{eqnarray}
  \rho_n(i) &=& \min_{j=1\ldots N_n, j \ne i} \| X_n^i - X_n^j \|_\infty \\
  \nu_{n-1}(i) &=& \min _{j=1\ldots N_{n-1}} \| X_n^i - X_{n-1}^j \|_\infty
\end{eqnarray}


\section*{Appendix B: Information-theoretic sensitivity analysis}
\label{app:MI_sens}

Mutual information is a convenient way to quantify the statistical dependence
between two random variables, $X_1$ and $X_2$:
\begin{eqnarray}
\label{mutual_information}
I(X_1, X_2) = \int f(x_1, x_2) \log \frac{f(x_1, x_2)}{f_{X_1}(x_1) f_{X_2}(x_2)} 
\mathrm{d}x_1 \mathrm{d}x_2
\end{eqnarray}

A more formal relation between statistical dependence and mutual information
can be shown with the help of copula functions as presented in
\cite{Calsaverini2009}, and included here for completeness.
\cite{Sklar1959} proved that the joint distribution of two random
variables, $F( x_1, x_2 )$, can be represented using a \textit{copula} function, 
$C: [0,1]^2 \rightarrow [0,1]$, and the marginals distributions, $F_{X_i}$:
\begin{eqnarray}
\label{copula}
F( x_1, x_2 ) = C( F_{X_1}(x_1), F_{X_2}(x_2) )
\end{eqnarray}

Given that $F_{X_1}$ and $F_{X_2}$ are continuous, then $F_{X_1}(X_1)$ and
$F_{X_2}(X_2)$ are two uniformly distributed random variables. Thus, the copula 
function can be regarded as a joint cumulative distribution of two uniformly
distributed random variables. It separates the study of the dependence between
random variables and their marginal distributions. By denoting $u_i = F_{X_i}(x_i)$, 
Eq. \eqref{copula} can be written as follows:
\begin{eqnarray}
F( F_{X_1}^{-1}(u_1), F_{X_2}^{-1}(u_2) ) = C( u_1, u_2 )
\end{eqnarray}

Assuming that the densities $f_{X_i}$ exist, then the joint probability density 
function is given by:
\begin{eqnarray}
\label{joint_pdf_mi}
f( x_1, x_2 ) &=& \frac{\partial^2}{\partial x_1 \partial x_2} F( x_1, x_2 ) \nonumber \\
&=& c( F_{X_1}(x_1), F_{X_2}(x_2) ) f_{X_1}(x_1) f_{X_2}(x_2)
\end{eqnarray}
where $c( F_{X_1}(x_1), F_{X_2}(x_2) ) = \frac{\partial^2}{\partial x_1 \partial x_2} 
C( F_{X_1}(x_1), F_{X_2}(x_2) )$ is the copula density. 

By substituting Eq. \eqref{joint_pdf_mi} in Eq. \eqref{mutual_information}, and using that
$u_i = F_{X_i}(x_i)$, the mutual information can be rewritten in terms of copula density function. 
\begin{eqnarray}
I(X_1, X_2) &=& \int c( F_{X_1}(x_1), F_{X_2}(x_2) ) f_{X_1}(x_1) f_{X_2}(x_2) \log 
c( F_{X_1}(x_1), F_{X_2}(x_2) ) \mathrm{d}x_1 \mathrm{d}x_2 \nonumber \\
\label{mi_entropy_copula}
&=& \int c( u_1, u_2 ) \log c( u_1, u_2 ) \mathrm{d}u_1 \mathrm{d}u_2
\end{eqnarray}

Eq. \eqref{mi_entropy_copula} reveals that mutual information is the negative copula
entropy. Thus it does not depend on the marginal distributions, it only quantifies
the dependence between the random variables which is contained in the copula function. 
This has been pointed out in literature in \cite{Calsaverini2009}.

\section*{Acknowledgments}

We thank Dr. Jochen Marschall from SRI International for valuable discussions
regarding the nitridation experiment.
This material is based upon work supported by the Department of 
Energy [National Nuclear Security Administration] under Award Number [DE-FC52-08NA28615].


\bibliographystyle{elsarticle-num-names}
\bibliography{NitroExpDes_main}

\clearpage

\begin{table}[htbp]
  \centering
  \begin{tabular}{|c|cc|ccc|ccc|ccc|}
    \hline
    \multicolumn{3}{|c|}{Sample Size $\rightarrow$} & 
    \multicolumn{3}{|c|}{$500$} &
    \multicolumn{3}{|c|}{$1,000$} &
    \multicolumn{3}{|c|}{$2,000$} \\
    \hline
    \multicolumn{3}{|c|}{Dimensionality} & 
    \multicolumn{3}{|c|}{$k$NN} &
    \multicolumn{3}{|c|}{$k$NN} &
    \multicolumn{3}{|c|}{$k$NN} \\
    \hline
    dim$(\mathbf{X},\mathbf{Y})$ & dim$(\mathbf{X})$ & dim$(\mathbf{Y})$ & 
    $3$ & $6$ & $9$ & $3$ & $6$ & $9$ & $3$ & $6$ & $9$\\
    \hline
    $2$ & $1$ & $1$ & $16\%$ & $10\%$ & $~8\%$ & $11\%$ & $~7\%$ & $~5\%$ & $~8\%$ & $~5\%$ & $~3\%$ \\
    $3$ & $2$ & $1$ & $21\%$ & $13\%$ & $11\%$ & $16\%$ & $~9\%$ & $~7\%$ & $12\%$ & $~7\%$ & $~5\%$ \\
    $4$ & $2$ & $2$ & $15\%$ & $11\%$ & $11\%$ & $10\%$ & $~7\%$ & $~6\%$ & $~7\%$ & $~5\%$ & $~4\%$ \\
    $5$ & $3$ & $2$ & $18\%$ & $14\%$ & $13\%$ & $13\%$ & $~9\%$ & $~9\%$ & $~9\%$ & $~6\%$ & $~9\%$ \\
    \hline
  \end{tabular}
  \caption{Absolute Relative Error for MI averaged over $1000$ trial runs}
  \label{tbl:rmseMI}
\end{table}

\begin{table}[htbp]
  \centering
  \begin{tabular}{|c|c|}
    \hline
    {\bf Strategy} & {\bf Sequence} \\
    \hline
    \multicolumn{2}{|c|}{Repeated measurements not allowed} \\
    \hline
    ME Sampling & $10~~1~~9~~2~~3~~8~~5~~7~~4~~6$ \\
    \hline
    IM Sampling & $~1~10~~9~~2~~3~~8~~7~~4~~6~~5$ \\
    \hline
    RND Sampling & $~6~10~~3~~1~~5~~9~~8~~2~~4~~7$ \\
    \hline
    \multicolumn{2}{|c|}{Repeated measurements allowed} \\
    \hline
    ME Sampling & $10~~1~10~~1~~1~10~10~~1~~1~~1$ \\
    \hline
    IM Sampling & $~1~10~~1~10~10~~1~10~10~~1~~1$ \\
    \hline
  \end{tabular}
  \caption{Results for example $1$, the additive noise case. Design sequences yielded by the three strategies for one particular Monte Carlo run}
  \label{tab:ex1_add_seq}
\end{table}

\begin{table}[htbp]
  \centering
  \begin{tabular}{|c|c|}
    \hline
    {\bf Strategy} & {\bf Sequence} \\
    \hline
    \multicolumn{2}{|c|}{Repeated measurements not allowed} \\
    \hline
    ME Sampling & $~5~~6~~7~~4~~1~~3~~8~~9~~2~10$ \\
    \hline
    IM Sampling & $~1~10~~9~~7~~2~~8~~4~~3~~5~~6$ \\
    \hline
    RND Sampling & $~6~~1~~8~~5~10~~9~~4~~2~~3~~7$ \\
    \hline
    \multicolumn{2}{|c|}{Repeated measurements allowed} \\
    \hline
    ME Sampling & $~5~~6~~6~~5~~5~~5~~6~~5~~6~~5$ \\
    \hline
    IM Sampling & $~1~10~~1~10~10~~1~~1~~1~~1~10$ \\
    \hline
  \end{tabular}
  \caption{Results for example $1$, the multiplicative noise case. Design sequences yielded by the three strategies for one particular Monte Carlo run}
  \label{tab:ex1_mult_seq}
\end{table}

\begin{table}[htbp]
  \centering
  \footnotesize
  \begin{tabular}{|c|c|}
    \hline
    {\bf Strategy} & {\bf Sequence} \\
    \hline
    ME Sampling & $14~21~6~20~13~12~27~19~5~11~22~4~18~17~10~3~9~16~15~26~2~7~23~8~24~25~1~0$ \\
    \hline
    IM Sampling & $21~5~20~0~1~13~27~14~2~12~15~16~11~22~19~9~10~3~4~6~26~25~23~7~8~24~18~17$ \\
    \hline
    ASC Sampling & $0~1~2~3~4~5~6~7~8~9~10~11~12~13~14~15~16~17~18~19~20~21~22~23~24~25~26~27$ \\
    \hline
  \end{tabular}
  \caption{Results for example $2$ (simulated data). Design sequences yielded by the three strategies.}
  \label{tab:ex2_seq}
\end{table}


\begin{table}[htbp]
  \centering
  \footnotesize
  \begin{tabular}{|c|c|}
    \hline
    {\bf Strategy} & {\bf Sequence} \\
    \hline
    ME Sampling & $14~21~12~13~6~4~20~5~27~11~3~22~19~16~15~18~17~10~9~2~26~7~8~23~24~1~0~25$ \\
    \hline
    IM Sampling & $21~1~20~12~27~14~13~11~19~22~26~9~10~25~23~7~24~8~18~17~0~5~2~15~16~3~4~6$ \\
    \hline
    ASC Sampling & $0~1~2~3~4~5~6~7~8~9~10~11~12~13~14~15~16~17~18~19~20~21~22~23~24~25~26~27$ \\
    \hline
  \end{tabular}
  \caption{Results for example $3$ (real data). Design sequences yielded by the three strategies.}
  \label{tab:ex3_seq}
\end{table}

\clearpage


\begin{figure}[!htp]
  \begin{center}
    \includegraphics[width=5.0in]{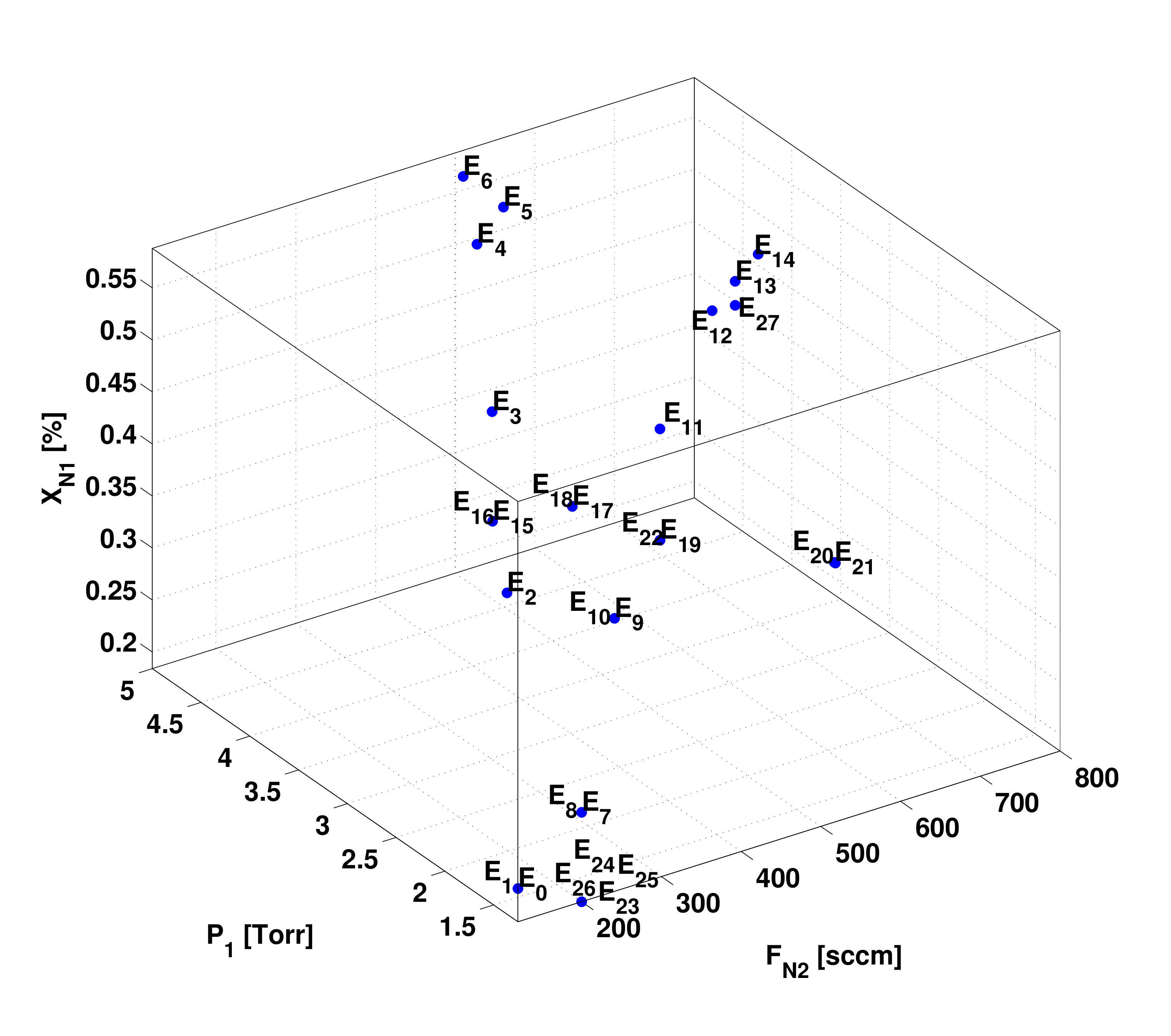}
  \end{center}
  \caption{Results for example $2$ (simulated data). 3D plot for inflow 
    scenarios used in Ref. \cite{Marschall_nitridation_AIAA}.}
  \label{fig:ex2_inflow}
\end{figure}

\begin{figure}[!htp]
  \begin{center}
    \includegraphics[width=5.0in]{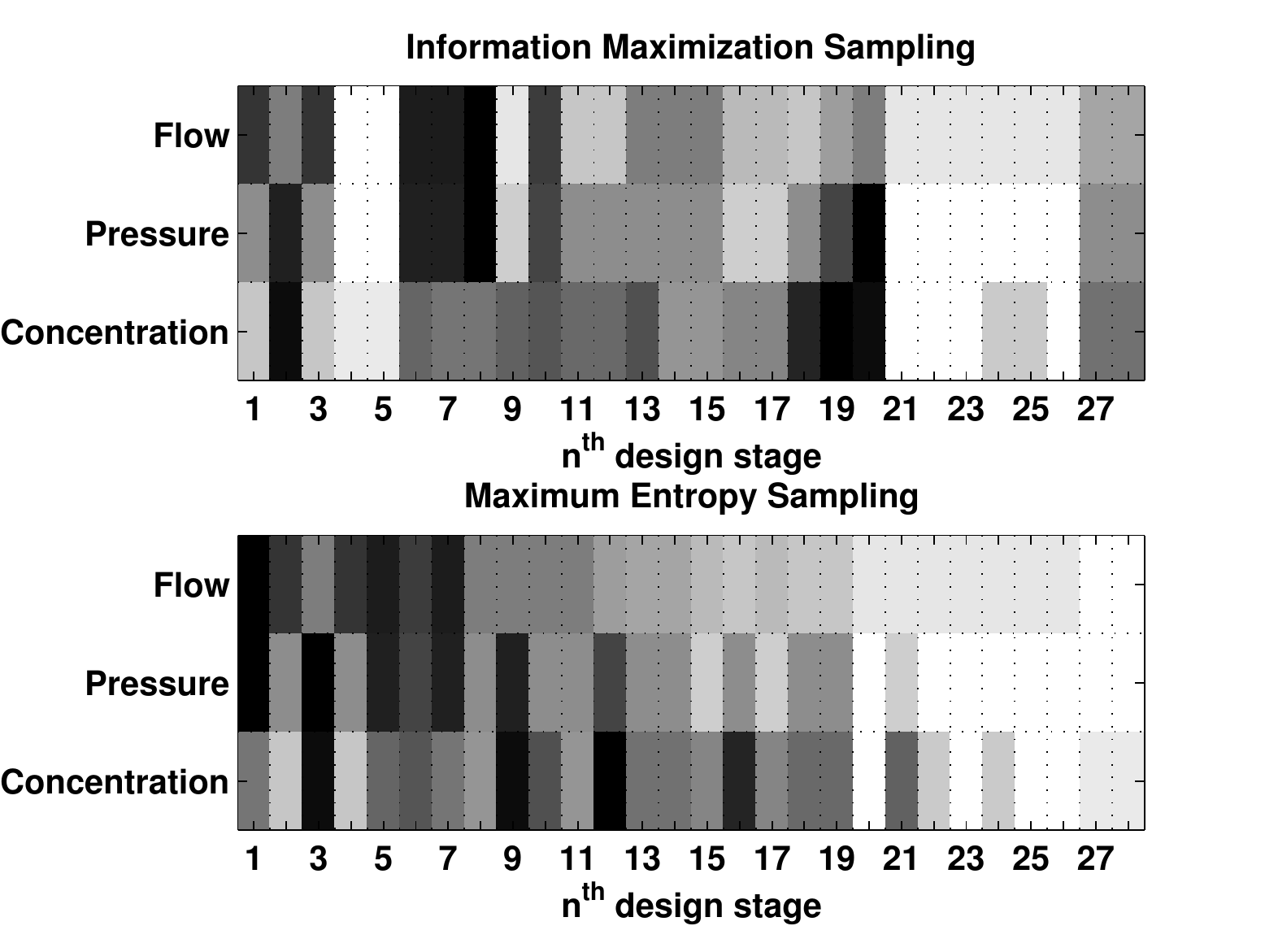}
  \end{center}
  \caption{Results for example $2$ (simulated data). Graphical representation for the
    inflow conditions selected by the two information theoretic strategies. Darker shadings
    indicate large inflow conditions such as flow, pressure, and concentration.}
  \label{fig:ex2_inflow}
\end{figure}

\end{document}